A Unified Framework for Producing CAI Melting, Wark-Lovering Rims and Bowl-Shaped CAIs

Kurt Liffman[1], Nicolas Cuello[2], David A Paterson[3],

1. Centre for Astrophysics and Supercomputing, Swinburne University of Technology, Hawthorn, Victoria, Australia.
2. Université de Lyon, F-69003 Lyon, France; Université Lyon 1, Observatoire de Lyon, 9 avenue Charles André, F-69230 Saint-Genis Laval, France; CNRS, UMR 5574, Centre de Recherche Astrophysique de Lyon; Ecole Normale Supérieure de Lyon, F-69007 Lyon, France.
3. Fluid Dynamics Laboratory, CSIRO, Highett, Victoria, Australia.

# Abstract

Calcium Aluminium Inclusions (CAIs) formed in the Solar System, some 4,567 million years ago. CAIs are almost always surrounded by Wark-Lovering Rims (WLRs), which are a sequence of thin, mono/bi-mineralic layers of refractory minerals, with a total thickness in the range of 1 to 100 microns. Recently, some CAIs have been found that have tektite-like bowl-shapes. To form such shapes, the CAI must have travelled through a rarefied gas at hypersonic speeds. We show how CAIs may have been ejected from the inner solar accretion disc via the centrifugal interaction between the solar magnetosphere and the inner disc rim. They subsequently punched through the hot, inner disc rim wall at hypersonic speeds. This re-entry heating partially or completely evaporated the CAIs. Such evaporation could have significantly increased the metal abundances of the inner disc rim. High speed movement through the inner disc produced WLRs. To match the observed thickness of WLRs required metal abundances at the inner disc wall that are of order ten times that of standard solar abundances. The CAIs cooled as they moved away from the protosun, the deduced CAI cooling rates are consistent with the CAI cooling rates obtained from experiment and observation. The speeds and gas densities required to form bowl-shaped CAIs are also consistent with the expected speeds and gas densities for larger, ~ 1 cm, CAIs punching through an inner accretion disc wall.

# 1 Introduction

Calcium Aluminium-rich Inclusions (CAIs) are sub-millimetre to centimetre sized, light grey, irregular to spherical-shaped rocks that are found in most primitive meteorites, but are most commonly found in carbonaceous chondrites. They are amongst the oldest rocks to form in the Solar System with an estimated age of 4567.30 ± 0.16 million years (Connelly et al. 2012). They formed over a relatively short time interval of 16,000 to 40,000 years (Jacobsen et al. 2008) and were partially formed via sub-micron condensates (Berg et al. 2009) from an $^{16}$O-rich gas of approximately solar composition in a high temperature environment of about 1600 K (ibid.) at a probable total pressure around $10^{-4}$ atm in a region close to the protosun (MacPherson et al. 2005, Liffman et al. 2012). The high formation temperatures, the presence of $^{10}$Be (Gounelle et al. 2013), and $^{16}$O enrichment close to solar values (McKeegan et al. 2011) are results consistent with CAI formation occurring at the interface region between the inner solar disc and the solar magnetosphere.

Although CAIs tend to have irregular or somewhat spherical shapes, Ivanova et al. (2014) provided examples of bowl-shaped CAIs from the CV chondrites North West Africa (NWA) 3118 and NWA 5950. As noted by Lorenz, Ivanova & Shuvalov (2012), the shape of these CAIs is consistent with the shapes obtained from molten droplets that are subject to hypersonic interaction with a rarefied gas. Ivanova et al. (2014) suggest that such high speed interaction may be due to shock waves generated by X-ray flares.

The shock wave formation mechanism is popular candidate for CAI melting and processing. It is a hypothesis that is often used to explain the formation of high temperature components that are found in meteorites, e.g. chondrules. However, theoretical analysis suggests that shock waves are too variable in their potential heating and cooling rates to reliably replicate the deduced heating and cooling rates of chondrules (Liffman 2009, Stammler & Dullemond 2014). Although shock waves are expected to exist in young stellar systems, none have yet been observed.

The only known, observed, transformation of protoplanetary dust into crystalline silicate: forsterite ($Mg_2SiO_4$), is due to radiative heating during a stellar outburst of the young stellar object (YSO) EX Lupi (Ábrahám et al. 2009). Ábrahám et al. were unable to find the expected infrared signature for shock-wave produced forsterite crystals. Intriguingly, later observations suggest that the newly formed forsterite crystals were moving radially away from EX Lupi, over the face of the protoplanetary disc, at speeds approaching 40 km s$^{-1}$ (Juhász et al. 2012).

This lack of observational evidence and negative theoretical analysis suggests that caution is required in adopting shock wave heating as the default formation mechanism for chondrules, CAIs, silicate dust and other high temperature materials that are found in meteorites. Indeed, it is more scientifically reasonable to use actual observations from protoplanetary systems in developing a theory for CAI processing. In this study, we will use the EX Lupi observations of radiative heating and subsequent ejection to provide us with an observationally-based mechanism for the formation of bowl-shaped CAIs.

In our proposed model, CAIs are ejected from the neighbourhood of the hot, inner edge of the Solar Protoplanetary Disc (or protosolar disc) where they are exposed to sufficient solar radiation to undergoing melting, evaporative loss and subsequent cooling as they move away from the protosun. This process of particle ejection from the inner protosolar disc may arise due to a jet flow – as has been observed by astronomers using the Spitzer Space Telescope (Poteet et al. 2011) and discussed by a number of theoreticians (Liffman & Brown 1995, Shu, Shang & Lee 1996, Bans & Königl 2012). However, there is another possible mechanism for ejecting material from near the protosun that occurs, somewhat paradoxically, when gas and dust accretes from the protosolar disc onto the protosun.

Observations indicate that the accretion of gas from a protoplanetary disc onto a protostar is typically accomplished via the stellar magnetosphere (Adams & Gregory 2012). Gas in the inner regions of the disc travels away from the midplane of the disc via the stellar magnetosphere onto the star. A fair portion of the gas is ionised and so is guided by and coupled to the stellar magnetic field. Dust and larger particles, however, have a much smaller charge to mass ratio relative to the gas and may not be highly coupled to the stellar magnetosphere. As the gas travels in a perpendicular direction away from the midplane of the disc the radial component of the stellar gravitational field decreases and is typically smaller than the required centripetal force to keep the particle co-moving with the infalling gas. As a consequence, particles are centrifugally flung from the accretional gas flow. We shall discuss this effect in Section 2.6.

A fair proportion of the ejected CAIs are likely to have a speed lower than the escape velocity of the YSO and they subsequently re-enter the protosolar disc either at the inner disc wall or at a radial distance further away from the Sun. These CAIs would experience heating during atmospheric re-entry into the protosolar disc. For sufficiently dense gas or high re-entry speeds the CAIs may again experience remelting, evaporative loss and/or surface modification.

The re-entry surface heating and the passage of the CAI through the hot gas at the inner disc wall are the reasons, we suggest, why most CAIs are surrounded by Wark-Lovering Rims (WLRs), which are a sequence of thin, mono/bi-mineralic layers of hibonite $((Ca,Ce)(Al,Ti,Mg)_{12}O_{19})$, perovskite $(CaTiO_3)$, spinel $(MgAl_2O_4)$, melilite $((Ca,Na)_2(Al,Mg,Fe^{2+})[(Al,Si)SiO_7].)$, anorthite $(CaAl_2Si_2O_8)$, pyroxene and olivine $((Mg, Fe)_2SiO_4)$, where the total thickness of the rim is in the range of 10 to 100 microns (Wark & Lovering 1977).

A schematic of our proposed CAI formation and processing scheme is shown in Fig. 1. A quantitative, theoretical description of our model is given in Section 2 with the results provided in Section 3. The Discussion section outlines the formation pathway for CAI processing in the inner disc. The Conclusions summarise the main results of this study.

## 2 Model

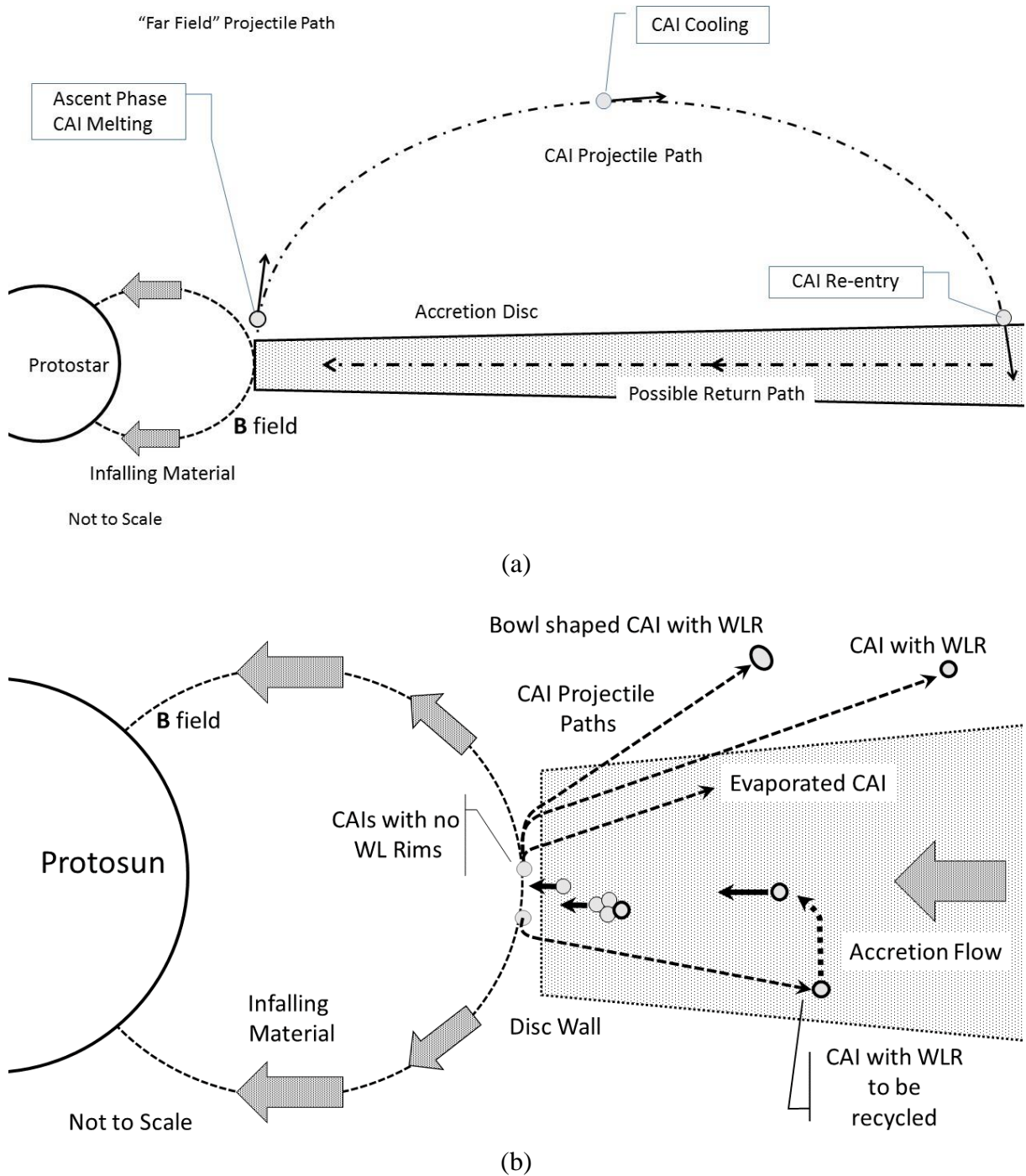

Fig. 1: Schematic depiction of the CAI formation model discussed in this paper. (a) CAIs are formed at the inner edge of the protosolar disc and ejected from the disc. The initial temperatures of the ejected CAIs may be above the CAI melting point, but they cool as they travel away from the Sun. Upon re-entry, the CAIs are subjected to a heat pulse. After re-entry, the CAI may return to the inner disc for further reprocessing. (b) The CAIs are formed near or in the inner rim of the protosolar disc. They are ejected from this region via the accretional inflow onto the protosun. As they pass through the hot inner disc region, they undergo evaporation and WLR formation. Some particles travel through the inner disc wall, while others are destroyed or sink to the midplane. The latter CAIs undergo recycling where they can agglomerate with other condensed/processed CAIs and be subsequently ejected or accrete onto the protosun.

In our formation model, CAIs begin as condensate by-products from the dust and gas in an optically thick region moving from the protosolar disc onto the protosun. A tentative, detailed model of how CAIs are gradually produced from condensates is provided in Liffman et al. (2012). Once CAIs are formed, the accretion process of moving gas and dust along the solar magnetosphere from the protosolar disc to the protosun (Adams & Gregory 2012) has the effect of moving the CAIs away from the midplane of the disc. This allows the radial centrifugal acceleration to move the CAIs away from the Sun, exposing them to direct solar radiation plus diffuse radiation from the disc and surrounding coronal atmosphere.

As the particles move along and then separate from the accretional flows, they are subject to radiative heating, which may be high enough to melt the CAIs. If the CAIs are ejected far enough from the inner disc region, then they will gradually cool as they move away from the Sun until they re-enter the protosolar disc (Fig. 1a). Depending on the re-entry distance from the protosun, the CAIs can move back to the inner disc wall and be subject to further reprocessing and ejection. As discussed in section 2.4, the return timescale due to disc accretion flow is upto of order hundreds of thousands of years. During this return phase, CAIs and other particles, on their way to the ejection region, may agglomerate with newly condensed CAI and/or other processed CAIs.

Prior to the CAIs leaving the inner disc region, they will tend to pass through the inner disc wall (Fig. 1b). During this process, the frictional heating can destroy the CAIs. As the CAIs pass through this hot, dense, gaseous inner disc region then they will be subject to evaporative processing plus condensation processes. It is this condensation of refractory materials that produces the WLRs observed around most CAI.

We note that in this paper the "outer regions of the disc" refers to any disc region that is not the inner disc wall region. We take this very general definition, because the ejection distance for the CAI moving away from the protosun may range from anywhere very close to the inner disc wall up to several au, or even outside the Solar System.

## 2.1 Length Scales

In attempting to understand the structure of the inner protosolar disc and the timescales for CAI formation, it is helpful to obtain some length scales of the inner protosolar disc and the protosun.

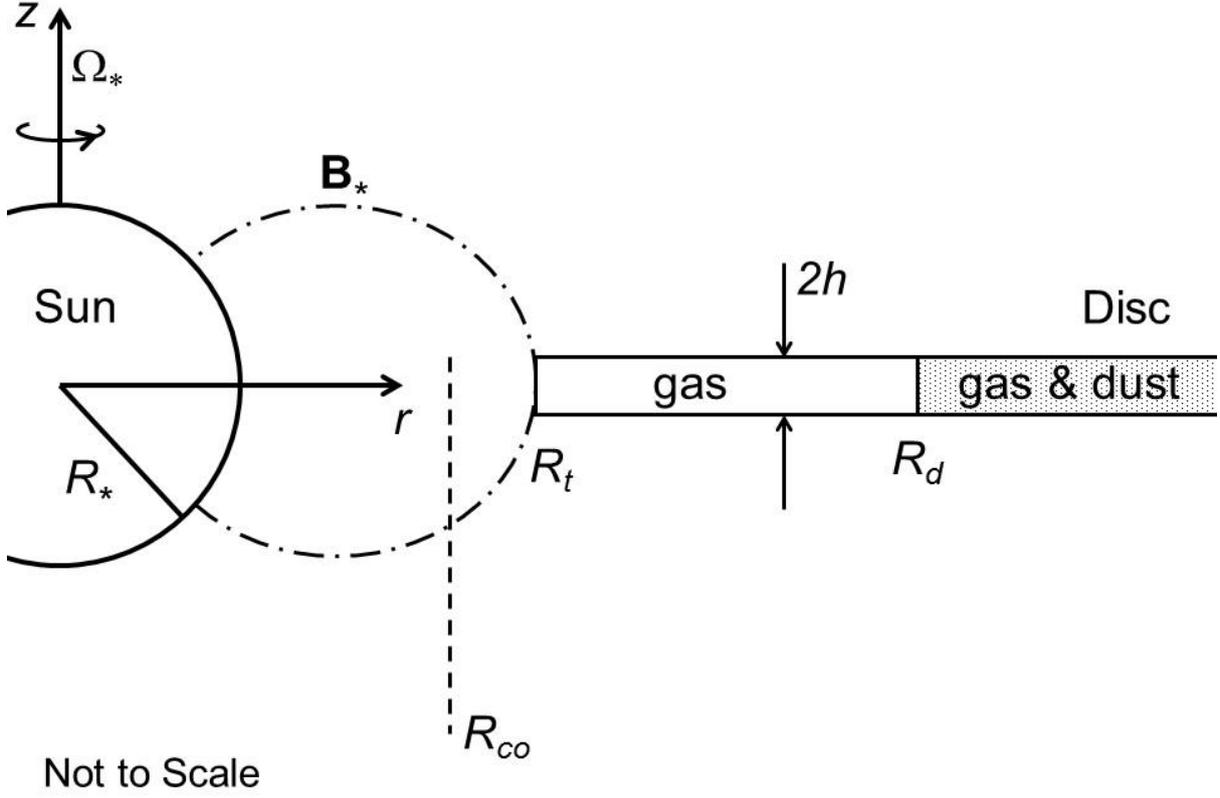

Fig. 2: Structure of the inner protosolar disc with $R_t$ - disc truncation radius, $R_*$ - radius of a star/protosun, $R_{co}$ - co-rotation radius, $R_d$ – silicate dust sublimation radius, $h$ – the isothermal scale height, B* - dipole component of the protosolar magnetosphere and $\Omega_*$ - angular rotational frequency of the Sun. CAIs are formed at or near $R_t$ and ejected over the disc or accreted onto the Sun. For illustrative purposes, the case of $R_t > R_{co}$ is shown.

A schematic summary of the structure of the inner protosolar disc is given in Fig. 2. Material from the inner disc accretes onto the star via stellar magnetic field lines (Königl 1991). Here, $R_t$ is the distance of the inner edge of the disc from the centre of a star, in the radial direction, $r$, where the stellar magnetosphere truncates the accretion disc (Ghosh & Lamb 1978)

$$R_t \approx \left( \frac{4\pi}{\mu_0} \frac{B_*^2 R_*^6}{\dot{M}_a \sqrt{GM_*}} \right)^{2/7} = 0.067 \left( \frac{(B_*(R_*)/0.1\,\mathrm{T})^2 (R_*/2R_\odot)^6}{(\dot{M}_a/10^{-8}\,\mathrm{M}_\odot\,\mathrm{year}^{-1})(M_*/M_\odot)^{1/2}} \right)^{2/7} \mathrm{au}, \quad (1)$$

$\mu_0$ is the permeability of free space, $B_*$ is the stellar magnetic field strength, $R_*$ is the stellar radius, $\dot{M}_a$ is the accretion rate of material onto the star, $G$ the universal gravitational constant and $M_*$ the mass of the star.

We assume that the mass accretion rate onto the star has the form:

$$\dot{M}_a(t) \approx \dot{M}_a(t_0) \left( \frac{t}{t_0} \right)^{-\eta}, \quad (2)$$

with $t > 10^4$ year, $t_0 = 10^6$ year, $\eta = 1.5$ and $\dot{M}_a(t_0) \approx 4 \times 10^{-8}\,\mathrm{M}_\odot\,\mathrm{yr}^{-1}$ (Hartmann et al. 1998).

The co-rotation radius, $R_{co}$, is the radial distance from the star where the angular rotation frequency of the star, $\Omega_*$, is equal to the keplerian angular frequency, $\Omega_{Kep}$, of the accretion disc. If we assume keplerian rotation for the disc then

$$\Omega_{Kep} = \left(\frac{GM_*}{R^3}\right)^{1/2} = 1.78 \times 10^{-5} \left(\frac{(M_*/M_\odot)}{(R/0.05\,\text{au})^3}\right)^{1/2} \text{Hz}, \quad (3)$$

and

$$R_{co} = \left(\frac{GM_*}{\Omega_*^2}\right)^{1/3} = 0.0716 \left(\left(\frac{M_*}{M_\odot}\right)\left(\frac{P_*}{7\,\text{days}}\right)^2\right)^{1/3} \text{au}. \quad (4)$$

In Eq. (4), the rotation period of the Sun, $P_*$, is normalised to a period of seven days, which is the median rotation period of 1 Myr old protostars listed in Bouvier et al. (2014).

An additional, radial length scale is the silicate dust sublimation radius, $R_d$. This is the distance between the inner disc rim and the centre of the protostar required to obtain a particular dust temperature, in this case 1500K, which is the approximate temperature for the sublimation of silicate dust. Of course, more refractory dust that has higher sublimation temperatures will still be able to exist at $r < R_d$. $R_d$ has the approximate formula (Espaillat et al., 2010):

$$R_d \approx \sqrt{\frac{3(L_* + L_a)}{16\pi\sigma_{SB}T_d^4}} \approx 0.06\,\text{au} \sqrt{\frac{(L_* + L_a)/L_\odot}{(T_d/1500\,\text{K})^4}}, \quad (5)$$

where, $\sigma_{SB}$ is the Stefan-Boltzmann constant, $T_d$ is the temperature of the dust, $L_*$ is the luminosity of the protosun (normalized to solar luminosity, $L_\odot$) and $L_a$ is the accretion luminosity:

$$L_a = \frac{GM_* \dot{M}_a}{R_*}\left(1 - \frac{R_*}{2R_t}\right)$$
$$= 3.1\,L_\odot \frac{(M_*/M_\odot)(\dot{M}_a/10^{-6}M_\odot\text{yr}^{-1})}{(R_*/10R_\odot)}\left(1 - \frac{R_*}{2R_t}\right). \quad (6)$$

In Eq. (6), we have normalised the equation for the case of the early protosun with a high accretion rate during the "Hayashi Track" stage of the Sun's evolution. The expected accretion luminosity of the protosun is greater than the standard solar luminosity. At the time of CAI formation, the Sun's luminosity was significantly higher than it is now. This can be seen in Fig. 3, where we show the expected luminosity and radius of the Sun during its first ten million years.

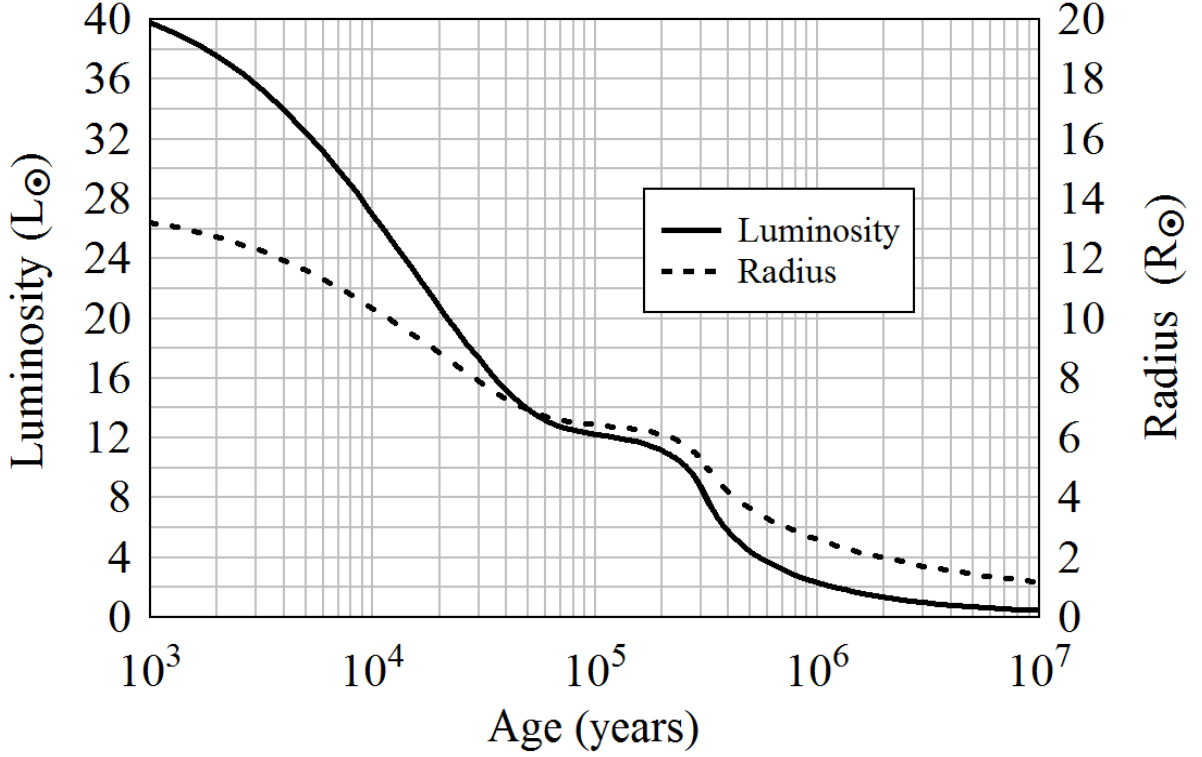

Fig. 3: Luminosity and radius of a Sun-like star (X = 0.703, Y = 0.277, Z = 0.020) for the first 10 million years (Siess, Dufour & Forestini 2000).

The isothermal scale height of the disc is given by

$$h(r) = \sqrt{\frac{2r^3 k_B T_D}{GM_* \bar{m}}} = 0.0013\,\text{au}\sqrt{\frac{(r/0.05\,\text{au})^3 (T_D/1500\,\text{K})}{(M_*/M_\odot)(\bar{m}/m_{H_2})}}, \quad (7)$$

with $k_B$ – Boltzmann's constant, $T_D$ the temperature of the inner disc, $r$ the cylindrical radial distance from the centre of the star and $\bar{m}$ the mean molecular mass of the gas. For this paper we have assumed that $\bar{m} \approx m_{H_2}$.

In calculating the isothermal scale height, the gas temperature is assumed to be approximately the same as the disc surface temperature, where the temperature of an optically thick, flat disc subject to stellar radiation (Friedjung 1985, Hartmann 1998) and differential friction in the accretion disc (Frank et al. 2002 ) is

$$T_{disc}(r) \approx \left( \frac{(L_* + L_a)}{4\pi^2 \sigma_{SB} R_*^2} \left( \sin^{-1}\left(\frac{R_*}{r}\right) - \frac{R_*}{r}\sqrt{1-\left(\frac{R_*}{r}\right)^2} \right) + \frac{3GM_* \dot{M}_a}{8\pi r^3 \sigma_{SB}}\left(1 - \left(\frac{R_*}{r}\right)^{1/2}\right) \right)^{1/4}. \quad (8)$$

We can use the isothermal scaleheight and the disc temperature to deduce the disc gas mass density $\rho$ as a function of $r$ and $z$:

$$\rho_{disc}(r,z) = \rho_c(r)\exp\left(-\left(\frac{z}{h(r)}\right)^2\right), \quad (9)$$

with $\rho_c(r)$ the midplane mass density of the disc gas, which has the definition

$$\rho_c(r) = \frac{\Sigma(r)}{h(r)\sqrt{\pi}}. \quad (10)$$

$\Sigma(r)$ is the surface column density of the disc. In this study, we use the minimum mass protosolar disc surface density of Hayashi (1981):

$$\Sigma(r) \approx 1700\left(\frac{\mathrm{au}}{r}\right)^{3/2}\,\mathrm{g\,cm}^{-2}. \quad (11)$$

These equations allow us to compute the pressure of the gas in the disc, $P_{disc}$, via the ideal gas equation:

$$P_{disc} = \frac{k_B \rho_{disc} T_{disc}}{\overline{m}}. \quad (12)$$

The resulting disc pressures and temperatures are displayed in Fig. 4.

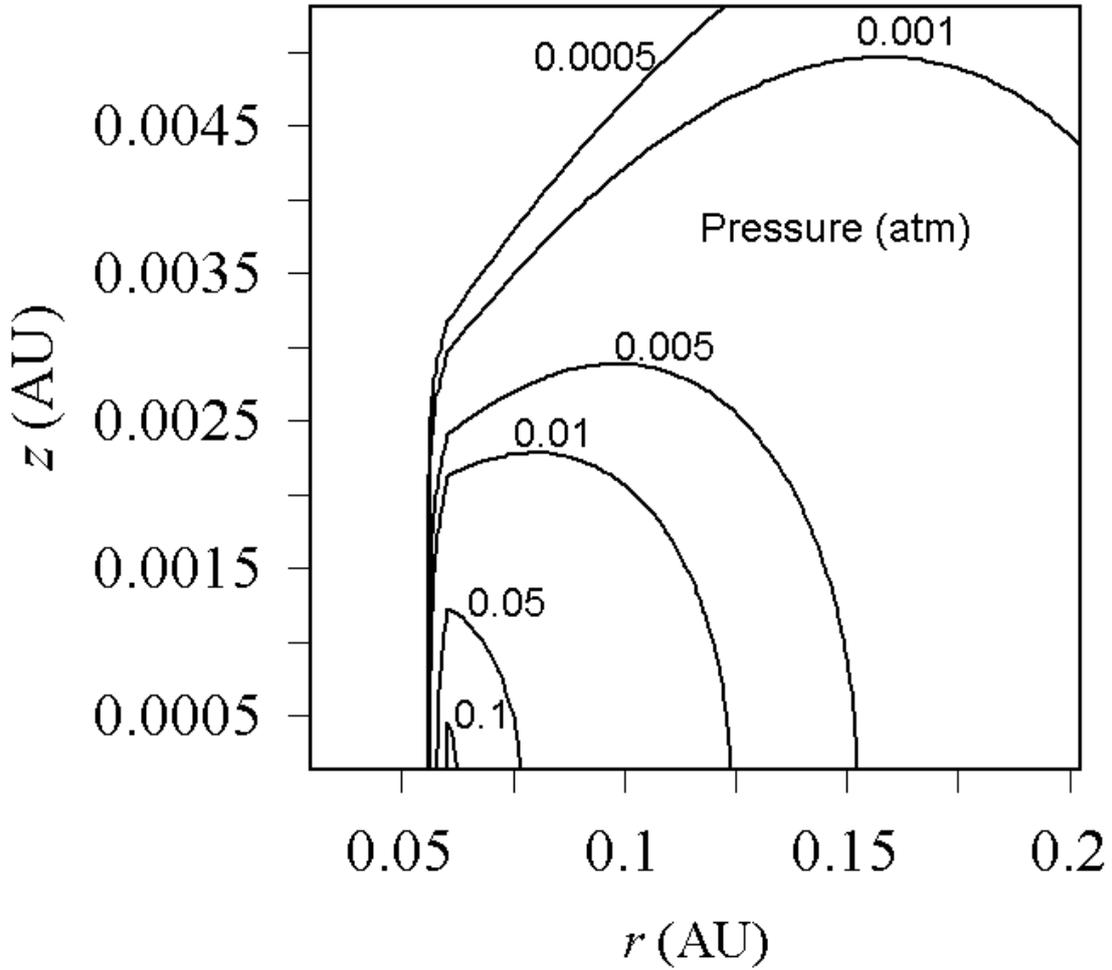

(a)

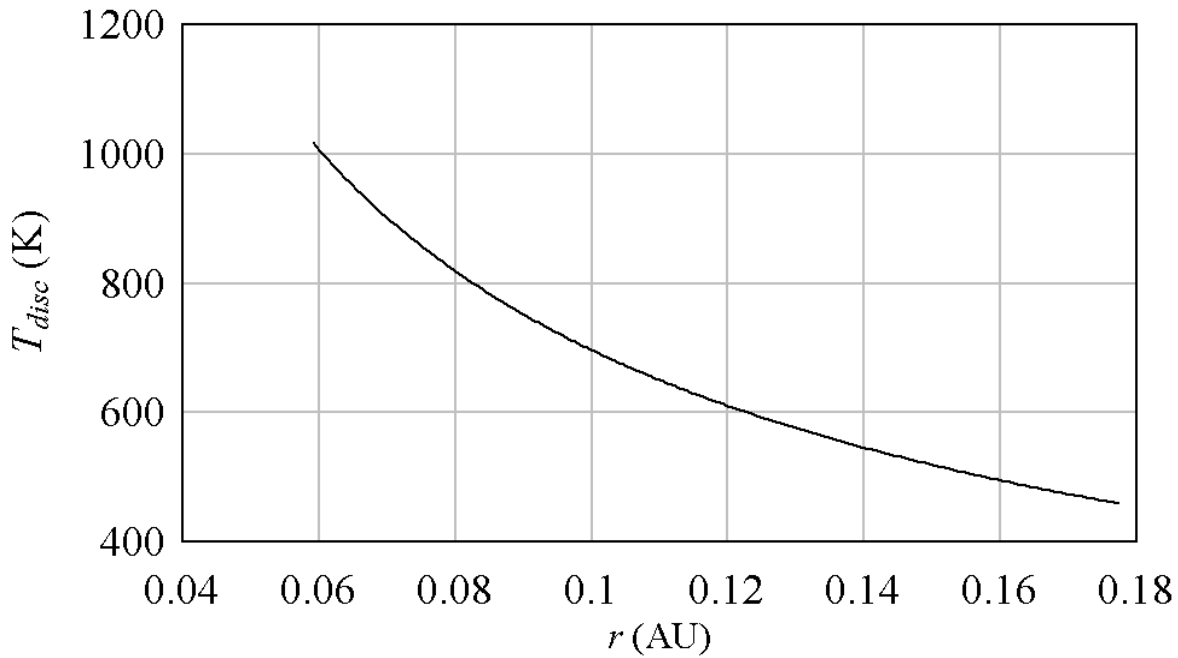

(b)

Fig. 4: (a) Gas pressure in the protosolar disc at a solar age of $1.2 \times 10^6$ years and a mass accretion rate from the disc to the protosun of $3 \times 10^{-8}$ $M_\odot yr^{-1}$ (b) Gas temperature of the protosolar disc. The highest gas temperatures occur at the inner rim of the disc.

## 2.2 CAI Formation Timescales

When discussing the formation of CAIs, it is important to determine when the CAIs were produced as this will determine the expected luminosity and size of the protosun when CAIs formed. Any coherent model of CAI formation should give this timescale. We claim this can be done by applying the model that CAIs are due to the accretion of dust and gas from the protosolar disc onto the protosun. In particular, this model should allow us to determine when this accretional gas had a pressure in the range of $10^{-5}$ to $10^{-3}$ atmospheres, which are the canonical pressures for optimal CAI formation.

The infalling gas and dust will tend to flow along the stellar dipole field lines with an initial speed, $v_a$, in the z direction in an (assumed) axisymmetric channel of initial width $\Delta$. The expected values for $v_a$ and $\Delta$ are uncertain at present, as these quantities are dependent on how the stellar magnetic fields diffuse into an accretion disc, the physics of which is still controversial. To obtain some representative values, we have adopted the standard boundary layer value for $\Delta$, where the stellar magnetosphere at $R_t$ replaces the surface of a compact object (Frank, King & Raine 2002):

$$\Delta \approx \frac{h(R_t)^2}{R_t} = 2 \times 10^{-5} \text{au} \frac{(h/0.001\,\text{au})^2}{(R_t/0.05\,\text{au})} . \quad (13)$$

Such a definition implies that $\Delta \ll h(R_t)$. Campbell (2010) suggests $\Delta \lesssim h(R_t)$. So by adopting Eq. (13), we are probably underestimating the width of the accretion flow channel and overestimating the accreting gas density, $\rho_a$.

The conservation of mass implies that

$$\frac{\dot{M}_a}{2} = 2\pi R_t \Delta \rho_a v_a. \tag{14}$$

Combining Eqs (13) and (14) gives:

$$\rho_a = \frac{\dot{M}_a}{4\pi h (R_t)^2 v_a}, \tag{15}$$

The initial infall velocity, $v_a$, in the $z$ direction is uncertain and in this study we will use a range of parameterisations from 1 to 10 km s$^{-1}$. To compute the pressure, we also need the temperature of the infalling gas. As a surrogate for the gas temperature, we use the effective particle temperature as the particle has to condense from the gas. From Eq. (5) the particle temperature $T_p$ at the inner rim of the protosolar disc is

$$T_p \approx \left( \frac{3(L_* + L_a)\varepsilon_a}{16\pi R_t^2 \sigma_{SB} \varepsilon_e} \right)^{1/4} = 1645\,\mathrm{K} \left( \frac{((L_* + L_a)/L_\odot)\varepsilon_a}{(R_t/0.05\,\mathrm{au})^2\,\varepsilon_e} \right)^{1/4}, \tag{16}$$

with $\varepsilon_a$ the radiation absorption coefficient and $\varepsilon_e$ is the particle radiation emission coefficient. In this formula, we have set $R_d = R_t$ and, in general, we will set $\varepsilon_a \approx \varepsilon_e$.

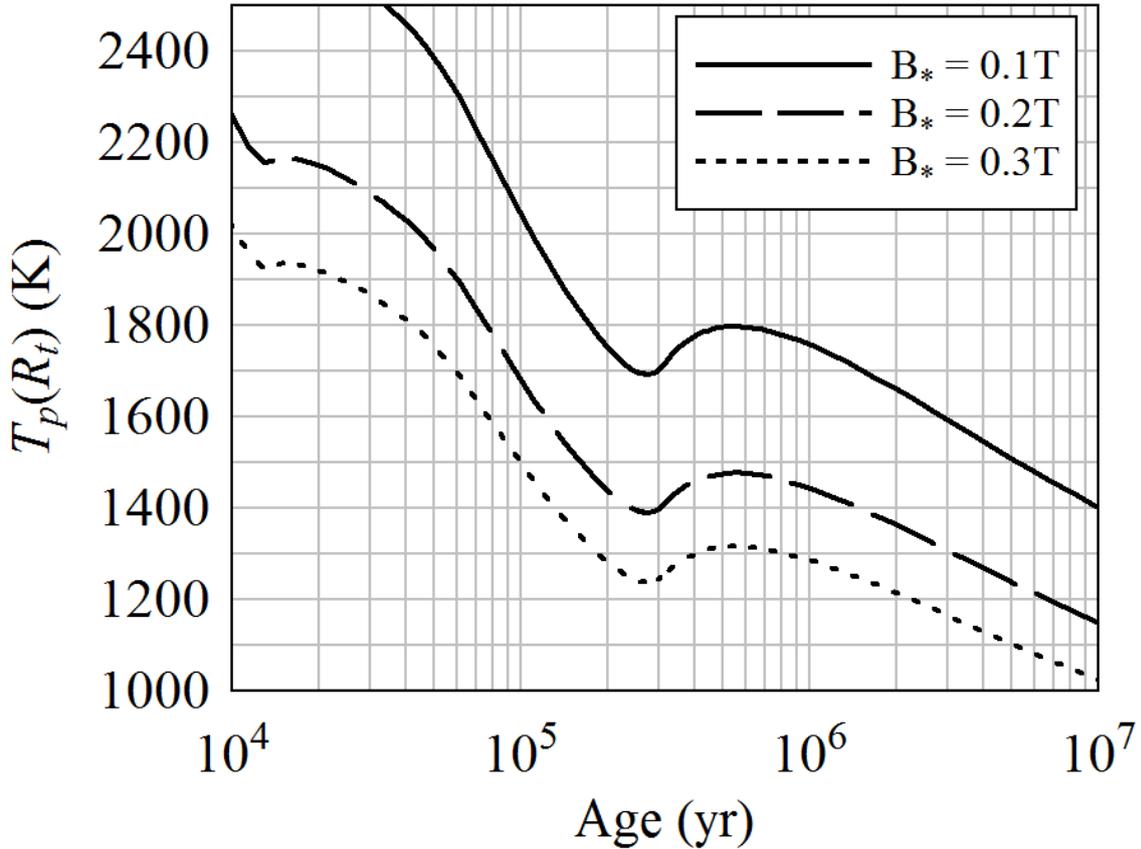

Fig. 5: $T_p(R_t)$ is the approximate, average temperature of a particle, in Kelvin, due to direct exposure to radiation from the Sun and reflected radiation from the inner rim of the protosolar disc. Temperatures are shown for three different values of the protosolar magnetic field, $B_*$, at the surface of the protosun. The strength of the protosolar field, given in Teslas, determines the stand-off distance of the inner disc rim from the protosun. The stronger the field the further away is the inner rim and the lower the particle temperatures. Temperatures

close to CAI formation temperatures tend to occur in the age range of 20,000 to 2,000,000 years – depending on the value of the stellar magnetosphere.

The results of these calculations are shown in Fig. 5. As can be seen, the temperatures are dependent on the stellar magnetic field strength. Between $10^4$ and $10^6$ years, the particle temperatures are close to the CAI formation temperatures (Berg et al. 2009). For these calculations, we have taken the radius and luminosity values for the pre-main sequence Sun as a function of time as shown in Fig. 3 and as given by Siess et al. (2000). For our calculation of the inner radius of the disc as given by Eq. (1), we have used the mass accretion rates given by Eq. (2) and also calculated the increasing mass of the Sun over time as determined by the accretion rates (Fig. 6). We have also assumed a solar rotational period of 8 days and a magnetic field strength in the range of (0.1 to 0.3 T) (Güdel 2007).

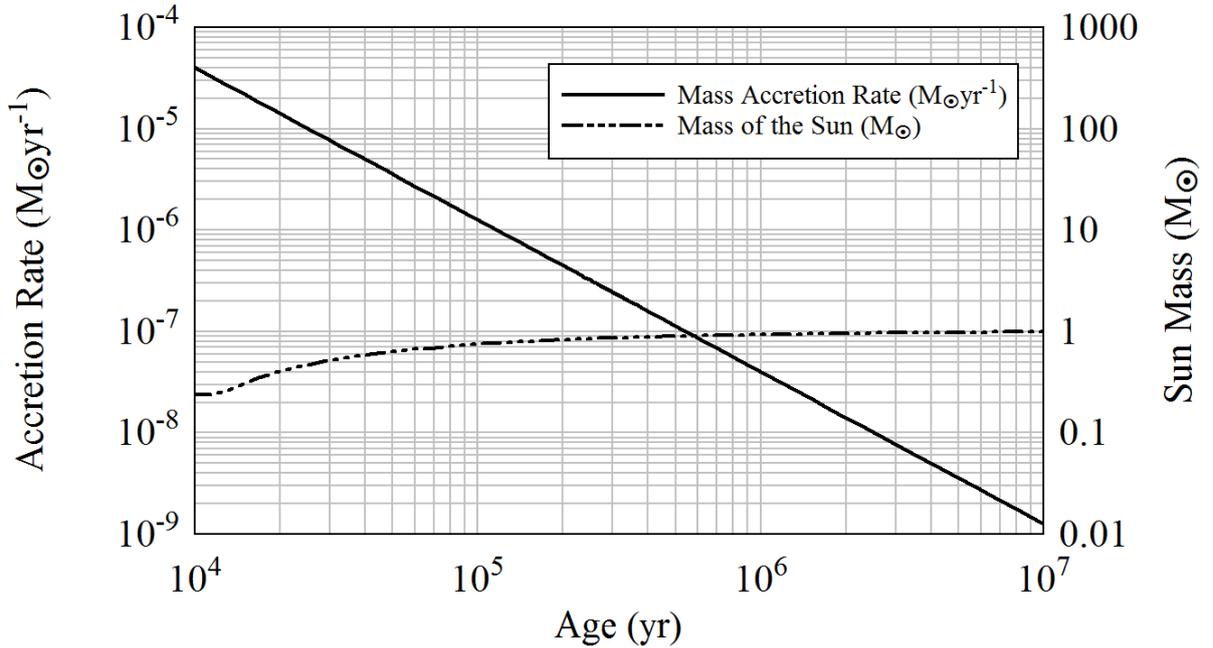

Fig. 6: Mass accretion rate from the protosolar disc onto the protosun and the mass of the protosun as a function of time.

Combining Eqs (15) and (16), we can obtain an expression for the approximate pressure, $P_a$, for the infalling, accreting, ideal gas in the near neighbourhood of the disc:

$$P_a = \frac{k_B \rho_a T_p}{\bar{m}} \approx \frac{k_B \dot{M}_a \left(3(L_* + L_a)\varepsilon_a\right)^{1/4}}{4\pi h (R_t)^2 v_a \bar{m} \left(16\pi R_t^2 \sigma_{SB} \varepsilon_e\right)^{1/4}}. \quad (17)$$

In Fig. 7, $P_a$ is shown as a function of time, the strength of the stellar magnetosphere at the surface of the protosun, $B_*(R_*)$, and as a function of the initial accretion infall speed, $v_a$. Depending on the values of these quantities, the pressures obtained in the accreting gas around the 10,000 to 3,000,000-year mark are similar to the $10^{-4}$ atmospheres (10 Pa) suggested for CAI formation (Berg et al. 2009, Blander et al. 1979). This result shows that CAIs could have potentially formed in the accretion columns between the protosolar disc and the protosun during the entire lifetime of the protosolar disc.

However, in our model, the production rate of CAIs is dependent on the mass accretion rate from the protosolar disc to the protosun. As we show in the next section, the production rate of CAIs decreases significantly with time, so even if CAIs were produced over a million year time frame there would be a bias towards CAIs that were formed at earliest times when the mass accretion rates were higher.

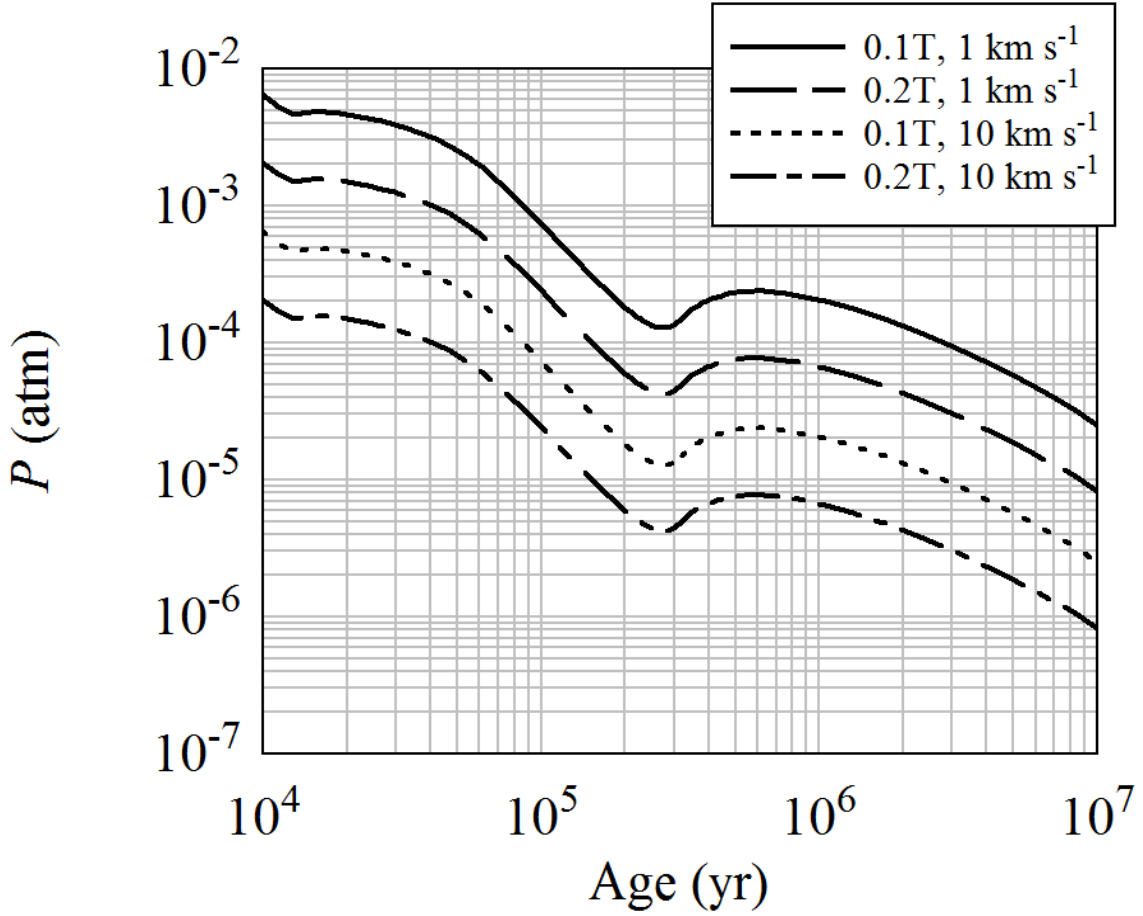

Fig. 7: The pressure, $P_a$, of the infalling accreting gas as it leaves the inner radius of the accretion disc on its journey to the protosun as a function of the age of the protosolar disc. The pressures are shown as a function of the infall speed from the disc surface and the stellar magnetic field strength at the surface of the Sun. The infall gas pressures in the first three million years appear to be roughly consistent with the deduced CAI condensation pressure of around $10^{-4}$ atmospheres.

## 2.3  The Total Mass of CAIs

In our model, CAIs are formed in or near the inner rim of the protosolar disc. The CAIs are ejected to the outer regions of the protosolar disc possibly via the centrifugal ejection mechanism outlined in Section 2.6. Using this simple idea, it is possible to derive a general formula for the total mass of CAIs that were ejected from the inner protosolar disc. This will provide us with an upper estimate for the total mass of CAIs that were incorporated into planetesimals.

The mass ejection rate of CAIs from the inner protosolar disc, $\dot{M}_{CAI}$, is given

$$\dot{M}_{CAI}(t) \approx f_{CAI} f_e \dot{M}_a(t),  \tag{18}$$

where $f_e$ is the fraction of the accreted mass that is ejected from the inner protosolar disc, while $f_{CAI}$ is the proportion of this ejected material that can, theoretically, finish up incorporated into a CAI.

The value of $f_e$ is somewhat uncertain. Typically, authors set $f_e \approx 0.01$ to $0.1$, a value range that is based on observations of young stellar systems, e.g., Hartigan, Morse & Raymond (1994), Hartigan, Edwards & Ghandour (1995). We wish to derive an upper bound for the total mass of CAIs in the protosolar disc, so we will set $f_e \approx 0.1$.

To estimate $f_{CAI}$ or an upper bound to this quantity, we note that CAIs consist of minerals that are rich in Ca, Al, Mg, Si and O. If we take Ca as our base element then in terms of solar abundances the mass proportion of Ca: $m_{Ca}$ is $m_{Ca} = 6.69 \times 10^{-5}$ (Anders & Ebihara 1982). From the minerals that tend to be found in CAI (MacPherson et al. 2005), we can make the following rough estimates: $m_{Al} \sim m_{Ca}$; $m_{Si} \sim 3 \times m_{Ca}$; $m_O \sim 6 \times m_{Ca}$ and $m_{Mg} \sim 2 \times m_{Ca}$. So the mass abundance consumed by CAIs will have the approximate upper bound of

$$f_{CAI} \approx m_{Ca} + m_{Al} + m_{Si} + m_{Mg} + m_O \approx 13 m_{Ca} < 20 m_{Ca} = 1.34 \times 10^{-3}. \qquad (19)$$

CAIs were formed over a time period, which we denote by $\tau_{CAI}$. This formation time period is thought to be around 20,000 years (Jacobsen et al. 2008). So the total mass of CAIs produced during this time is

$$M_{CAI} \approx \int_{t_s}^{t_s + \tau_{CAI}} f_{CAI} f_e \dot{M}_a(t) dt, \qquad (20)$$

where $t_s$ is starting time for CAI formation with $t = 0$ the time zero for the formation of the protosun. Using Eq. (2), we have the general formula:

$$M_{CAI} \approx \frac{f_{CAI} f_e \dot{M}_a(t_0) t_0^\eta}{\eta - 1} \left( \frac{1}{t_s^{\eta-1}} - \frac{1}{(t_s + \tau_{CAI})^{\eta-1}} \right). \qquad (21)$$

For the case $\tau_{CAI} \ll t_s$

$$M_{CAI} \approx f_{CAI} f_e \dot{M}_a(t_s) \tau_{CAI}. \qquad (22)$$

In Fig. 8, we plot the expected total production of CAIs as a function of time as determined from Eq. (21) with $f_e = 0.1$, $f_{CAI} = 1.34 \times 10^{-3}$ and $\tau_{CAI} = 20,000$ years. For example, if the formation of CAIs started at $t_s = 10^5$ years then the total production of CAIs during a 20,000 year time frame is approximately $3 \times 10^{-6}$ $M_\odot$, which is about 3% of the total rock mass of the planets: $10^{-4}$ $M_\odot$, (Hubbard & Marley 1989). So, in this scenario, CAIs make up 3% of the rock mass of the Solar System. At a formation age of $10^6$ years, total CAI production was approximately 0.1% of the total rock mass of the planets

The observed abundances for CAIs in meteorites are < 3% for carbonaceous chondrites and nearly 0% for all other primitive meteorites (Hezel et al. 2008). This may indicated that $10^{-6}$ $M_\odot$ of CAIs is an upper bound on the total production of CAIs.

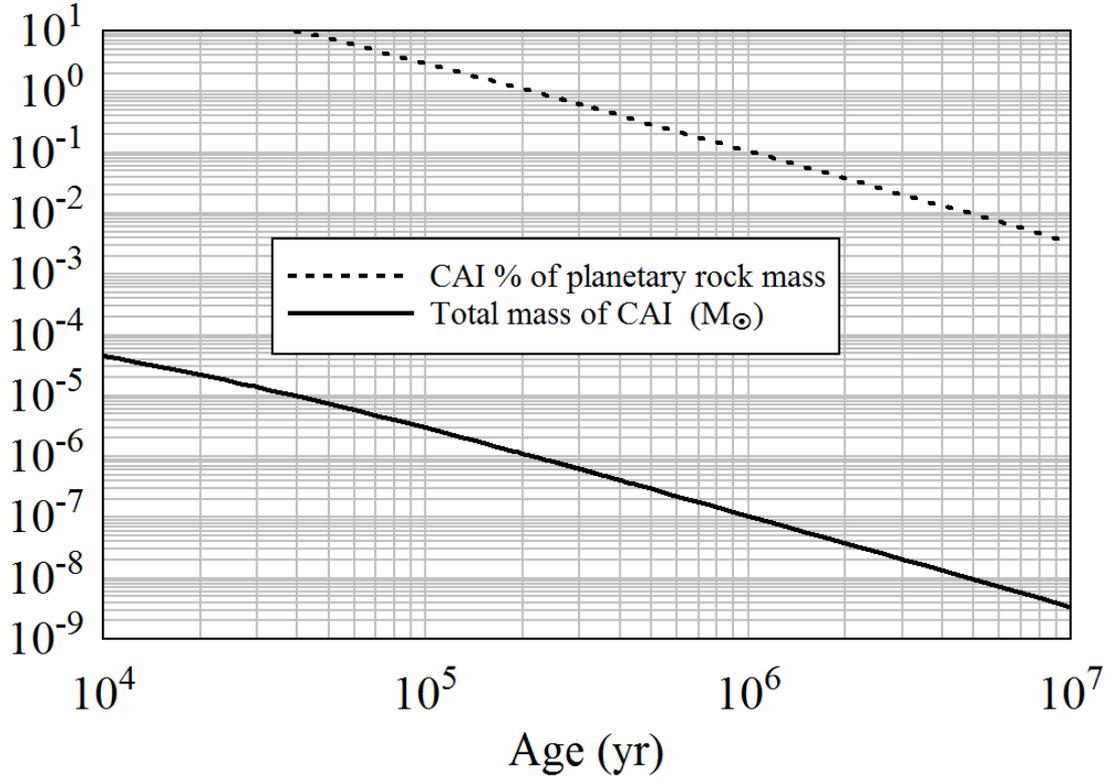

Fig. 8: Total mass of CAIs produced over a 20,000 year time period in absolute value (solid line) or in percentage terms relative to the total mass of rock in the Solar System, where the time is start time of production. So, for example, at $10^5$ years the production of CAIs started at 100,000 years and stopped at 120,000 years, where about $3\times10^{-6}$ $M_\odot$ of CAIs were produced, which is about 3% of the total rock mass of the planets: $10^{-4}$ $M_\odot$. At a formation age of $10^6$ years, total CAI production decreases to 0.1% of the total rock mass of the planets.

## 2.4 Disc Gas Accretion Timescales

As illustrated in Fig. 1, the major thesis of this paper is that the CAIs were launched from the inner rim of the protosolar disc and landed in regions of the protosolar disc that were further away from the protosun. Implicit in this model is that the ejected particles could return from the outer regions of the protosolar disc to the inner rim via radial particle drift and/or gas accretion. Thus, via this ejection and return process, CAIs, and other particles, were subject to cyclic reprocessing.

Inward radial particle drift occurs when the pressure gradient in the disc causes the gas to have a lower angular velocity relative to the keplerian angular velocity of the particle. The subsequent particle-gas drag causes the particles to lose angular momentum and they fall towards the protosun. Particle drift has been well documented in the literature for many years (e.g., Weidenschilling 1977) and is only briefly mentioned here.

Such hypothetical particle drift assumes that the disc gas pressure decreases as one moves away from the star. This is a reasonable, first order assumption that may be true for a fair section of the disc, but is certainly not true at the inner disc edge. Also, if a local pressure maximum can be induced within the disc then a particle trap arises, where the particles, on average, move towards the protostar at the accretion rate. Such particle traps have been discussed for decades and have recently been observed (e.g., van der Marel *et al.* 2013). We also note that CAIs range in diameter from less than 100 microns to more than a centimetre (Chaumard et al. 2014). The smaller CAIs will be less affected by particle drift and will be more coupled to the disc gas (Laibe et al. 2012). For both these reasons, it is reasonable to use the gas accretion timescale as a lower bound on the migration timescale for a CAI.

The mass accretion rate in an accretion disc is given by

$$\dot{M}_a = -2\pi r \Sigma v_r, \tag{23}$$

where $v_r$ is the height-averaged, radial infall speed of the disc gas. If we assume that the surface column density $\Sigma$ has the form (e.g., see Eq. (11))

$$\Sigma(r) = \Sigma_0 \left(\frac{r_0}{r}\right)^\beta \tag{24}$$

then it is possible to combine these two equations to compute the accretion timescale, $\tau_a$, for gas to travel from a distance $r$ in the disc to the inner truncation radius $R_t$:

$$\tau_a = \frac{2\pi \Sigma_0}{\dot{M}_a (2-\beta)} \left( r^2 \left(\frac{r_0}{r}\right)^\beta - R_t^2 \left(\frac{r_0}{R_t}\right)^\beta \right) \quad \text{for } \beta \neq 2 \tag{25}$$

or

$$\tau_a = \frac{2\pi \Sigma_0 r_0^2}{\dot{M}_a} \ln\left(\frac{r}{R_t}\right) \quad \text{for } \beta = 2. \tag{26}$$

From Eq. (11), $\Sigma_0 = 1700 \text{ g cm}^{-2}$, $r_0 = 1 \text{ au}$ and $\beta = 1.5$. The resulting gas accretion timescales as a function of accretion rates are shown in Fig. 9, where we see that the gas accretion timescales decrease for increasing accretion rates. From this figure, if the particle is located at a distance of 0.2 au from the protosun then it takes a maximum of 5,000 years for the particle to arrive at the inner truncation radius (~0.05 au) assuming an accretion rate of $10^{-7}$ M$_\odot$ year$^{-1}$. In obtaining this timescale, we have ignored any other particle drift mechanism. If the accretion rate is, say, $10^{-9}$ M$_\odot$ year$^{-1}$ then it takes a maximum of 300,000 years to travel from 0.3 au to the inner truncation radius located at ~0.2 au).

These results open the possibility that a CAI may have been ejected to the outer regions of the protosolar disc and then returned some tens to hundreds of thousands of years later for further processing.

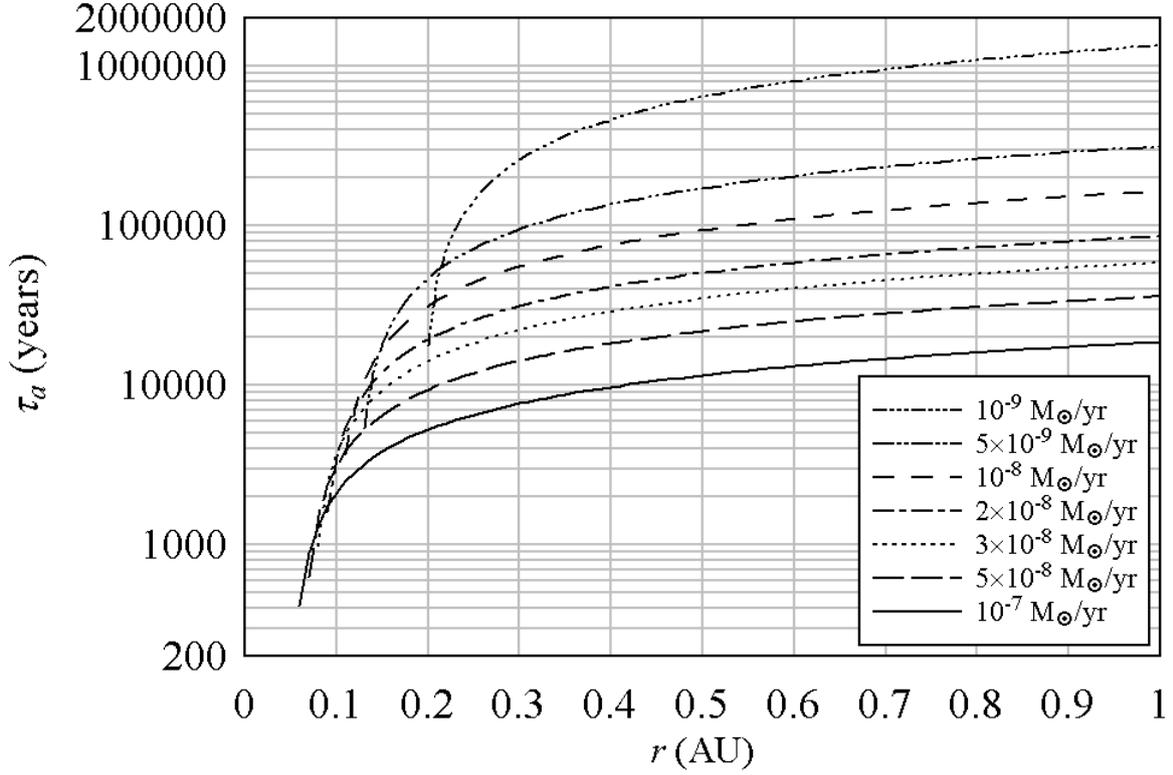

Fig. 9: Time taken for gas to flow from a distance *r* to the inner truncation radius of the protosolar accretion disc. The travel timescales are smallest for the highest accretion rates. As an example, for a particle that is located at 0.2 au from the protosun then it will take a maximum of 5,000 years for the particle to arrive at the inner truncation radius assuming an accretion rate of $10^{-7}$ $M_\odot$ year$^{-1}$.

## 2.5 Particle/CAI Equation of Motion

In our model the CAI particles are, initially, in a circular keplerian orbit at or near the inner truncation radius of the protosolar disc. As discussed in the Section 1 and shown schematically in Fig. 10, the accretional inflow and/or the protostellar jet flow gives the particles an initial "boost" velocity that is assumed to be primarily in the *z* direction as this is perpendicular to the disc midplane. If we assume that the self-gravity of the disc is negligible compared to the gravity of the protostar then the equations of motion for a particle in the cylindrical coordinate $r, \phi$ and *z* directions are:

$$m_p \ddot{r}_p = m_p r_p \dot{\phi}_p^2 - \frac{GM_* m_p r_p}{[r_p^2 + z_p^2]^{3/2}} - \frac{C_D}{2} \rho_g \pi a_p^2 v_{pg}^2 \hat{v}_{pg} \cdot \hat{r} , \qquad (27)$$

$$m_p \left( r_p \ddot{\phi}_p + 2 \dot{r}_p \dot{\phi}_p \right) = -\frac{C_D}{2} \rho_g \pi a_p^2 v_{pg}^2 \hat{v}_{pg} \cdot \hat{\phi}, \qquad (28)$$

and

$$m_p \ddot{z}_p = -\frac{GM_* m_p z_p}{[r_p^2 + z_p^2]^{3/2}} - \frac{C_D}{2} \rho_g \pi a_p^2 v_{pg}^2 \hat{v}_{pg} \cdot \hat{z}, \tag{29}$$

With $r_p$, $\phi_p$ and $z_p$ the cylindrical coordinates of the CAI particle, $\rho_g$ the average mass density of the gas, $C_D$ the drag coefficient for the interaction between the gas and the dust particle, $\boldsymbol{v}_g$ and $\boldsymbol{v}_p$ are the gas flow velocity and dust velocity, respectively, $\boldsymbol{v}_{pg} = \boldsymbol{v}_p - \boldsymbol{v}_g$. Symbols with a caret are unit vectors, while $m_p$ is the mass of an individual, approximately spherical, CAI, so

$$m_p \approx \frac{4}{3} \pi a_p^3 \rho_p, \tag{30}$$

with $a_p$ the average CAI radius and $\rho_p$ the average mass density of the CAI. The azimuthal gas speed $v_{g\phi}$ is approximately $r\Omega_*$, as we assume that the gas flow arises at the inner truncation radius and the gas flow is coupled to the protostellar magnetosphere, which is, to a first approximation, co-rotating with the protostar.

The drag coefficient, $C_D$, is given by (Probstein 1968, Hayes & Probstein 1959)

$$C_D(S) = \frac{2}{3S}\sqrt{\frac{\pi T_p}{T_g}} + \frac{2S^2 + 1}{\sqrt{\pi} S^3} \exp(-S^2) + \frac{4S^4 + 4S^2 - 1}{2S^4} \operatorname{erf}(S), \tag{31}$$

where $T_p$ is the temperature of the CAI particle, erf the error function, exp the exponential function, $S$ is the thermal Mach number:

$$S = |\boldsymbol{v}_{pg}|/v_T, \tag{32}$$

with $v_T$ the thermal gas speed:

$$v_T = \sqrt{2k_B T_g / \bar{m}}. \tag{33}$$

This gas drag coefficient is applicable in the free-molecular flow regime, where the collisional mean free path of the gas is larger than the diameter of the CAI particle. It can be shown that, for this study, free molecular flow is a good approximation for the gas/particle interaction due to the expected, very low gas densities in the protosolar disc (Fig. 2 of Liffman 1992).

## 2.6 Radial Motion

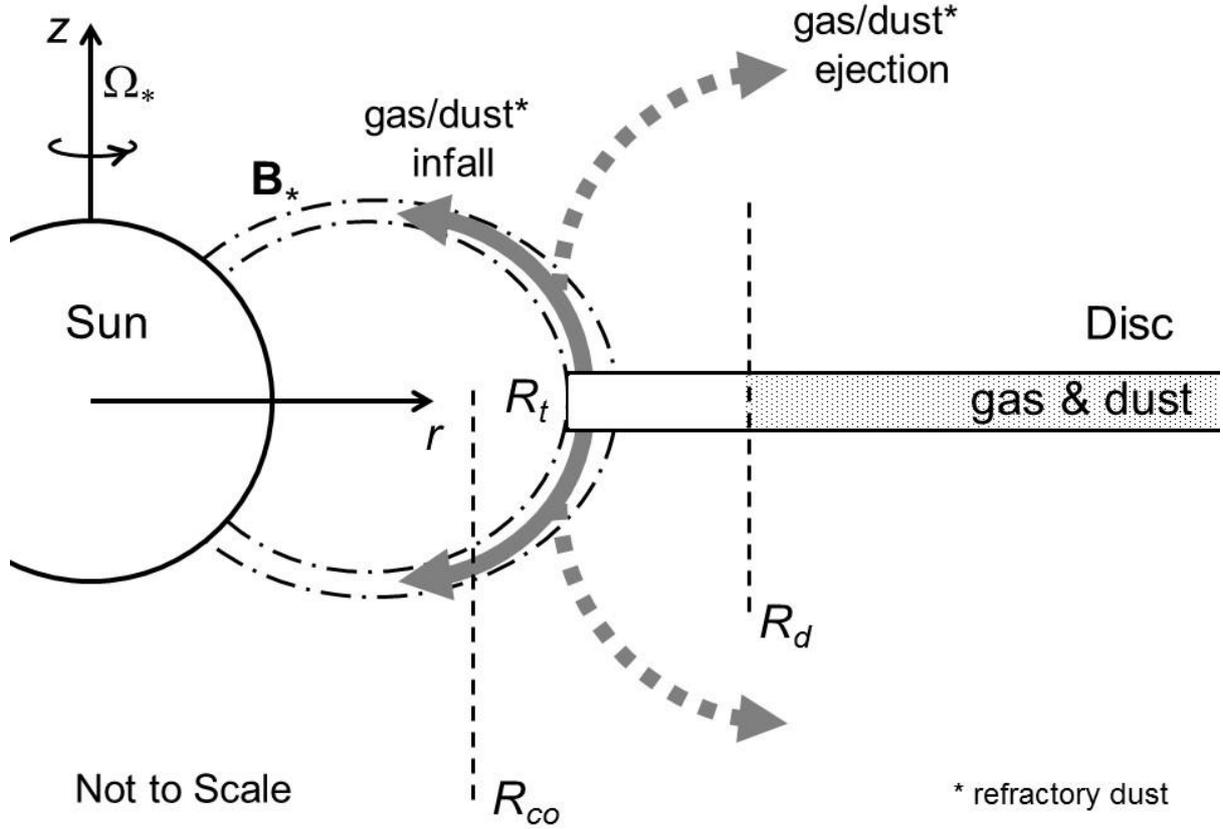

Fig. 10: A schematic of refractory particle movement as they flow along the accretion stream from the disc to the star. The centrifugal forces on the particles may eject them from the accretion stream and away from the protostar as they move away from the disc plane. This is particularly true if $R_t > R_{co}$, as the co-rotating angular speed of the magnetic field at the truncation radius will be greater than the keplerian angular speed.

Particles may undergo significant radial motion away from the protosun if they are lifted a sufficient distance away from the midplane of the protosolar disc (Fig. 10). To see this, we initially consider the angular acceleration equation Eq. (28) for the case where the particle is above the disc and the effect of gas drag can be neglected. In this case, we have

$$r_p \ddot{\phi}_p + 2\dot{r}_p \dot{\phi}_p = 0. \tag{34}$$

This equation has the solution

$$r_p^2 \dot{\phi}_p = \text{constant} = \ell, \tag{35}$$

i.e., the specific angular momentum of the particle, $\ell$, when it is not subject to a torque, is a constant. It follows that

$$v_{p\phi} = r_p \dot{\phi}_p = \frac{r_0^2 \dot{\phi}_0}{r_p} = \frac{\ell_0}{r_p}, \tag{36}$$

where, in this case, $r_0 = R_t$ and $\dot{\phi}_0 = \Omega_*$, i.e., the gas and particles are co-rotating with the star at the inner truncation radius. Note that even though the dust particle would start at or near the keplerian angular speed, gas drag would likely accelerate or decelerate the particle to the co-rotational angular gas speed of the stellar magnetosphere, hence the assumption $\dot{\phi}_0 = \Omega_*$. Using Eq. (36), approximate, analytic solutions are available for the radial speed when one considers the radial equation of motion, Eq. (27), without gas drag:

$$\ddot{r}_p = v_{pr} \frac{dv_{pr}}{dr} = r_p \dot{\phi}_p^2 - \frac{GM_* r_p}{[r_p^2 + z_p^2]^{3/2}} = \frac{\ell_0^2}{r_p^3} - \frac{GM_* r_p}{[r_p^2 + z_p^2]^{3/2}} , \quad (37)$$

Suppose the particle is given a significant boost velocity in the $z$ direction such that $z \to \infty$ ( or $z$ becomes comparable to $r$) then

$$\ddot{r}_p \approx \frac{\ell_0^2}{r_p^3} > 0 , \quad (38)$$

and the particle starts to accelerate in the radial direction with the subsequent radial speed

$$v_{pr} \approx \frac{\ell_0}{r_0} \sqrt{1 - \left(\frac{r_0}{r_p}\right)^2} . \quad (39)$$

So, in this scenario, the particle increases in radial speed and we have

$$v_{pr} \to \frac{\ell_0}{r_0} \text{ as } r_p \to \infty . \quad (40)$$

It could be argued that setting $z \to \infty$ is a trifle unrealistic, so let us assume that the boost in the $z$ direction is small and that $z \ll r$. For such a case, Eq. (37) becomes

$$\ddot{r}_p = v_{pr} \frac{dv_{pr}}{dr} = r_p \dot{\phi}_p^2 - \frac{GM_* r_p}{[r_p^2 + z_p^2]^{3/2}} \approx \frac{\ell_0^2}{r_p^3} - \frac{GM_*}{r_p^2} . \quad (41)$$

The particle will move outward if $\ell_0^2 > GM_* r_p$. In that case, Eq. (41) has the solution

$$v_{pr} \approx \sqrt{r_0^2 \dot{\phi}_0^2 \left(1 - \left(\frac{r_0}{r_p}\right)^2\right) - \frac{2GM_*}{r_0}\left(1 - \frac{r_0}{r_p}\right)} . \quad (42)$$

As $r_p \to \infty$ then

$$v_{pr} \to \sqrt{r_0^2 \dot{\phi}_0^2 - \frac{2GM_*}{r_0}} = \frac{\ell_0}{r_0} \sqrt{1 - \frac{2GM_* r_0}{\ell_0^2}} . \quad (43)$$

It is clear from these results that as the $z$ boost speed of the particle increases then the radial outward motion becomes more likely and the final outward radial speed of the particle increases.

At the truncation radius, the angular speed of any co-rotating gas located at or near the truncation radius will have an angular speed, $v_{co}$,

$$v_{co} = r\Omega_* = 78 \left(\frac{r}{0.05\,\text{au}}\right)\left(\frac{7\,\text{days}}{P_*}\right) \text{km s}^{-1}, \tag{44}$$

while the keplerian speed, $v_{Kep}$, of the gas near the inner truncation radius is

$$v_{Kep} = \left(\frac{GM_*}{r}\right)^{1/2} = 133 \left(\frac{M_*/M_\odot}{r/0.05\,\text{au}}\right)^{1/2} \text{km s}^{-1}, \tag{45}$$

If the truncation radius were less than the co-rotation radius then the particle would leave the disc with an angular speed that is lower than the keplerian angular speed. The tendency for such a particle would be to fall towards the star. However, as the particle rises above the midplane of the disc and moves radially towards the star then the radial component of the star's gravitational force decreases, but the angular momentum of the particle remains the same. As such, the particle may be ejected away from the star if, for example $\ell_0^2 > GM_* r_p$.

Similarly, if the truncation radius is greater than the co-rotation radius then the angular speed of the particle is greater than the keplerian speed and the particle will move away from the star. The only point where the co-rotation speed would equal the keplerian speed would be at or near the co-rotation radius.

## 2.7 CAI Temperature, Cooling Rate and Frictional Reheating

To compute the temperature of the CAIs, we assume that for their flight path above the protosolar disc, they are subject to unobstructed solar radiation. As such, their temperature is given by

$$T_p \approx \left(\frac{(L_* + L_a)\varepsilon_a}{16\pi(r^2 + z^2)\sigma_{SB}\varepsilon_e}\right)^{1/4} = 1250\,\text{K} \left(\frac{((L_* + L_a)/L_\odot)\varepsilon_a}{\left(\sqrt{r^2+z^2}/0.05\,\text{au}\right)^2 \varepsilon_e}\right)^{1/4}. \tag{46}$$

This equation is an approximation, because the geometry of the disc (where the disc can block out some of the direct radiation from the protosun) and the radiation from the disc also play a part in heating or cooling the particle. This is discussed in some detail in Liffman et al. (2012). However, for the purposes of this study, we simplify the analysis to the base assumption of direct solar radiative heating until the CAI particle re-enters the protosolar disc.

As the CAIs move away from the protosun, they will cool, where their cooling rate can be deduced by differentiating Eq. (46):

$$\frac{dT_p}{dt} \approx -\frac{1}{2}\left(\frac{(L_* + L_a)\varepsilon_a}{16\pi\sigma_{SB}\varepsilon_e}\right)^{1/4} \dot{R} R^{-3/2}$$

$$= -8.5\,\text{K hr}^{-1}\left(\frac{\varepsilon_a}{\varepsilon_e}\right)^{1/4}\left(\frac{L_* + L_a}{L_\odot}\right)^{1/4} \frac{\left(\dot{R}/40\,\text{km s}^{-1}\right)}{(R/0.1\,\text{au})^{3/2}}, \tag{47}$$

where we have set $R = \sqrt{r^2 + z^2}$. This cooling rate is consistent with the 0.5 to 50 K hr$^{-1}$ cooling rate range deduced for type B CAIs (Stolper & Paque 1986). The 50 km s$^{-1}$ normalisation speed is also consistent with the observed speed of ejected forsterite grains from observations of the protostellar system Ex Lupi (Juhász et al. 2012).

Upon protosolar disc re-entry the CAI is subject to gas heating due to the hypersonic interaction between the CAI particle and the surrounding gas. For the case of free molecular flow, the rate of temperature change for the CAI due to the interaction with the dilute gas is given by the energy balance equation (Probstein 1968, Gombosi et al. 1986, Combi et al. 1997), where we follow the Gombosi et al. prescription:

$$\frac{4}{3}\pi\rho_p a_p^3 C_p \frac{dT_p}{dt} \approx 4\pi a_p^2 \rho_g v_{pg}(T_r - T_p)C_H, \qquad (48)$$

where $C_p$ is the specific heat of the CAI particle and $C_H$ is a heat transfer coefficient:

$$C_H = \frac{(\gamma+1)k_B}{8(\gamma-1)\overline{m}S^2}\left[\frac{S\exp(-S^2)}{\sqrt{\pi}} + \left(S^2 + \frac{1}{2}\right)\mathrm{erf}(S)\right], \qquad (49)$$

with $\gamma$ the ratio of specific heats. $T_r$ is the adiabatic recovery temperature (with correction to the formula from Kitamura (1986)):

$$T_r = \frac{T_g}{\gamma+1}\left[2\gamma + 2(\gamma-1)S^2 - \frac{\gamma-1}{\frac{1}{2} + S^2 + \frac{S\exp(-S^2)}{\sqrt{\pi}\,\mathrm{erf}(S)}}\right], \qquad (50)$$

and $T_g$ the temperature of the gas.

If the CAI re-enters the protostellar accretion disc at high speed then it will be subject to significant re-heating due to frictional drag, where the equilibrium temperature of the particle will be (Appendix A)

$$T_p \approx \left(\frac{\rho_g v_{pg}^3}{8\sigma_{SB}\varepsilon_e}\right)^{1/4} = 1938\,\mathrm{K}\left(\left(\frac{\rho_g}{10^{-7}\,\mathrm{kg\,m^{-3}}}\right)\left(\frac{v_{pg}}{40\,\mathrm{km\,s^{-1}}}\right)^3\right)^{1/4}, \qquad (51)$$

where we note that this temperature is independent of the particle's size. As a consequence of this re-entry process, the surface of the CAI may be subject to significant frictional reheating and may possibly be the source of the ubiquitous WLRs. The resulting drag forces and reheating may also produce the observed bowl shaped CAIs.

## 2.8 Stopping Time

As the quasi-spherical CAI traverses the upper regions of the Solar Protoplanetary Disc at hypersonic speeds, it will be subject to gas drag with the subsequent deceleration described by the equation

$$\frac{dv_{pg}}{dt} \approx -\frac{3\rho_g C_D}{8a_p\rho_p}v_{pg}^2, \qquad (52)$$

where for hypersonic flow, from Eq. (31), $C_D \approx 2$. Using this value for $C_D$, Eq. (52) has the solution

$$v_{pg}(t) = \frac{v_{pg}(0)}{1 + t/\tau_{\mathrm{stop}}}, \qquad (53)$$

where $\tau_{\mathrm{stop}}$ is the time required for the particle to reach half of its initial impact speed of $v_{pg}(0)$. The parameterized form of $\tau_{\mathrm{stop}}$ is

$$\tau_{\text{stop}} = \frac{4a_p\rho_p}{3\rho_g v_{pg}(0)} = 5\times 10^3 \text{ s} \frac{\left(\dfrac{a_p}{0.5\text{ cm}}\right)\left(\dfrac{\rho_p}{3000\text{ kg m}^{-3}}\right)}{\left(\dfrac{\rho_g}{10^{-7}\text{ kg m}^{-3}}\right)\left(\dfrac{v_{pg}(0)}{40\text{ kms}^{-1}}\right)}. \tag{54}$$

For this parameterisation, the stopping time is about 1.4 hours. For particles that are a tenth and one hundredth of the size, the stopping times are approximately 8.3 minutes and 50 seconds, respectively. The stopping distance, $l_s$, is simply:

$$l_s \approx \frac{4a_p\rho_p}{3\rho_g} = 0.00133 \text{ au} \frac{\left(\dfrac{a_p}{0.5\text{ cm}}\right)\left(\dfrac{\rho_p}{3000\text{ kg m}^{-3}}\right)}{\left(\dfrac{\rho_g}{10^{-7}\text{ kg m}^{-3}}\right)}. \tag{55}$$

This is a small distance relative to relevant length scales, such as the truncation radius of the inner disc (~0.05 au).

## 2.9 Gas Frictional Heating Cooling Rate

As discussed in the previous sections, when a particle interacts at hypersonic speeds with a gas, it is subject to gas drag and to frictional heating. The frictional heating will decrease over time due to the decreasing relative speed between the particle and the gas. To obtain an estimate of this cooling rate, we differentiate Eq. (51) with respect to time and use Eq. (52) to obtain

$$\begin{aligned}\frac{dT_p}{dt} &\approx -\frac{9C_D\rho_g^{5/4}v_{pg}^{7/4}}{32a_p\rho_p(8\sigma_{SB}\varepsilon_e)^{1/4}} \\ &= -5,200\text{ K/hr}\,\frac{\left(\rho_g/10^{-7}\text{kgm}^{-3}\right)^{5/4}\left(v_{pg}/100\text{kms}^{-1}\right)^{7/4}}{\left(a_p/0.5\text{cm}\right)\left(\rho_p/3000\text{kgm}^{-3}\right)}.\end{aligned} \tag{56}$$

So, the expected frictional cooling rates for frictional gas heating are quite significant. We will return to this estimate in section 3.2, where simulations show the CAIs obtain cooling rates of this order of magnitude.

## 2.10 CAI evaporation

Frictional and radiative heating, if high enough, will cause the CAI to evaporate. To model the evaporation of CAIs, we use the Langmuir evaporation formula, which, for a spherical particle in a vacuum, gives the rate of mass loss due to evaporation, $\dot{m}_{ev}$, as:

$$\dot{m}_{ev} \approx -\alpha_{ev}4\pi a_p^2 p_{ev}\sqrt{\frac{\bar{m}_{ev}}{2\pi k_B T_p}}, \tag{57}$$

with $\bar{m}_{ev}$ the mean molecular mass of molecules produced from the evaporating particle, $\alpha_{ev}$ the evaporation coefficient and $p_{ev}$ the vapour pressure from the CAI. The vapour pressure term often has the form

$$p_{ev} \approx 10^{A_{ev} - B_{ev}/T_p} \text{ Pa},\qquad(58)$$

where $A_{ev}$ and $B_{ev}$ are empirically derived, constants.

For an approximately spherical particle, Eq. (57) gives the rate of change in the CAI's radius as

$$\frac{da_p}{dt} \approx -\frac{\alpha_{ev} p_{ev}}{\rho_p}\sqrt{\frac{\bar{m}_{ev}}{2\pi k_B T_p}}.\qquad(59)$$

So the particle radius as a function of time and the time required for the particle to evaporate, $\tau_{ev}$, are

$$a_p \approx a_p(0) - \frac{\alpha_{ev} p_{ev} t}{\rho_p}\sqrt{\frac{\bar{m}_{ev}}{2\pi k_B T_p}},\qquad(60)$$

and

$$\tau_{ev} \approx -\frac{a_p(0)\rho_p}{\alpha_{ev} p_{ev}}\sqrt{\frac{2\pi k_B T_p}{\bar{m}_{ev}}} \approx 4.6\,\text{days}\,\frac{\left(\dfrac{a_p(0)}{0.5\,\text{cm}}\right)\left(\dfrac{\rho_p}{3000\,\text{kg m}^{-3}}\right)}{\left(\dfrac{\alpha_{ev}}{0.01}\right)\left(\dfrac{p_{ev}}{5\,\text{Pa}}\right)}\sqrt{\frac{\left(\dfrac{T_p}{1500\,\text{K}}\right)}{\left(\dfrac{\bar{m}_{ev}}{45\,\text{amu}}\right)}}.\qquad(61)$$

For this paper, we assume that the mass density of a CAI is approximately that of the mineral melilite: 3000 kg m$^{-3}$. We take $\bar{m}_{ev} \approx 45\,\text{amu}$ (Love & Brownlee 1991), $A_{ev} \approx 9.6$ and $B_{ev} \approx 1.35\times10^4$ K. The latter two values are from Allen et al. (1965) and are the given values for the evaporation of meteoritic-like stone for values of $p_{ev}$ in SI units of Pa. The resulting evaporation times for the CAI-like particles with diameters of 1 cm, 1 mm and 0.1 mm are given in Fig. 11. The very smallest particles exposed to temperatures in excess of 1500 K have lifetimes of order hours or less.

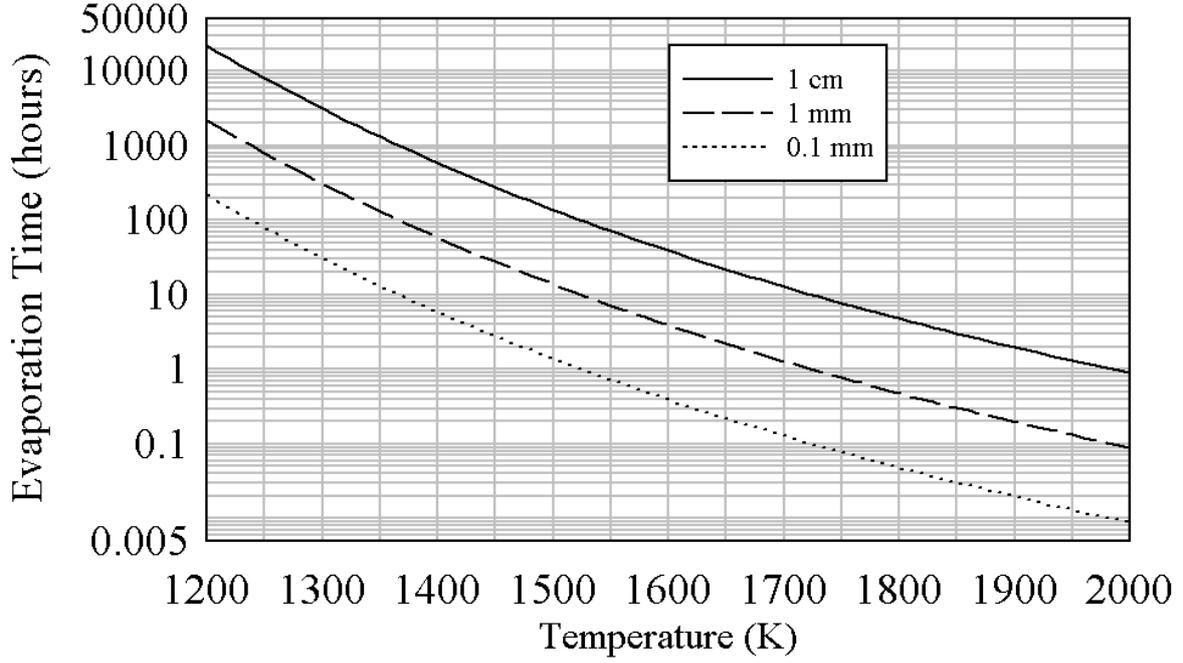

Fig. 11: Approximate evaporation times for spherical, CAI-like particles of different diameters over a range of temperatures with $\alpha_{ev} \approx 0.01$.

The rate of energy consumed for the evaporation of CAI-like materials is

$$\dot{E}_{ev} = \zeta_{ev}\dot{m}_{ev} \approx 4\pi a_p^2 \zeta_{ev}\alpha_{ev} 10^{A_{ev}-B_{ev}/T_p}\sqrt{\frac{\overline{m}_{ev}}{2\pi k_B T_p}}, \qquad (62)$$

with $\zeta_{ev} \approx 8\times 10^6$ Jkg$^{-1}$ is the energy for the evaporation of a unit mass of CAI-like material (ibid.). The evaporation coefficient is usually set at $\alpha_{ev} \approx 0.1$ (e.g., Boss et al. 2012), but we have taken the lower limit of $\alpha_{ev} \approx 0.02$, as suggested by some of the experimental results from Nagahara & Ozawa (1996).

Including evaporation, the energy equation we use to describe the temperature evolution of the CAI is

$$\frac{4}{3}\pi \rho_p a_p^3 C_p \frac{dT_p}{dt} \approx 4\pi a_p^2 C_H \rho_g v_{pg}(T_r - T_p) + \frac{a_p^2(L_* + L_a)\varepsilon_a}{4R^2} - \\ 4\pi a_p^2 \varepsilon_e \sigma_{SB} T_p^4 - \pi a_p^2 \zeta_{ev}\alpha_{ev} 10^{A_{ev}-B_{ev}/T_p}\sqrt{\frac{\overline{m}_{ev}}{2\pi k_B T_p}}. \qquad (63)$$

The first term on the right hand side calculates the rate of energy input due to gas-particle interaction, the second term is the radiative input from the protosun, the third term is the radiative loss from the particle and the fourth term is the rate of energy loss due to evaporation.

## 2.11 Ram Pressure

In addition to high temperatures and evaporative mass loss, a CAI's hypersonic interaction with the disc gas will subject it to ram pressure, $P_{ram}$, given by

$$P_{ram} = \frac{\rho_g v_{pg}^2}{2} = 4.93 \times 10^{-3} \, \text{atm} \left( \frac{\rho_g}{10^{-7} \, \text{kg m}^{-3}} \right) \left( \frac{v_{pg}}{100 \, \text{km s}^{-1}} \right)^2. \tag{64}$$

As can be seen via Eq. (53), the effect of gas drag should rapidly decrease the ram pressure:

$$P_{ram} = \frac{\rho_g}{2} \left( \frac{v_{pg}(0)}{1 + t/\tau_{stop}} \right)^2. \tag{65}$$

A CAI that entered the inner disc wall at hypersonic speeds would suffer high temperatures and reasonable ram pressures. Both of which would be decreasing on timescales of tens of minutes. In addition, the inner disc would have a high gas temperature where the silicate materials may have evaporated. Such a physical scenario would appear to be a recipe for the brief condensation of refractory, high temperature materials onto the surface of the CAI, provided that the ram pressure could overcome the surface evaporation of the CAI.

## 2.12 Rim Accretion versus Evaporation

As the CAI travels through the disc gas (assumed to be mainly molecular hydrogen) where the gas has a number density, $n_g$, then the rate at which the gas particles are colliding with the CAI, $J_c$, is

$$\begin{aligned} J_c &= n_g v_{pg} \\ &\approx 3 \times 10^{25} \, \text{molecules m}^{-2} \text{s}^{-1} \left( \frac{\rho_g}{10^{-6} \, \text{kg m}^{-3}} \right) \left( \frac{v_{pg}}{100 \, \text{km s}^{-1}} \right). \end{aligned} \tag{66}$$

The CAI will be subject to gas frictional temperatures that may be high enough to evaporate the CAI, where the evaporation flow, $J_{ev}$ is

$$\begin{aligned} J_{ev} &\approx \frac{p_{ev}}{\sqrt{2\pi \bar{m}_{ev} k_B T_p}} \\ &\approx 9.03 \times 10^{26} \, \text{molecules m}^{-2} \text{s}^{-1} \frac{\left( p_{ev}(3000 \, \text{K}) / 1.24 \, \text{atm} \right)}{\sqrt{\left( \bar{m}_{ev} / 45 \, \text{amu} \right) \left( T_p / 3000 \, \text{K} \right)}}, \end{aligned} \tag{67}$$

where the evaporant particles are assumed to have a mass of 45 amu.
In our computer code, we have made the semi-plausible assumption that evaporation occurs when $J_{ev} \geq J_c$, while condensation or rim formation occurs when $J_{ev} < J_c$.

## 2.13 Rim Accretion

The major thesis of this study is that CAIs were formed and processed in the inner disc region of the protosolar disc (Fig. 1) prior to possibly being ejected to the outer regions of the protosolar disc. As discussed in sections 2.6 and 3.2, some of these CAIs may have been propelled out of this region due to interaction between the solar magnetosphere and the inner disc. In particular, in section 3.2 we discuss how the CAIs can travel through the inner disc wall at speeds of order 10 to 100 km s$^{-1}$. For such a case, the CAIs will be heated by gas friction to temperatures far in excess of 1000K (e.g., Eq. (51) and Fig. 14) and pass through a relatively dense gas (gas densities ~ 10$^{-6}$ kg m$^{-3}$). Given that these conditions are somewhat analogous to commercial hypersonic, high temperature ceramic coating processes, e.g., High

Velocity Oxygen Fuel (HVOF) coating (Bolelli et al. 2007), it is appropriate to determine the possible thickness of rim material that may accrete/condense onto the CAI.

We consider a spherical CAI with radius, $a_p$, and mass density, $\rho_p$, that is travelling through a gaseous medium with speed, $v_{pg}$. The relevant gaseous atoms/molecules/crystals that will make up the rim have a number density, $n_g$, and a mass, $m_g$, where $Q$ is the probability that a gas particle will stick to the CAI. From this description, we can write that the growth in the radius of the CAI as

$$\frac{da_p}{dt} \approx \frac{Q n_g m_g v_{pg}}{4 \rho_p} \approx \frac{Q \rho_g f_g m_g v_{pg}}{4 \rho_p m_{H_2}} , \qquad (68)$$

where $\rho_g$ is total gas density, $f_g$ is the proportion of rim mineral in the gas and $m_{H_2}$ is the mass of a hydrogen molecule.

After a time, $t$, the thickness of the rim, $\Delta a_p$, is

$$\Delta a_p \approx \frac{Q \rho_g f_g m_g v_{pg} t}{4 \rho_p m_{H_2}}$$

$$= 10.7\,\mu\text{m} \frac{\left(\rho_g / 10^{-6}\,\text{kg m}^{-3}\right)\left(v_{pg}/50\,\text{km s}^{-1}\right)\left(m_g / m_{Mg_2SiO_4}\right)\left(t/10^3\,\text{s}\right)}{\left(1/Q\right)\left(\rho_p / 3000\,\text{kg m}^{-3}\right)\left(f_{Mg_2}/f_g\right)} , \qquad (69)$$

For this normalisation, we have chosen the rim material to be the mineral forsterite ($Mg_2SiO_4$), which is an observed component of pristine WLRs that surround CAIs (e.g., Bodénan 2014). We have taken $m_{Mg_2SiO_4}$ = 140.767 amu and $f_{Mg_2}$ = 1.8×10[-5] as based on solar atomic abundances (Anders & Ebihara 1982). From this normalisation it is apparent that it is semi-plausible to produce 10 micron forsterite rims on a timescale of about 20 minutes. However, rim sequences in WLRs tend to be in the range of 1 to 100 microns with different mineral layers. Eq. (69) suggests that it would be difficult to produce these rim sequences on the expected timescales, speeds and gas densities in our model. One possible solution is that we have underestimated the expected atomic abundances of "rock-like" minerals at the inner accretion disc wall.

It should be noted that our model has refractory material ejected from the accretional flow between the disc and star, where a portion of this refractory material re-enters into the inner disc wall, where the frictional temperatures are such that the refractory material evaporates. As such, it is plausible that there is an increase in "rock-like" minerals in the very inner accretion disc. This idea is analysed in more detail in Appendix B, where it is shown that the mass abundance of a particular element in the disc inner wall, $f_{Wi}$, is given by

$$f_{Wi} \approx \frac{f_{\odot i}}{(1-\chi_i)}, \qquad (70)$$

with $f_{\odot i}$ the solar mass abundance for the element and $\chi_i$ the portion of the element that is returned to the inner disc wall via the centrifugal ejection process. This equation shows that to increase the abundance of a particular "metal" element by a factor of 10 then $\chi_i = 0.9$, i.e., 90% of a particular element has to be recycled back into inner disc wall.

We will consider this in more detail in section 3.2

## 2.14 Particle Shape and Weber Number

In this paper, we wish to investigate whether the hypersonic re-entry of CAIs into the solar accretion disc can potentially produce the observed bowl-shaped CAIs. The deformation of liquid drops in shock waves has been widely discussed in the literature. In particular, as shown in Hsiang & Faeth (1995), the deformation of a liquid drop is dependent on the Weber number (We) and the Ohnesorge number (Oh).

The Weber number is indicative of drag forces that disrupt liquid and surface tension that will keep a droplet from breaking into small droplets. It is given by:

$$\text{We} = \frac{2\rho_g v_{pg}^2 a_p}{\sigma_p} = 3.2 \frac{\left(\frac{\rho_g}{10^{-7}\,\text{kgm}^{-3}}\right)\left(\frac{v_{pg}}{40\,\text{kms}^{-1}}\right)^2\left(\frac{a_p}{5\,\text{mm}}\right)}{\left(\frac{\sigma_p}{0.5\,\text{Nm}^{-1}}\right)}, \quad (71)$$

where $\sigma_p$ is the surface tension of the CAI. The representative value of 0.5 Nm$^{-1}$ for the surface tension is consistent with the values chosen by Ivanova et al. (2014).

The Ohnesorge number has the definition:

$$\text{Oh} = \frac{\mu_p}{\sqrt{2\rho_p a_p \sigma_p}} = 0.13 \frac{\left(\frac{\mu_p}{0.5\,\text{Pas}}\right)}{\sqrt{\left(\frac{\rho_p}{3000\,\text{kgm}^{-3}}\right)\left(\frac{a_p}{5\,\text{mm}}\right)\left(\frac{\sigma_p}{0.5\,\text{Nm}^{-1}}\right)}}, \quad (72)$$

with $\mu_p$ is the dynamic viscosity of the CAI melt, where we have taken 0.5 Pas as a representative value for the CAI viscosity. This value is obtained from viscosity measurements of 1500 °C, Al$_2$O$_3$-CaO-MgO-SiO$_2$ slags that are used in blast furnaces (Zhao et al. 2012).

The Ohnesorge number measures the degree of difficulty of deforming a liquid particle. A higher Ohnesorge number implies that viscous forces are becoming larger relative to surface tension forces and the drop becomes more difficult to deform. As a consequence, higher values of Oh may imply that higher values of We are required to enable the deformation of the droplet. For "small" values of the Ohnesorge number, a liquid drop tends to be deformed into a bowl-like shape for Weber numbers between 1 and 10.

A relation linking We and Oh can be deduced from Hsiang & Faeth (1995):

$$\text{We}_0 = \left(\frac{\text{We}_{cr}}{4}\right)\left(1 + \sqrt{1 + \frac{4K'\text{Oh}}{\text{We}_{cr}^{1/2}}\sqrt{\frac{\rho_g}{\rho_p}}}\right)^2, \quad (73)$$

where We$_0$ is the initial Weber number as the droplet is struck by the shock wave. It takes a certain amount of time for the droplet to show deformation and instability, and the droplet will start moving with the flow before that happens. As such, We$_{cr}$ is the critical Weber number for a particular characteristic behaviour when the droplet starts to deform or be disrupted. *K'* is an empirically determined constant. From the data given in Hsiang & Faeth (1995), we set *K'* ≈ 15. Using Eq. (73) with the characteristic values of particle and gas density (3,000 kg m$^{-3}$ and 10$^{-7}$ kg m$^{-3}$, respectively), shows that the Weber number remains approximately unchanged for values of the Ohnesorge number in excess of 1000. As such, for the purposes of this study, bowl-shaped CAI will tend to form for Weber numbers between 1 and 10. From Eq. (71), this result implies that bowl-shaped CAIs will mainly occur for larger CAI with diameters of around 1 cm. CAI with diameters of 1 mm or less will tend to have Weber numbers that are smaller than 1 and are thus less likely to form bowl-shapes when re-entering the protosolar disc.

This behaviour can be seen in more detail in Fig. 12, where we have plotted the log of the Weber number as a function of the log of the gas mass density, $\rho_g$, and the speed of the particle relative to the gas, $v_{pg}$. In this graph, the appropriate log$_{10}$(We) values for the formation of bowl-shaped CAIs occur between 0 and 1. For canonical gas-particle speeds of 40 km s$^{-1}$, this implies gas densities of around 10$^{-7}$ kg m$^{-3}$ for 1 cm diameter CAIs (Fig. 12a). For 1 mm diameter CAIs, the required gas densities are around 10$^{-7}$ kg m$^{-3}$ (Fig. 12b). Smaller particles are less likely to still have a high speed to reach these higher density regions due to the decreased stopping times and distances (Eq. (54) and Eq. (55)). As such, it is more likely that larger CAIs will obtain the appropriate Weber numbers to form a bowl shape if they reach the appropriate melt temperatures.

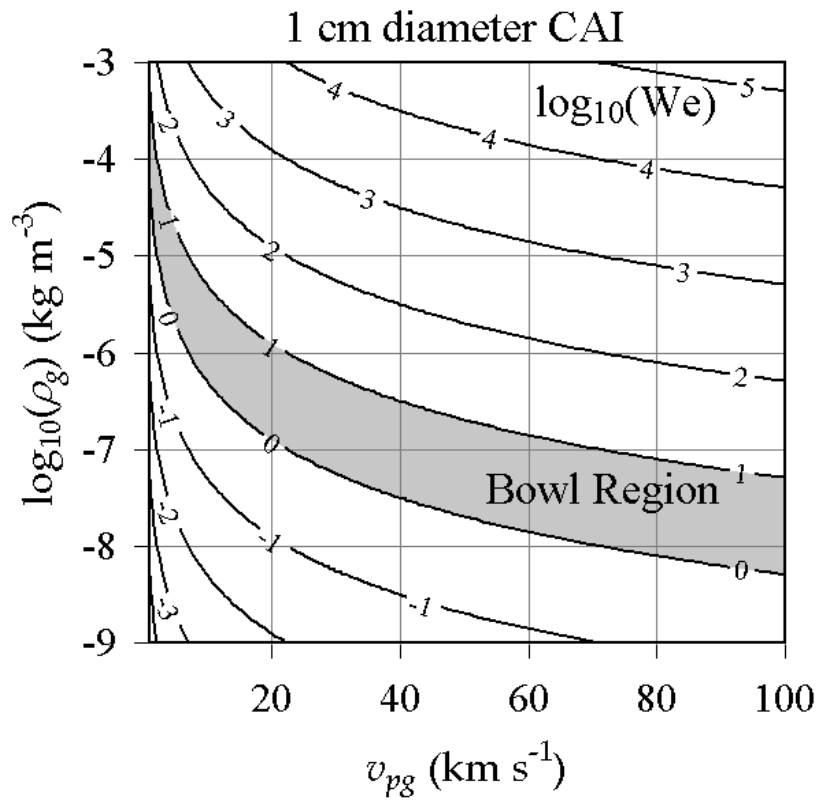

(a)

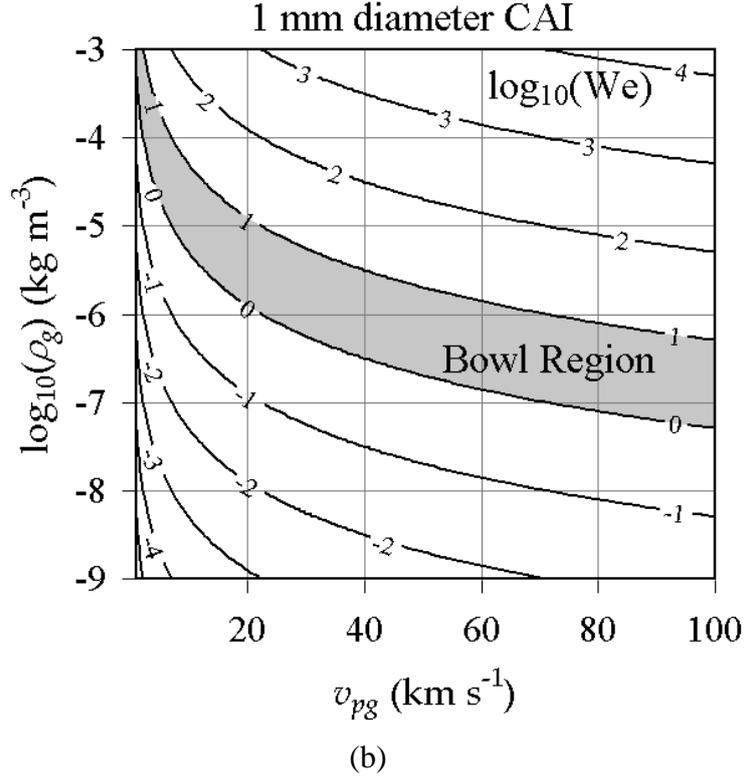

(b)

Fig. 12: (a) $\log_{10}(We)$ Weber numbers for 1 cm particles for a range of gas densities and relative speeds between the gas and the particle. Bowl shaped CAIs occur for $\log_{10}(We)$ values between 0 and 1; (b) Same as for (a) except now we compute the Weber numbers for 1 mm particles.

## 2.15 Numerics of the Monte Carlo model of Section 3

In Section 3 and Appendix C, we investigate the temperature changes in the CAIs as they travel along their flight paths. To do this, we simulate the ejection, projectile path and re-entry of each CAI particle. To determine the path of each particle, the code integrates the equations of motion given in Section 2.5, which have the vectorial form

$$\frac{d\mathbf{v}_p}{dt} = -\frac{GM_*}{r^2}\hat{\mathbf{r}} - \frac{3C_D\rho_g}{8a_p\rho_p}v_{pg}^2\,\hat{\mathbf{v}}_{pg}. \qquad (74)$$

The code uses cartesian coordinates so $\mathbf{r} = (x, y, z)$, where the origin is located at the centre of the Sun. The discretisation of the system is energy conserving and is obtained from Greenspan (1981). The resulting system of difference equations was integrated by using Newton's method of iteration (*ibid.*). The accuracy of the code was checked by computing the orbits of particles around the Sun and comparing the orbits with standard analytic solutions. In addition, a separate code was built using the engineering software package Mathcad to check the solution to all the equations built into the main code. Any errors found were corrected in the appropriate manner.

This code has been used and tested in a number of other astrophysics papers, e.g., Liffman & Brown (1995), Liffman & Toscano (2000) and Liffman (2005). A modified version of the code has also been used in the simulation of granular media, e.g., Liffman et al. (1997) and Liffman et al. (2001). Further information about the numerical code is given in these papers.

The particles were launched from or near the truncation radius, $R_t$, of the protosolar disc, where the position of the truncation radius relative to the co-rotation radius, $R_{co}$ is important in determining the subsequent projectile path of the particles (Appendix C). For example, if $R_t > R_{co}$ then the co-rotating magnetosphere would have a greater angular speed relative to the accretion disc and the CAIs would be ejected at higher radial speeds and be more likely to ram into the inner disc wall. The protosolar disc is modelled using the formulae given in Eqs (7) through to (11).

# 3 Results

## 1.2 Million Year Old Protosun

For our simulations, we assume a scenario that is similar to the observed behaviour of Ex Lupi, which is a million-year old star that underwent a significant mass accretion event with the subsequent formation and outward radial movement of forsterite dust (Juhász et al. 2012). As such, we consider the scenario where CAIs were formed in a series of mass accretion events over a relatively short time period.

The protosun is set to approximately 1.2 million years of age, which implies that the Sun had a radius of 2.4 $R_\odot$ a mass of 0.94 $M_\odot$ and a luminosity of 2 $L_\odot$ (Fig. 3 and Fig. 6). The magnetic field strength at the surface of the Sun is $B_*(R_*) = 0.1$ T. The expected, average accretion rate at 1.2 million years is, by Eq. (2), $3.1 \times 10^{-8}$ $M_\odot$ yr$^{-1}$. The rotation period of the star is set to be 7 days, which gives a co-rotation radius of 0.07 au.

For the first set of simulations (the results of which are given in Appendix C), we consider three particle diameters: 1 cm, 1 mm and 0.1 mm. For all the simulations in Appendix C, we allow the particles to evaporate, but we ignore WLR formation. We set the launch point at $0.99R_t$, i.e., just within the truncation radius of the inner disc. The launch point is on the disc midplane. A summary of the results from these simulations is given below. We modify these assumptions, regarding WLR formation and the launching point, in section 3.2.

## 3.1 1.2 Million Year Old Protosun Results Summary and Discussion

For the simulations given in Appendix C, the accretion rates were: $2 \times 10^{-8}$, $3 \times 10^{-8}$, and $5 \times 10^{-8}$ $M_\odot$yr$^{-1}$, where the larger accretion rates produced discs with smaller inner disc truncation radii as can be seen from Eq. (1). We assumed a stellar surface magnetic field of 0.1T, so the aforementioned accretion rates give inner truncation radii of 0.076, 0.067 and 0.058 au, respectively. As a consequence, the CAIs have corresponding initial temperatures of 1241K, 1330K and 1464K.

These initial temperatures are due to accretion heating (Eq. (6)) and direct radiation from the young Sun (Eq. (46)). However, if these particles were launched within view of the inner disc rim wall then there is an additional radiation component from the disc wall, which increases the particle temperatures by a factor of about $\sqrt[4]{3} \approx 1.3$ (Eq. (16)). So the potential initial or formation temperatures of the particles are 1630K, 1750K and 1930K, respectively. We have not taken into the account the disc radiation in our particle temperature calculations, because the geometry makes the radiative transfer calculations quite complicated and potentially very time consuming (e.g., Liffman et al. 2012).

However, if we note such higher initial temperatures are a possibility, then we can compare these temperatures to the expected melting temperatures of CAIs. As a first approximation, we can use the mineral melilite (a solid solution of åkermanite ($Ca_2MgSi_2O_7$) and gehlenite ($Ca_2Al_2SiO_7$)) as a proxy for CAI composition. As such, an initial temperature in excess of 1830K is sufficient to melt all of the CAI if the CAIs are composed of 0% or more of åkermanite by mass, i.e., Åk$_x$ for $x > 0$ (Mendybaev, Richter & Davis 2006). The minimum solidus temperature for melilite is approximately 1660K. As a consequence, CAIs processed in the high accretion ($5 \times 10^{-8}$ M$_\odot$yr$^{-1}$) case will be completely molten, while CAIs processed in the low accretion case ($2 \times 10^{-8}$ M$_\odot$yr$^{-1}$) will be completely solid. CAIs processed at accretion rates between these two extremes may be partially molten.

As the CAIs move away from the Sun, the subsequent cooling rates are dependent on the vertical ejection speed and the initial temperature of the particle. Particles with a larger initial vertical ejection speed and larger initial temperature have the highest cooling rates (Fig. 31). Such a result can be predicted analytically via Eq. (47) and the radial speed equations given in Section 2.6. However, without using mathematics, the dynamics of CAI cooling works as follows: a higher vertical ejection speed translates into a higher radial ejection speed. A higher radial speed implies that the particle moves away from the star at a higher speed and so the cooling rate increases. All the computed cooling rates in the above simulations plus the analytic equation Eq. (47) are consistent with the observed cooling rates for CAIs (Stolper & Paque 1986).

Upon re-entry into the disc, the CAIs are subject to frictional heating when the CAIs interact with the thin, upper atmosphere of the disc at hypersonic speeds. The amount of frictional heating is somewhat proportional to the particle size, where 1 cm diameter particles develop significantly higher temperatures compared to 1 mm diameter particles. This effect arises, because smaller particles have smaller stopping times (Eq. (54)) and larger surface area to mass ratios. As a consequence, smaller particles are heated for a shorter time and they can radiate away that heat more efficiently relative to larger particles. For the 1cm particles, the subsequent heating and cooling rates can be in the range of thousands of Kelvin per hour.

The relationship between the co-rotation speed and keplerian speed at the inner truncation radius has a dramatic effect on CAI processing. If the angular speed of the gas co-rotating with the solar magnetosphere is less than the keplerian speed (which occurs when $R_t < R_{co}$ – in this case $R_{co} = 0.07$ au) then some of the ejected particles end up on inbound orbits that bring them closer to the protosun and they are eventually destroyed via evaporation. Other particles may end up on outbound orbits that tend to re-enter the disc within 0.5 au of the protosun (Fig. 25 and Fig. 36).

The heating due to re-entry into the "far" disc (i.e., away from the inner disc wall) is dependent on size and position. Smaller particles reach lower re-entry temperatures relative to larger particles, while particles that reenter closer to the protosun tend to suffer larger temperature spikes (Fig. 27, Fig. 34, Fig. 37 and Fig. 39). This size-dependent heating result suggests that WLRs are not a consequence of this hypothetical, far disc, re-entry heating, because such rims tend to be ubiquitous around just about all CAIs, irrespective of size (MacPherson et al. 2005). However, re-entry heating may still produce the bowl shaped CAI. As shown by Eq. (71) and Fig. 12, for the given gas densities of $10^{-7}$ kg m$^{-3}$ and relative speeds of 40 km s$^{-1}$, bowl-shaped CAIs are more likely to occur for 1 cm diameter particles.

When $R_t > R_{co}$ then the gas co-rotating with the solar magnetosphere has an angular speed that is greater than the keplerian speed. For such a case, we have that particles are ejected on outbound orbits, where the particles tend to travel far greater distances relative to the $R_t < R_{co}$ case (Fig. 40). The temperature behaviour of the particles in this case is intriguing (Fig. 41). If the particles collide with the inner disc wall then they suffer significant gas frictional heating, which can vaporise the particles. Particles that survive this impact or miss the inner disc wall, may be ejected from the Solar System or re-enter the disc at significant distances away from the protosun. When the particles re-enter the far disc then the re-entry temperatures are relatively minor as the disc gas densities decrease significantly with distance from the protosun.

A point to note is the possible reprocessing of the particles that punch through the inner disc wall. Such particles are subject to high gas temperatures, frictional heating and significant gas drag. As such, there is a possibility of forming bowl-shaped CAIs, partial or full vaporisation of CAIs and the condensation of WLRs in a high temperature, high gas density environment. If the inner disc wall is high enough then we have a filtering mechanism for CAIs: CAIs that make it through will show some evidence of their hypersonic passage through a significant region of hot, dense gas. CAIs that do not make it through will be vaporised or recycled back into the inner disc region.

We will simulate this scenario in the next section.

## 3.2 Inner Disc Wall Path, 1 cm particles, 3×10⁻⁸ M☉yr⁻¹, $R_t > R_{co}$

In this simulation, we launched the particles close to the inner wall of the disc at 0.9999 $R_t$. The radius and luminosity of the Sun were left unchanged at 2.4 R☉ and 2 L☉, respectively. We set the solar rotation period to 4.5 days, so the protosun was rotating with a shorter period than the observed, average rotation period for solar-like stars. We assumed a solar magnetic field strength of 0.8 T and a disc accretion rate onto the Sun of $\dot{M}_a = 3 \times 10^{-8}$ M☉ yr⁻¹. These values set $R_t = 0.059$ AU and $R_{co} = 0.052$ au. Assuming a 0.94M☉ Sun, the keplerian rotation speed at the truncation radius/launch point is 118 km s⁻¹, while the co-rotation speed at that point is 144 km s⁻¹. We assumed that the gas at the inner wall of the disc is co-rotating with the stellar magnetic field. Hence the gas and any entrained particles/CAIs have an assumed initial angular speed of 144 km s⁻¹.

In this simulation, we turn on evaporation and WLR formation as per the theory discussed in Sections 2.13, 2.10 and 2.12.

Using these parameters, it turns out that launching such particles with such a high angular speed from the midplane of the disc causes the particles to be completely vaporised when they enter the inner disc wall. To stop this from occurring, we launched the particles at an altitude of 3.5 scale heights (0.00388 au) from the midplane of the disc. At this altitude, the gas density at the inner disc wall is reduced and the particles can survive the entry into the disc wall. A high launch point might be justified by the possibility that the CAIs will remain coupled to the infalling flow until the density of the gas flow decreases as it accelerates along the magnetospheric field lines. It will require a detailed simulation of dust and particle transport in the infalling gas to determine where the dust and CAIs would begin to separate from the infalling gas.

We launched the particles with vertical speeds of (1) 5.7, (2) 17.2, (3) 28.7, (4) 40.3, (5) 51,8, (6) 63.3, (7) 74.8, (8) 86.2, (9) 97.7 and (10) 109 km s⁻¹. The resulting projectile paths are shown in Fig. 13(a) and (b), where we see that the first three particles in this sequence are subject to considerable gas drag and remain close to the inner disc wall, while the latter seven particles are ejected from the inner rim region and land further away in the disc, with the exception of the last particle in the sequence which was ejected from the Solar System.

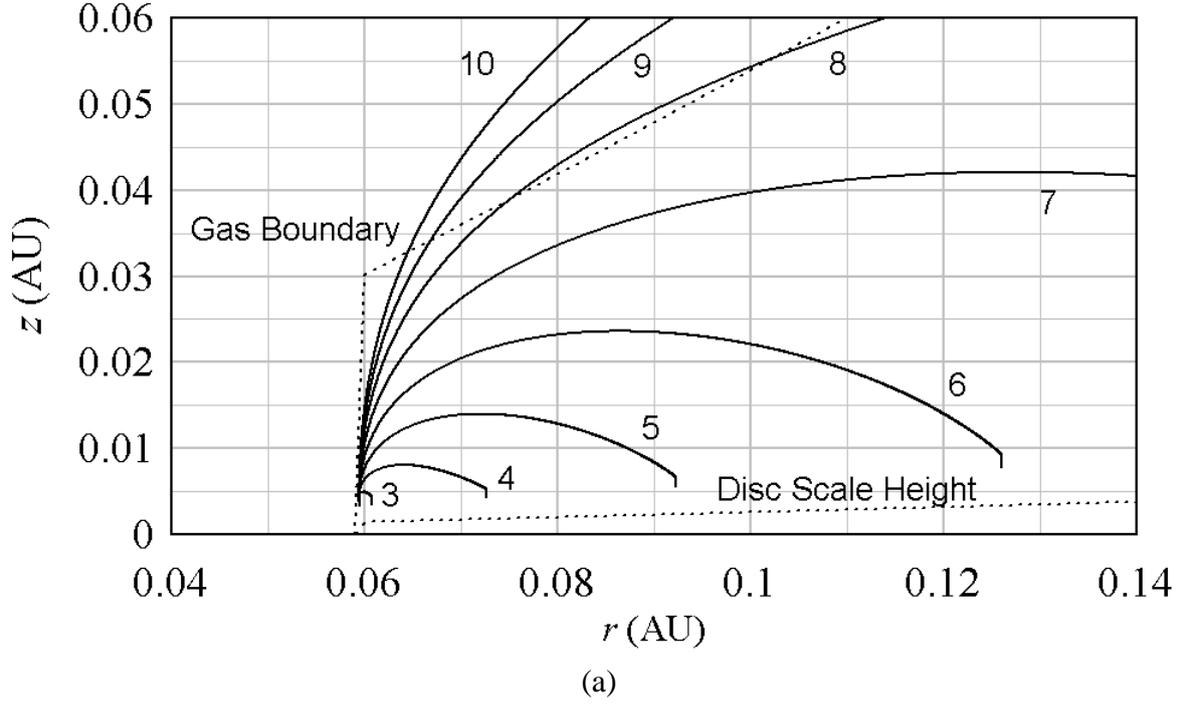

(a)

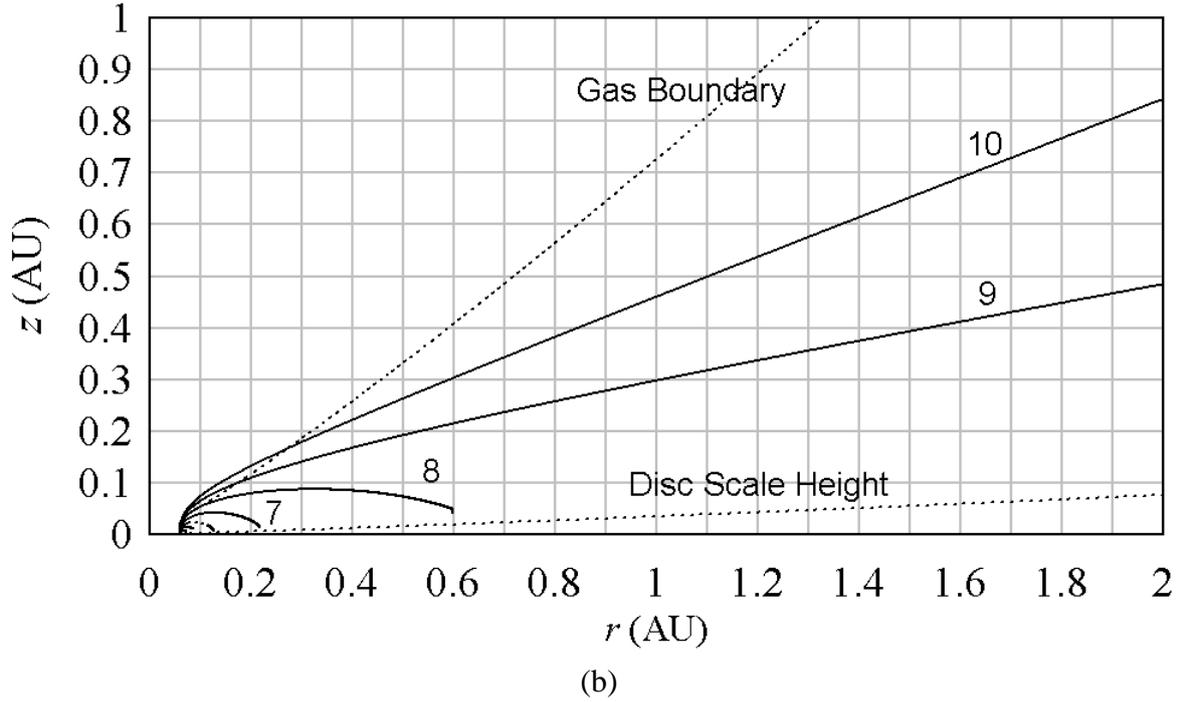

(b)

Fig. 13: Projectile paths for 1 cm CAIs for the case of $\dot{M}_a = 3\times10^{-8}$ $M_\odot yr^{-1}$ with a $1.2\times10^6$ year old protosun. The co-rotation angular speed of the stellar magnetic field is 144 km s$^{-1}$, while the keplerian orbital angular speed is 118 kms$^{-1}$. The particles are given the co-rotation angular speed and labelled as per the initial $z$ speed, where $v_{1z}$ = 5.8 (km s$^{-1}$), $v_{2z}$ = 17.3, $v_{3z}$ = 28.8, $v_{4z}$ = 40.3, $v_{5z}$ = 51.8, $v_{6z}$ = 63.3, $v_{7z}$ = 74.8, $v_{8z}$ = 86.2, $v_{9z}$ = 97.7 and $v_{10z}$ = 109 km s$^{-1}$. All the particles were launched from the inner truncation radius of $R_t$ = 0.059 au and an altitude of 3.5 scale heights (0.00388 au) from the midplane of the disc. As seen in (a), particles 1 to 3 do not move far from the inner disc wall. In (b), all the remaining particles re-enter the disc, apart from particle 10 which is completely ejected from the disc.

The subsequent temperatures of the particles are displayed in Fig. 14. There are two sections of the temperature profile for these particles. As the particles enter the inner disc wall, they are instantly heated by intense gas frictional heating from a radiative equilibrium temperature of 1400K to temperatures in the range of 2600K to 3200K (Fig. 14a). The particles are immersed in the inner disc and are subject to gas drag. The particles subsequently cool to 1400K on a timescale of around ten to twenty minutes. Some of the particles are able to move through the inner disc wall and launch on a ballistic path above the disc, where they cool down on timescales of days (Fig. 14b).

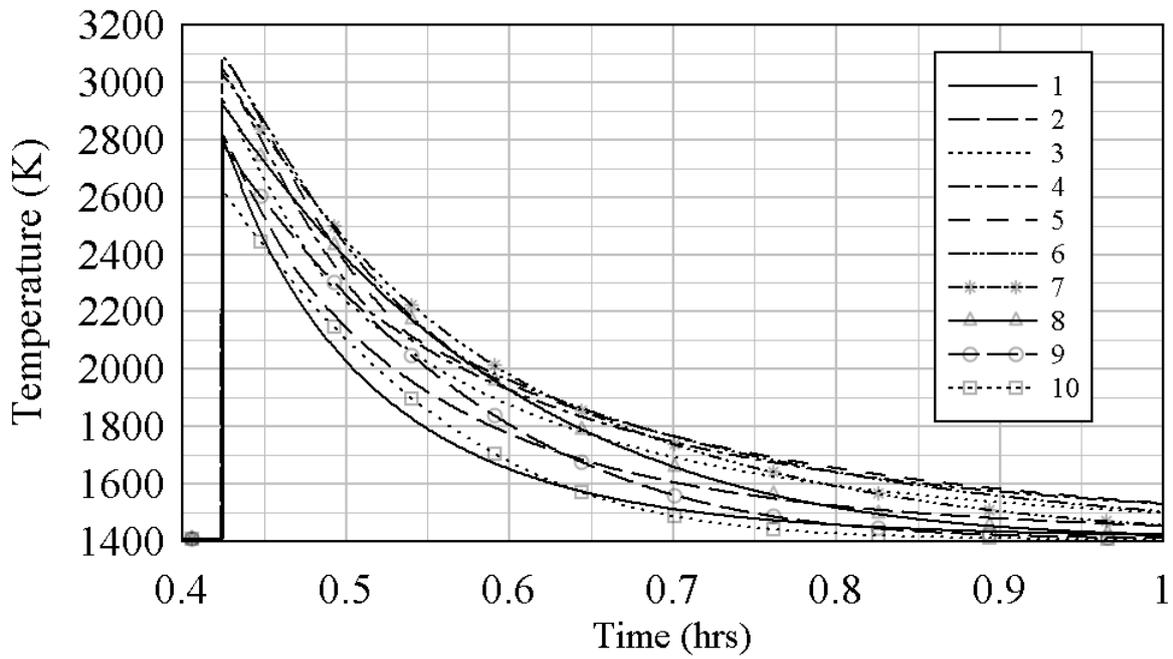

(a)

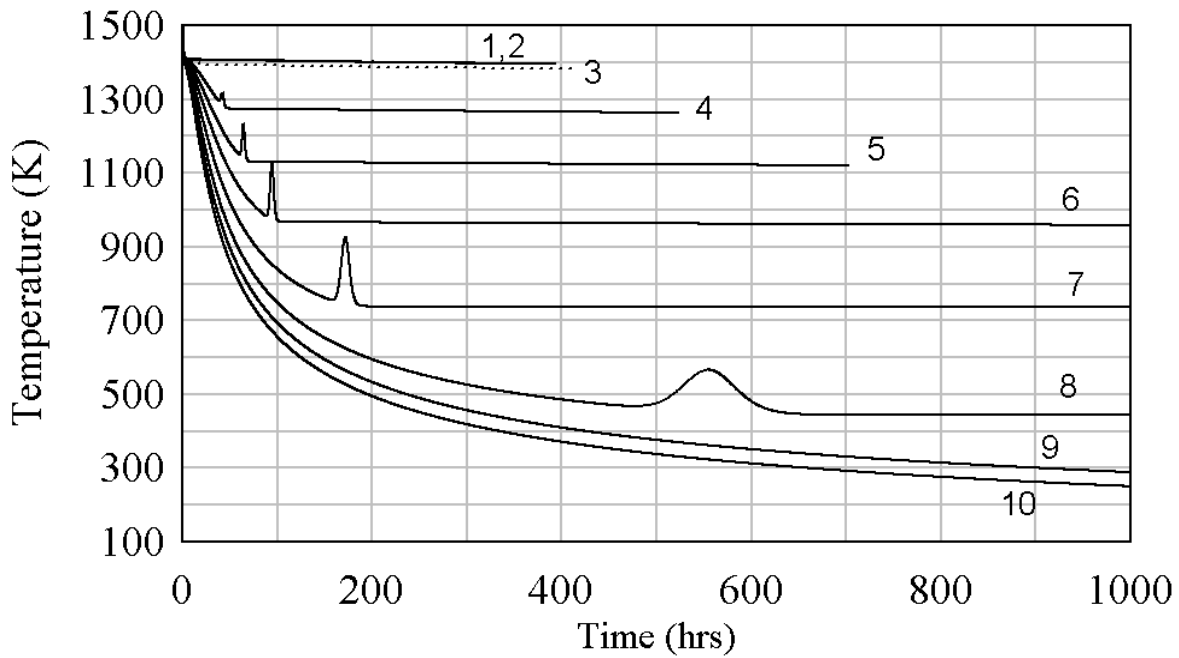

(b)

Fig. 14: Particle temperatures as a function of time. (a) The 1 cm diameter CAIs start with a temperature of around 1400K. As the particles enter the inner rim wall, they are subject to significant frictional heating, which heats the particles to a range of temperatures between 2,600 and 3,200 K. The particles cool to the equilibrium 1400 K temperatures on a timescale of approximately 20 minutes. (b) Particles 4 to 10 are ejected from the neighbourhood of the inner disc wall and subsequently cool on timescales of days. Particles 4 to 8 suffer a small temperature increase on re-entry into the disc. Particle 10 is ejected from the Solar System and cools to interstellar temperatures as it moves away from the protosun.

The time rate of cooling for the CAIs as they travel through the inner disc wall is shown in Fig. 15(a, b), where we see that the initial cooling rates are of order $10^4$ K hr$^{-1}$. From Fig. 13, particles 4 through 10 pass through the inner disc and move radially away from the protosun. Once these particles have been ejected from the disc they then have much lower cooling rates of order 10 K hr$^{-1}$ (Fig. 15(c)).

The initial high cooling rates shown in Fig. 15(a, b) are due to gas drag slowing the particle's speed relative to the gas as is deduced from Eq. (56). These high cooling rates are somewhat consistent with the "Flash Heating" scenario of Wark & Boynton (2001), who proposed that WLRs around CAIs were produced by an initial phase of high temperature in excess of 2,000 °C with subsequent rapid cooling. The lower cooling rates in Fig. 15(c) are due to the decrease in radiative heating as the particles move away from the Sun (Eq. (47)). These later cooling rates are consistent with those obtained from crystallisation experiments, e.g., Stolper & Paque (1986).

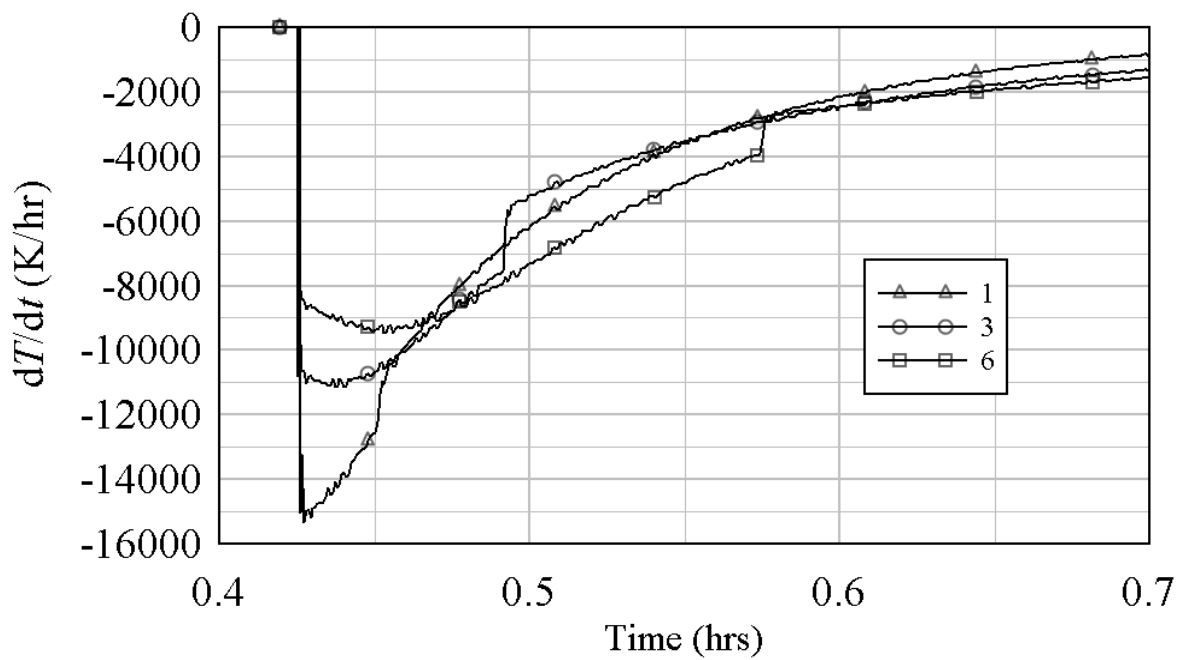

(a)

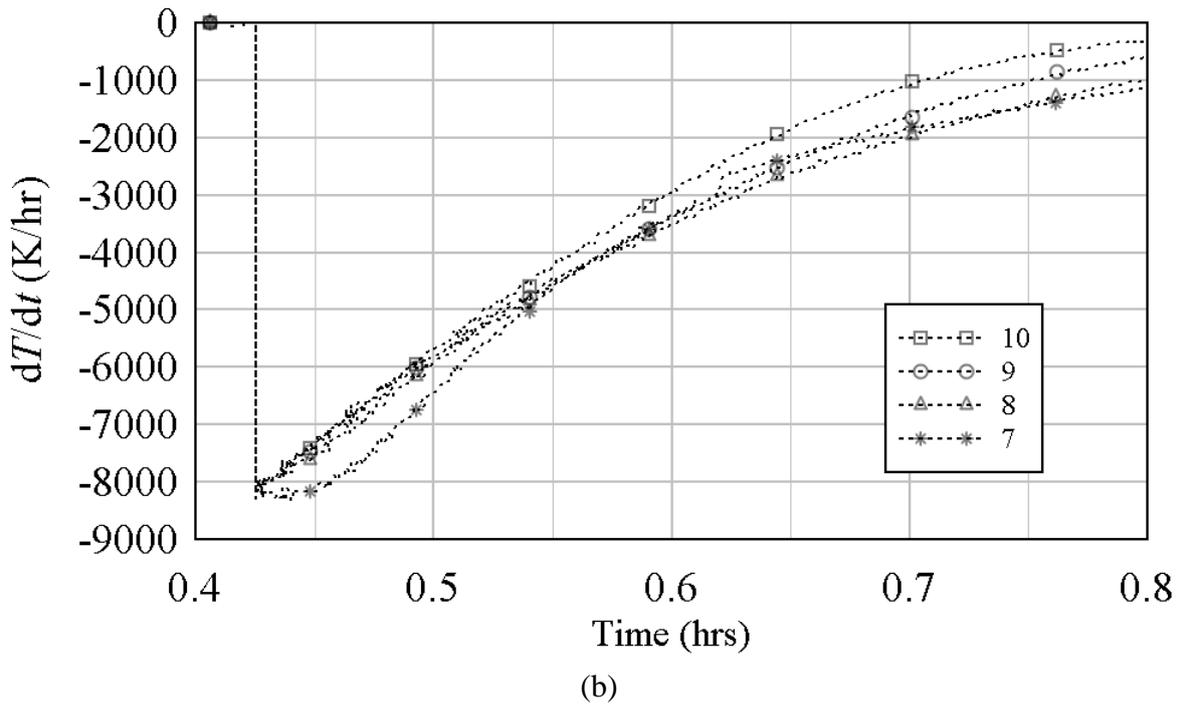

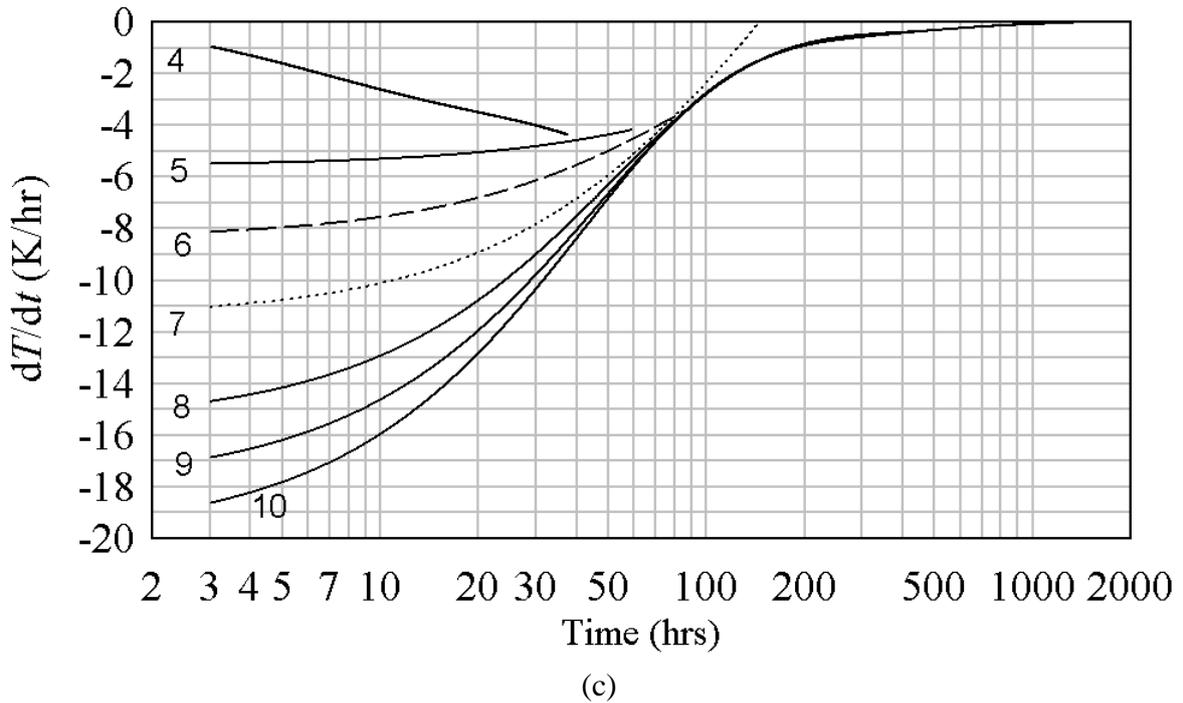

Fig. 15: CAI cooling rates for (a,b) passage through the inner disc wall and (c) above the disc. The cooling rates for passage through the inner disc wall are of order $10^4$ K hr$^{-1}$, while the path across the face of the disc gives cooling rates of around 10 K hr$^{-1}$. The discontinuities in the cooling rates shown in (a) are due to the change over from evaporation to condensation.

The discontinuities in the cooling rates of Fig. 15(a) is due to our assumption that CAIs are evaporating when $J_{ev}/J_c > 1$, where $J_{ev}$ is the evaporative gas flux and $J_c$ is the collision ram flux of the gas. When $J_{ev}/J_c < 1$, the CAIs are undergoing WLR formation (section 2.12). The ratio of $J_{ev}$ to $J_c$ is shown in Fig. 16.

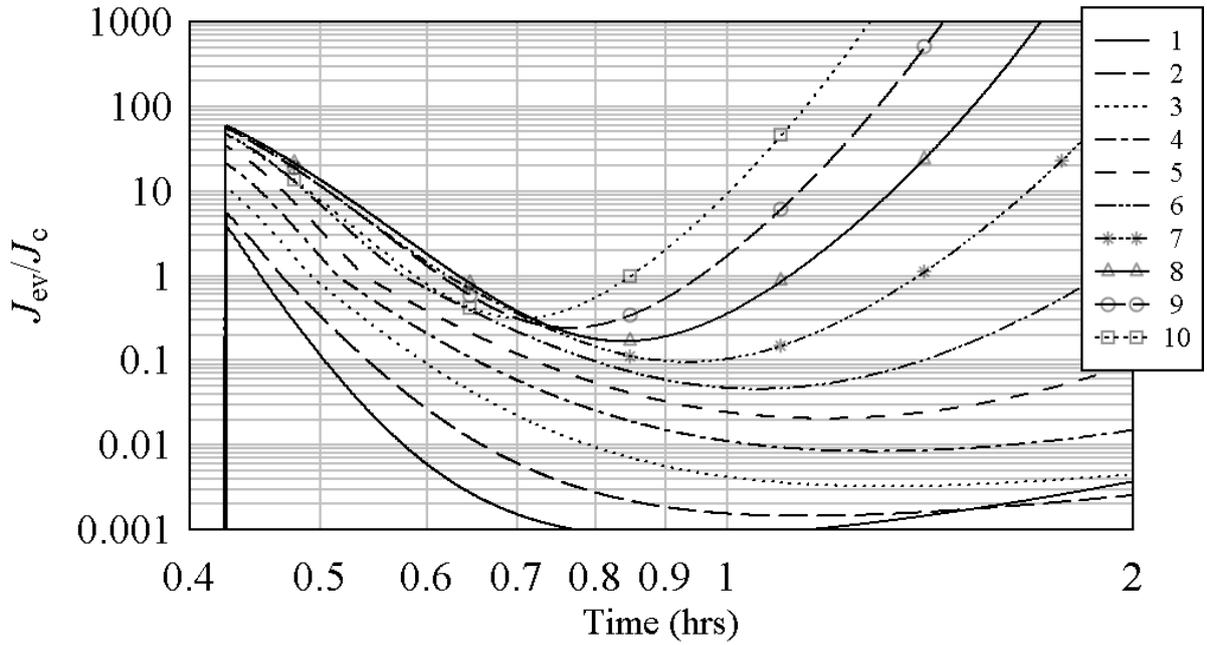

Fig. 16: The ratio of evaporative gas fluxes, to collisional gas flux, $J_c$, at the surface of a CAI. When $J_{ev}/J_c > 1$, the CAIs are evaporating, but when $J_{ev}/J_c < 1$ the CAIs are undergoing WLR formation.

Such high initial temperatures produce significant evaporative loss, which we show in Fig. 17. Similarly, the masses of the CAIs decrease by factors of 30% to 85%. It is apparent that the smallest evaporative losses occur for the particles that have the smallest and largest initial ejection speeds. CAIs with intermediate ejection speeds suffer the greatest evaporative mass loss.

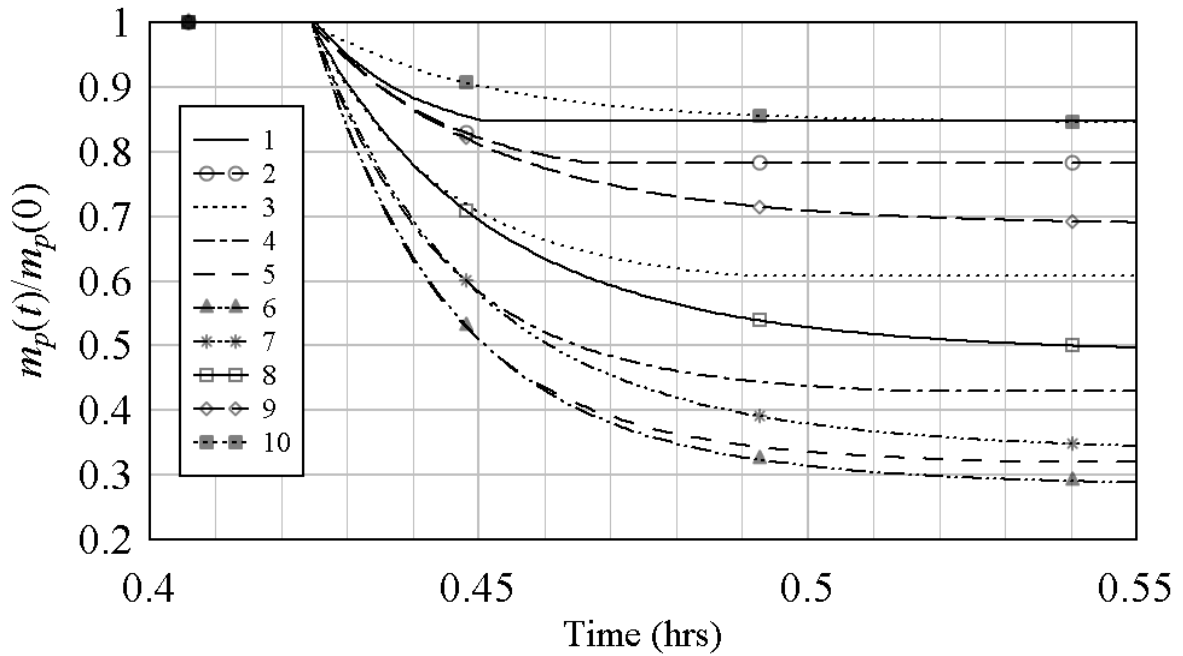

Fig. 17: Evaporative mass loss suffered by 1 cm CAIs that enter the inner wall of the protosolar disc. The mass decreases by a factor of 30% to 90%. CAIs that have the lowest or highest ejection speeds suffer the least mass loss. Conversely, CAIs with the midrange ejection speeds suffer the most mass loss.

Our evaporation mass loss model (Section 2.10) is an approximate model as it implicitly assumes that the constituents of the evaporating solid/liquid remain constant and well mixed with time. In reality, different elements or compounds will have different evaporation rates, so the parameters that describe evaporative mass loss will change with time. None-the-less, a mass loss of 30% is consistent with the deduced evaporative mass loss of magnesium in some Type A and B CAIs (Grossman et al. 2000). An evaporative mass loss of 90% is consistent with the deduced evaporative mass loss of magnesium from FUN (Fractionation and Unidentified Nuclear effects) CAIs (Williams et al. 2014).

In addition to high temperatures and evaporative mass loss, for the first twenty minutes to an hour of the CAI's hypersonic interaction with the inner disc gas, they will be subject to ram pressures (Eq. (64)) as the particles travel through the inner disc wall. Fig. 18 displays the calculated ram pressure for each particle. The initial ram pressures are in the range of $10^{-3}$ to $10^{-2}$ atmospheres, but these pressures quickly decrease on the aforementioned timescales.

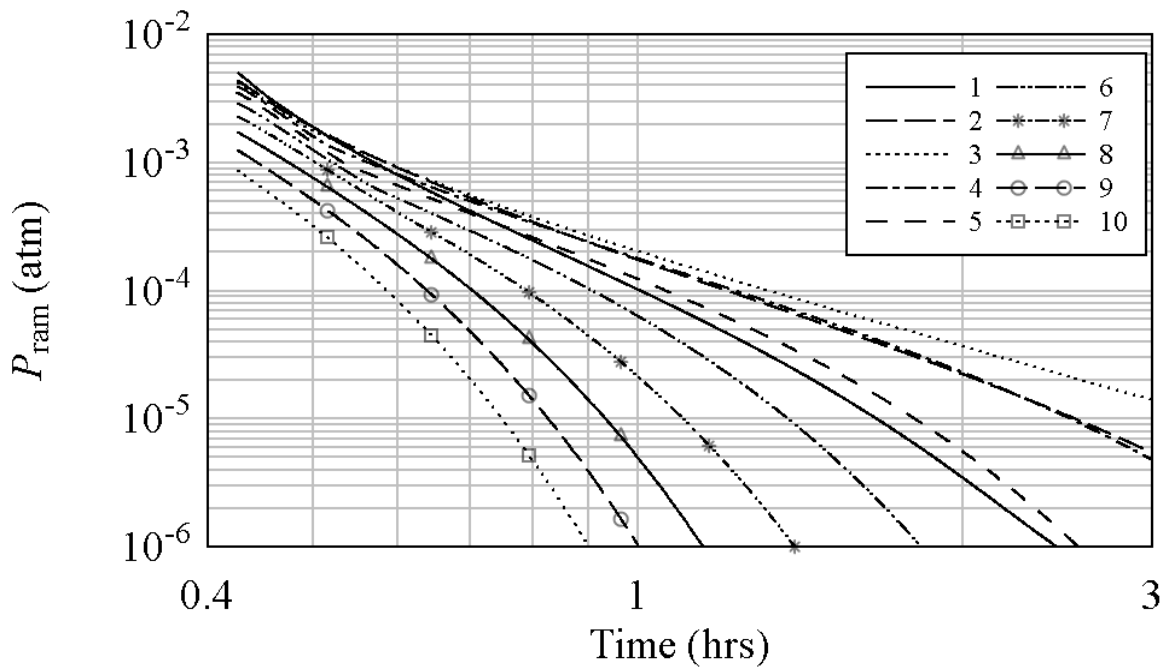

Fig. 18: Ram pressures experienced by the particles as they travel through the inner disc wall. The highest ram pressures are of order $10^{-3}$ atmospheres in magnitude and occur in the first half an hour of entering the disc wall.

The olivine/forsterite WLR thicknesses obtained by the CAIs are shown in Fig. 19. The rim thicknesses are less than equal to 35 microns, where the thickest rims occur for particles that are moving through the denser gas. To obtain these WLRs, we have assumed that the abundance of "metals" in the inner disc wall is ten times that of standard solar abundances and that WLRs only form for gas-particle speeds in excess of 1 km s$^{-1}$.

This high abundance of metals is explained via a simple recycling model as given in Appendix B which assumes that a portion of the CAIs that enter the inner disc wall are evaporated via the re-entry heating. The model shows that one can obtain a factor of ten increase in the mass abundance of an element provided that 90% of the element is returned via the CAIs that leave the disc wall to accrete onto the Sun and are then returned via the ejection/evaporation process.

The total thickness of WLRs can be up to 100 microns. The results given in Fig. 19 suggest that thicker WLRs are formed when the CAIs are travelling through denser gas. Higher gas densities occur closer to the disc midplane. However, in this simulation, we have our initial ejection point at 3.5 scale heights above the disc midplane. We did this, because a launch point closer to the midplane tends to vaporise all the CAIs as they enter the disc. However, there may be other values in the phase space of this particular simulation, which allow the survival of the CAIs and allows them to accrete larger WLRs. We will not consider these other possibilities in this study.

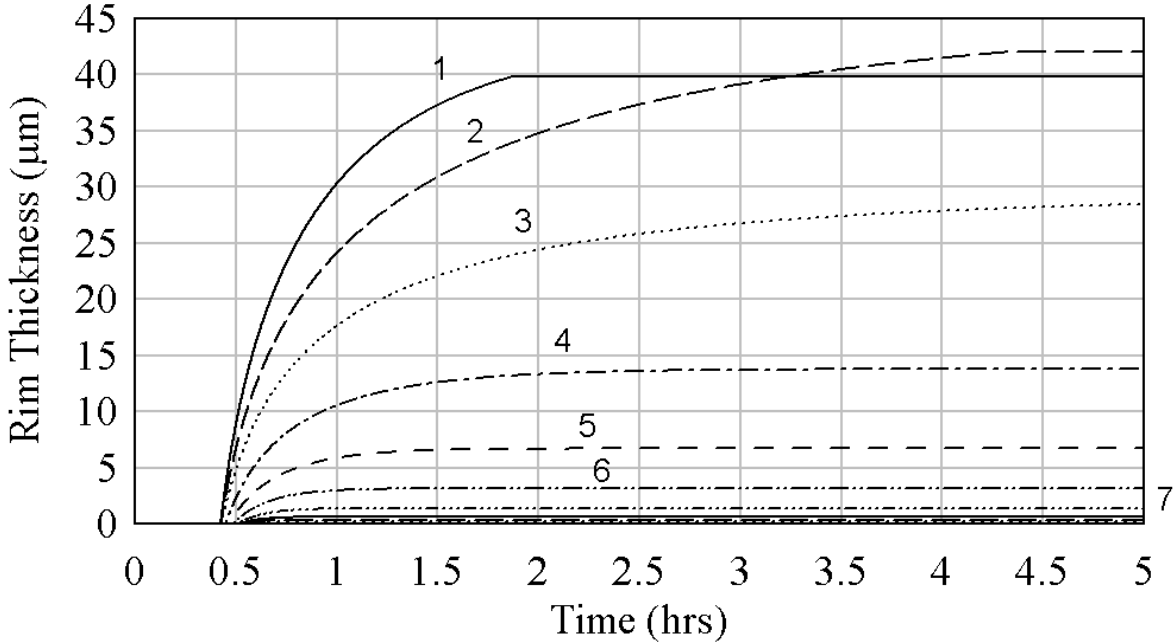

Fig. 19: Forsterite WLR thicknesses obtained for 1 cm CAIs. The WLR forms after the initial evaporation phase. The thickest olivine/forsterite rim is 35 microns in thickness. This occurs for the lowest speed CAI that is travelling through the densest gas. In this calculation, we have assumed that WLRs only form at particle speeds relative to the gas greater than 1 km s$^{-1}$. We have also assumed that the abundance of Mg and other relevant elements in the inner disc wall is 10 times standard solar abundances.

### 3.2.1  Inner Disc Wall Path, 1 mm particles, $3\times10^{-8}$ M$_\odot$yr$^{-1}$, $R_t > R_{co}$

We now consider the same parameter space as in Section 3.2, but use 1 mm diameter CAIs instead of 1cm CAIs. The resulting projectile paths are shown in Fig. 20. Relative to the 1 cm diameter CAIs (Fig. 13), the 1 mm CAIs travel shorter distances. None of the 1 mm CAIs are ejected from the protosolar disc and 0.4 au is the longest downrange radial distance obtained from the protosun. This is due to the greater relative drag force that acts on the 1 mm CAIs as the enter the inner disc wall.

The initial temperatures of the CAIs as they enter the disc are similar to those obtained for the 1 cm particles, but the cooling times are significantly shorter (Fig. 21). For example, particles 1 and 2, that have the lowest vertical ejections speeds of 5.8 and 17.3 km s$^{-1}$, cool from initial temperatures of around 2800K to 1400K in around five minutes. The resulting mass loss due to evaporation is shown in Fig. 22. As in the previous section, particles with low ejection speeds (e.g., particles 1 and 2) suffer little or no mass loss. Particles with the highest ejection speeds (particles 9 & 10) enter the disc at higher altitudes and are subject to lower gas densities, so their evaporative mass loss is small relative to the particles in the midrange of ejection speeds. Such particles suffer significant mass loss.

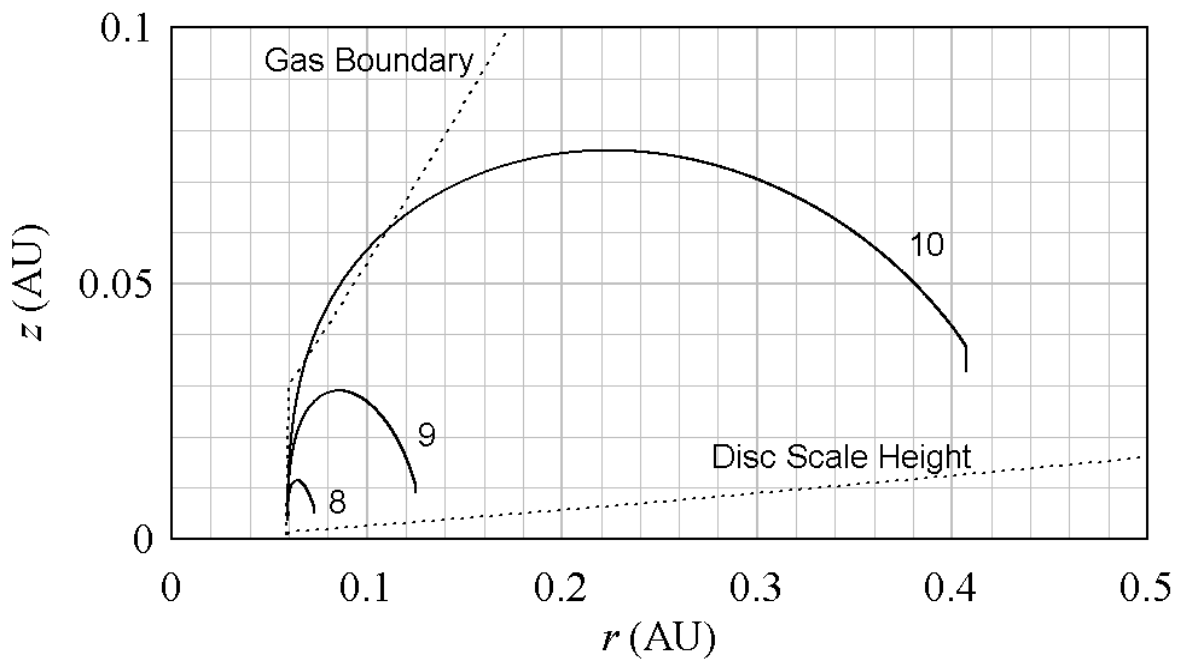

Fig. 20: Projectile paths for 1 mm diameter CAIs with the same initial parameters as the particles shown in Fig. 13.

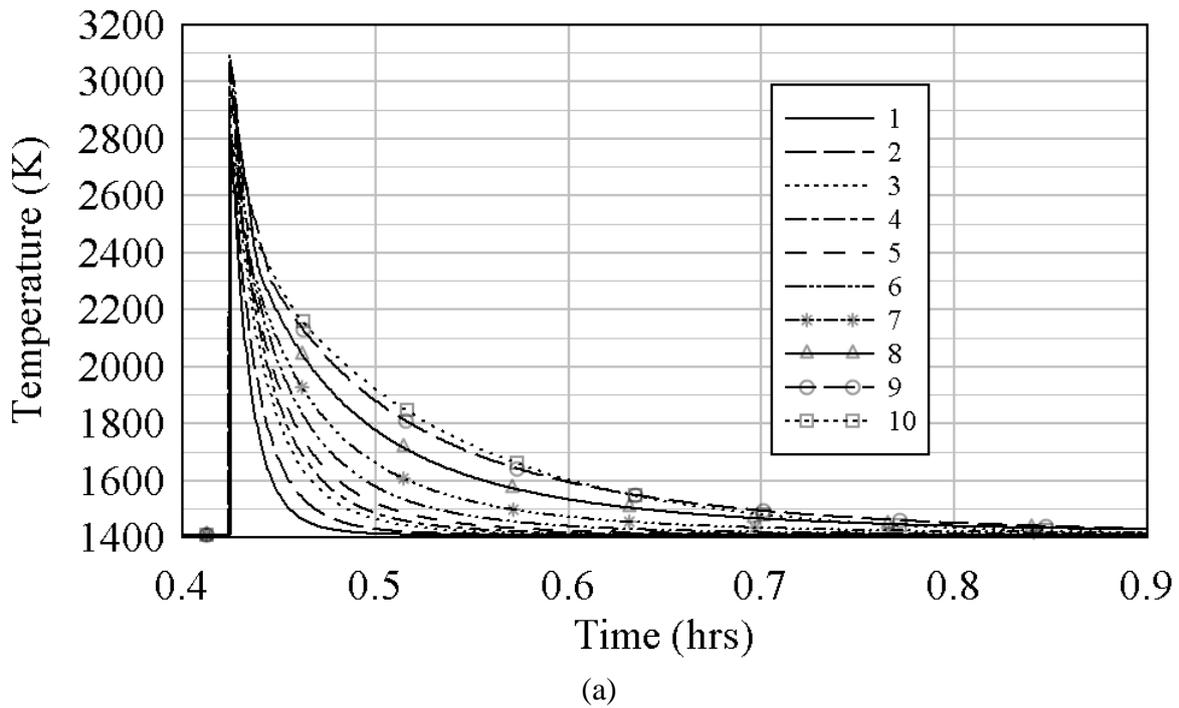

(a)

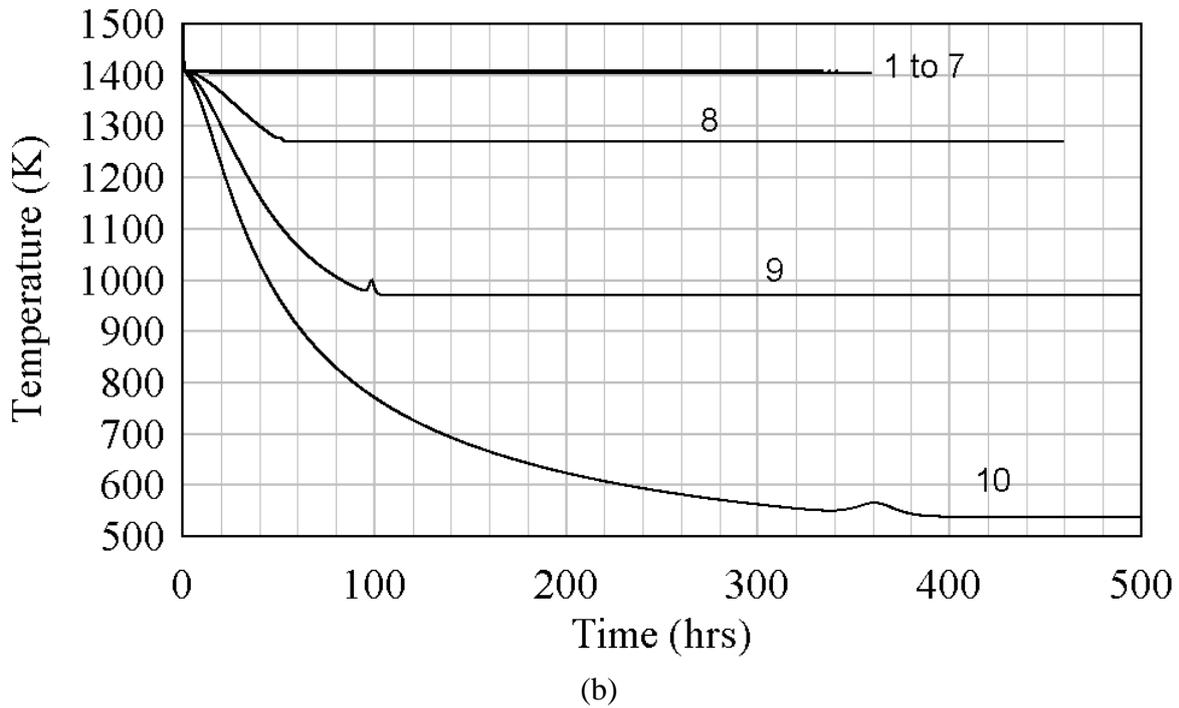

(b)

Fig. 21: Particle temperatures as a function of time. (a) The 1 mm diameter CAIs start with a temperature of around 1400K. As the particles enter the inner rim wall, they are subject to significant frictional heating, which heats the particles to a range of temperatures between 2,600 and 3,100 K. The particles cool to the equilibrium 1400K temperatures on a time scale of approximately 3 to 10 minutes. (b) Particles 8 to 10 are ejected from the neighbourhood of the inner disc wall and subsequently cool on timescales of days. These particles suffer a small temperature increase on re-entry into the denser regions of the disc away from the inner disc wall.

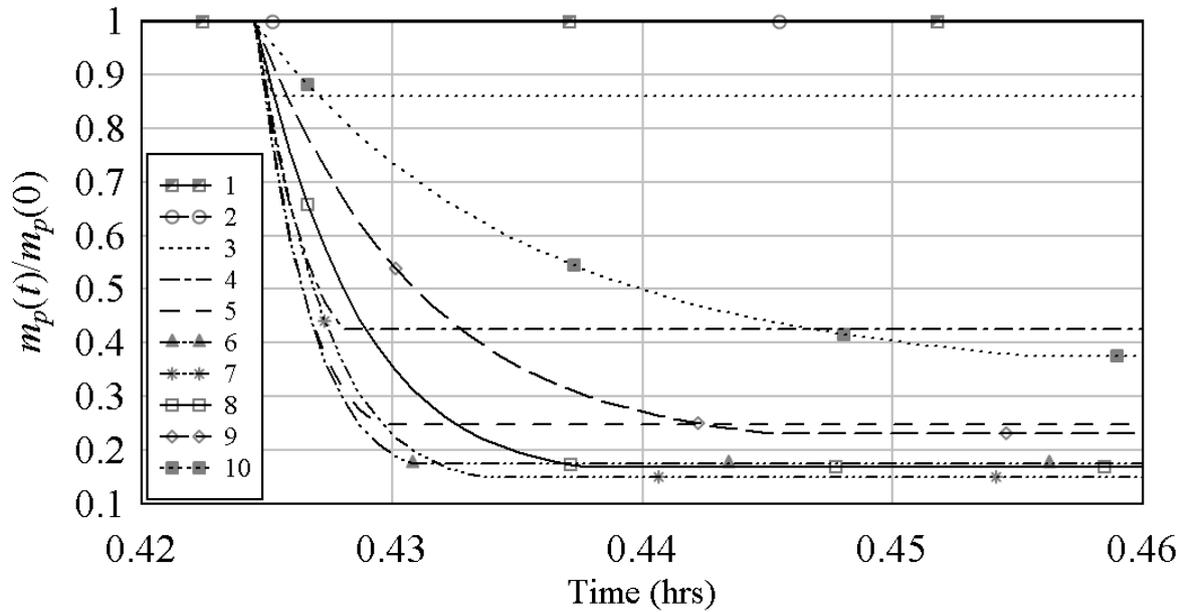

Fig. 22: Evaporative mass loss suffered by 1 mm CAIs that enter the inner wall of the protosolar disc. Particles 1 and 2 suffer no mass loss. The masses of the remaining particles decrease by a factor of 15% to 85%. CAIs that have the lowest or highest ejection speeds suffer the least mass loss. Conversely, CAIs with the midrange ejection speeds suffer the most mass loss.

Overall, with the exception of the first two particles, the 1 mm CAIs tend to suffer more evaporative loss compared to the 1 cm particles. Wark & Boynton (1984) undertook a brief, preliminary study regarding the possible relationship between size and composition for Allende CAIs. They found that smaller CAIs had a Rare Earth Element (REE) enrichment of 12 to 26 times solar abundance (a possible sign of very significant evaporation), which is comparable to the REE enrichment observed in larger CAIs. They also found that the smaller CAIs were enriched in the refractory corundum ($Al_2O_3$) relative to larger CAIs. They concluded that there is a possible, albeit, weak correlation between CAI size and composition, where smaller CAIs are more refractory compared to larger CAIs. This result is consistent with our numerical results, but with the usual caveat that these are preliminary experimental and numerical results.

Finally, in Fig. 23, we find that the thickness of WLRs for 1 mm CAIs is, approximately, a factor of four smaller than for the rims around 1 cm CAIs (Fig. 19). The reason why the smaller CAIs obtain thinner WLRs is due to their smaller initial momentum and the subsequent shorter time/distances the smaller particles move through the inner disc gas at hypersonic speeds. As a consequence, they encounter less condensable WLR material as compared to larger CAIs. To our knowledge, there has been no quantitative analysis of WLR thickness as a function CAI size. As a consequence, it is a prediction of this study that, on average, WLR thickness should be roughly proportional to the size of the CAI. That is larger CAIs will tend to have thicker WLRs.

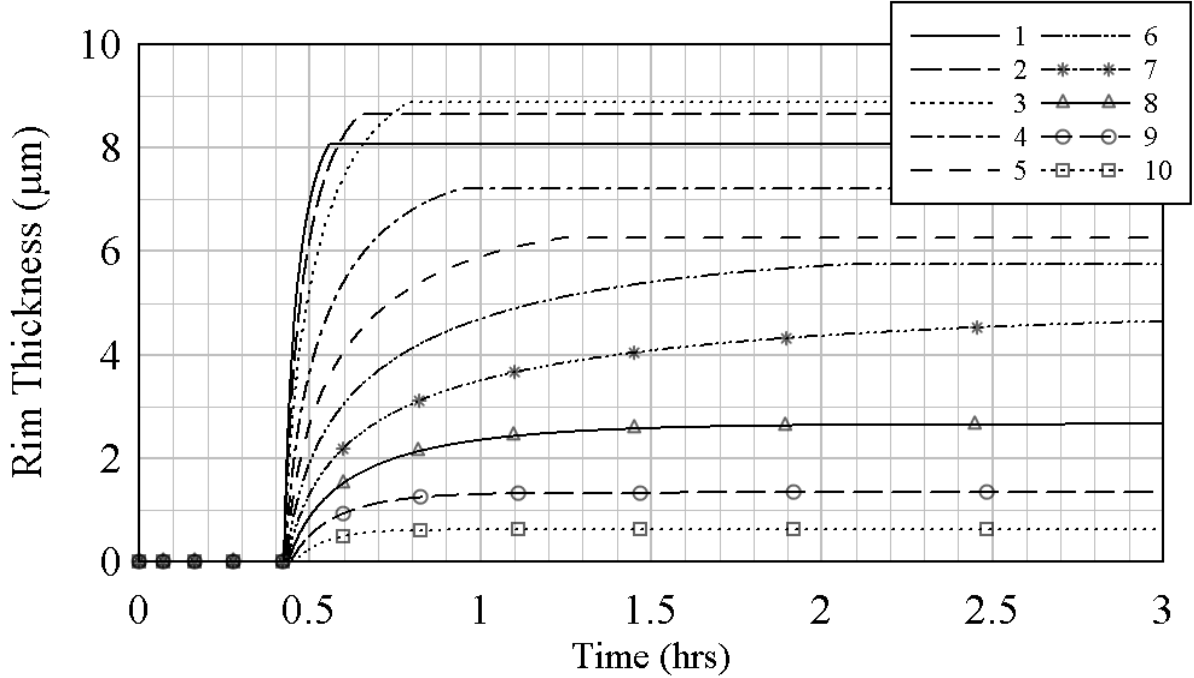

Fig. 23: Forsterite WLR thicknesses obtained for 1 mm CAIs. The WLR forms after the initial evaporation phase. The thickest olivine/forsterite rim is 9 microns in thickness. This occurs for the lowest speed CAI that is travelling through the densest gas. In this calculation, we have assumed that WLRs only form at particle speeds relative to the gas greater than 1 km s$^{-1}$. We have also assumed that the abundance of Mg and other relevant elements in the inner disc wall is 10 times standard solar abundances.

## 4 Discussion

In this study, we argue that CAIs were first formed in the very inner regions of the protosolar disc, where the protosun interacted with the inner disc via direct radiation and the solar magnetosphere. The approximate gas temperatures and densities in this region are described by Eq. (8) and Eq. (9), respectively. For the accretion disc and protosolar values given in Section 3.2, the resulting inner disc pressures and temperatures are shown in Fig. 4. These high pressures plus the complex interplay between the high gas temperatures and the higher dust temperatures (due to direct radiation) provide an environment where high temperature condensates can form. This is the starting point of our model, where these condensates then accrete towards the Sun via the solar magnetosphere, but are subsequently ejected via centrifugal forces (Section 2.6 and Fig. 10).

This ejection mechanism is based almost purely on standard Newtonian mechanics. It does not use exotic ejection mechanisms such as jet outflows (e.g., Shu et al. 1996). The mechanism requires accretion along the solar magnetosphere onto the protosun from the protosolar disc, where the accretion mechanism drags particles away from the midplane of the disc. Due to this motion, the centrifugal force due to the particle's initial keplerian angular velocity can become larger than the centripetal radial component of the gravitational force connecting the CAI to the Sun and hence the CAI is ejected from the inner disc.

The mass accretion flows from the protosolar disc to the protosun arise from the direct interaction between the solar magnetosphere and the disc. The geometry of this situation makes it likely that the CAIs re-entered the disc, possibly, at hypersonic speeds. In this scenario, the inner protosolar disc wall becomes the primary site for CAI processing. This is particularly true if the inner disc is outside the co-rotation radius of the Sun. In such a case, a co-rotating solar magnetosphere will have a higher angular speed than the keplerian speed at the inner disc wall. Particles that begin to co-move with the gas of the co-rotating magnetosphere will be more likely to be projected, via this "propeller" mechanism, in a radial direction.

CAI that re-enter via the inner disc will be subject to significant gas drag and frictional heating (Sections 2.7 and 3.2). For sufficient particle speed and gas density, it is possible for the molten CAI to be deformed into a bowl shape (Section 2.14). Such re-entry heating also produces significant evaporation of the CAI (Fig. 17), which is consistent with observations that CAI suffered major evaporative losses (Grossman et al. 2000, Williams et al. 2014). Preliminary simulation results suggest that smaller (1 mm diameter) CAIs tend to suffer more evaporation relative to larger (1 cm diameter) CAIs. This result is consistent with preliminary observational results where smaller CAI tended to suffer more evaporation relative to larger CAI (Wark & Boynton 1984).

The evaporation of CAI would have injected metal-rich gas into the disc inner wall. As a consequence, we predict that the inner walls of YSO discs will tend to be metal rich relative to the rest of the disc due to this recycling effect.

Our model for the formation of the nearly ubiquitous WLRs (Section 2.13) assumes that there are nanodroplets of mineral condensates in the atmosphere of the inner disc. The WLR forms as a coating due to the hypersonic passage of the CAI through this saturated medium. An analogous technique, called HVOF, is used in industry to create strong, low porosity coatings. The different layers of the WLR were, presumably, due to the passage of the CAI through the different micro-climates and corresponding condensate regimes within the inner disc wall. To simulate the ~100 micron thickness observed
in some WLRs, it was necessary to increase the abundance of "metals" in the gas by a factor of ten relative to standard solar abundances (Section 3.2 and Fig. 19). A simple recycling model given in Appendix B shows that one can obtain a factor of ten increase in the mass abundance of an element provided that 90% of the element is returned from the material that leaves the disc wall to accrete onto the Sun.

If the CAI had sufficient momentum to punch through the inner disc wall and traverse the upper atmosphere of the disc then it would slowly cool as it moves on a path away from the protosun (Fig. 13). This cooling rate is simply due to the decrease in solar radiation received by the CAI as it moves away from the Sun. Cooling rates of around 10 K hr$^{-1}$ are theoretically expected for such a process (Eq. 47). Our simulations give cooling rates of 1 to 20 K hr$^{-1}$ (Fig. 15(c)). These cooling rates are similar to what has been deduced from studying CAIs (Stolper & Paque 1986). The reprocessed CAI eventually re-enters the disc at distances far away from the inner disc rim. In this way, CAIs were samples of the inner protosolar disc that were transported across the entire disc.

# 5  Conclusions

In this study, we have provided a comprehensive theory to explain the processing and transport of CAIs. CAIs are the oldest known rocks to have formed in the Solar System (Connelly et al. 2012). They were formed at high temperatures (MacPherson et al. 2005), are enriched in $^{16}$O, with the ratios of the oxygen isotopes similar to solar values (McKeegan et al. 2011) and they once contained the short lived radionuclide $^{10}$Be (Gounelle et al. 2013). This evidence is consistent with the idea that CAIs formed close to the protosun in the inner most regions of the protosolar accretion disc. In addition, CAIs are found in most primitive meteorites and are also present in Comet Wild 2 (Brownlee 2014). CAIs were transported from the inner disc regions to the outermost reaches of the protosolar disc.

Spitzer Space Telescope observations of the protostars Ex Lupi (Ábrahám et al. 2009) and HOPS 68 (Poteet et al. 2011) indicate that refractory minerals, in this case forsterite, are formed close to the protostar and then ejected at high speed (~40 km s$^{-1}$) across the face of the disc from the inner to the outer regions of the disc (Juhász et al. 2012). Infrared observations of discs in young stellar systems show that high temperature, crystalline silicate material (e.g. olivine and pyroxene) is more likely to be found near the protostar than in the outer sections of the disc (van Boekel et al. 2004).

Given this observational evidence, we propose that CAIs were transported at hypersonic speeds from the inner protosolar disc to the outer regions of the disc. With this assumption, the deduced CAI cooling rates are derived almost immediately (Section 2.7) as they are a consequence of the CAI moving further away from the protosun.

The proposed transport mechanism is centrifugal ejection from the accretion column of gas that is accreting from the protosolar disc to the protosun (Section 2.6). The ejection mechanism uses standard, Newtonian physics and does not require more complex ejection mechanisms – such as protostellar jet flows.

The geometry of the accretion column along the solar magnetosphere connecting the Sun to the protosolar disc (Adams & Gregory 2012) raises the possibility that the centrifugally ejected CAIs re-enter the inner disc wall at hypersonic speeds. This re-entry process can deform liquid CAIs and thereby change the CAI droplets from a spherical shape to a bowl-shape (Section 2.14). The re-entry heating also produces evaporation from the CAIs (Section 2.10). Our simulation results give amounts of CAI evaporation that are consistent with observations (Section 3.2). As the CAI penetrates further into the disc, its speed and temperature decrease and the CAI accretes condensate material that has evaporated from other CAIs. We claim that this coating process produces the WLRs that are found around most CAIs (Section 2.13). As such, CAIs are unique samples of the inner disc from the very beginning of the Solar System.

# Acknowledgements

This research was partly supported by the Programme National de Physique Stellaire and the Programme National de Planétologie of CNRS/INSU, France. NC is grateful to the LABEX Lyon Institute of Origins (ANR-10-LABX- 0066) of the Université de Lyon for its financial support within the program "Investissements d'Avenir" (ANR-11- IDEX-0007) of the French government operated by the National Research Agency (ANR). KL acknowledges the support from Prof. Sarah Maddison and the Centre for Astrophysics and Supercomputing, Swinburne University of Technology.

# Appendix A  Gas Frictional Heating – No Evaporation

In our model, assuming no evaporation, the amount of heating a particle undergoes as it re-enters the disc atmosphere is given by

$$\frac{4}{3}\pi \rho_p a_p^3 C_p \frac{dT_p}{dt} \approx 4\pi a_p^2 C_H \rho_g v_{pg}(T_r - T_p) + \frac{a_p^2(L_* + L_a)\varepsilon_a}{4R^2} - 4\pi a_p^2 \varepsilon_e \sigma_{SB} T_p^4 . \quad (A.1)$$

Typically, the particles will be re-entering the disc at hypersonic speeds, so, $C_H$, the heat transfer coefficient (Eq. (49)) and, $T_r$, the recovery temperature (Eq. (50)) have the limits:

$$\lim_{S \to \infty} C_H = \frac{k_B(\gamma+1)}{8\bar{m}(\gamma-1)} \quad (A.2)$$

and

$$\lim_{S \to \infty} T_r = \frac{2T_g(\gamma-1)S^2}{\gamma+1} . \quad (A.3)$$

Dividing by $4\pi a_p^2$ and taking the large $S$ limit, the energy equation has the form

$$\frac{\rho_p a_p C_p}{3}\frac{dT_p}{dt} \approx \frac{(\gamma+1)k_B \rho_g v_{pg}}{8\bar{m}(\gamma-1)}\left(\frac{2T_g(\gamma-1)S^2}{(\gamma+1)} - T_p\right) + \frac{(L_* + L_a)\varepsilon_a}{16\pi R^2} - \sigma_{SB}\varepsilon_e T_p^4 . \quad (A.4)$$

$$\approx \frac{\rho_g v_{pg}^3}{8} + \frac{(L_* + L_a)\varepsilon_a}{16\pi R^2} - \sigma_{SB}\varepsilon_e T_p^4$$

For large $R$, this becomes:

$$\frac{\rho_p a_p C_p}{3}\frac{dT_p}{dt} \approx \frac{\rho_g |v_p - v_g|^3}{8} - \sigma_{SB}\varepsilon_e T_p^4, \quad (A.5)$$

which gives an indicative steady state temperature of

$$T_p \approx \left(\frac{\rho_g v_{pg}^3}{8\sigma_{SB}\varepsilon_e}\right)^{1/4} = 1938\,\mathrm{K}\left(\left(\frac{\rho_g}{10^{-7}\,\mathrm{kg\,m^{-3}}}\right)\left(\frac{v_{pg}}{40\,\mathrm{km\,s^{-1}}}\right)^3\right)^{1/4}, \quad (A.6)$$

where we have set $\varepsilon_e = 1$.

# Appendix B  Metal/Rock Enrichment of the Inner Disc

We consider a schematic model of the Protosolar disc inner disc, where gas and dust is feeding into the disc via mass accretion from the outer disc region. Material also leaves the inner disc to accrete onto the protosun. However, there is a return of metal/rock-like material to the inner disc due to the centrifugal ejection of such material as it leaves the midplane of the disc (Sections 2.6 and 3.2), where a significant proportion of the CAIs re-entering the inner disc wall suffers substantial evaporation (Sections 2.10 and 3.2). These mass flows are displayed schematically in Fig. 24, where mass accretes into the inner disc at rate $\dot{M}_a$. Similarly, material will leave the inner disc and accrete onto the protosun at a mass rate of $\dot{m}_{out}$, while the centrifugal ejection processes will inject the CAI back into the inner disc at a mass rate of $\dot{m}_{in}$. These ejected particles will penetrate a stopping distance $l_s$ (Eq. (55)) into the disc. The expected stopping distance is of order 0.001 au, which is small relative to the inner truncation radius, $R_t$, (Eq. (1)) which is of order 0.05 au in size.

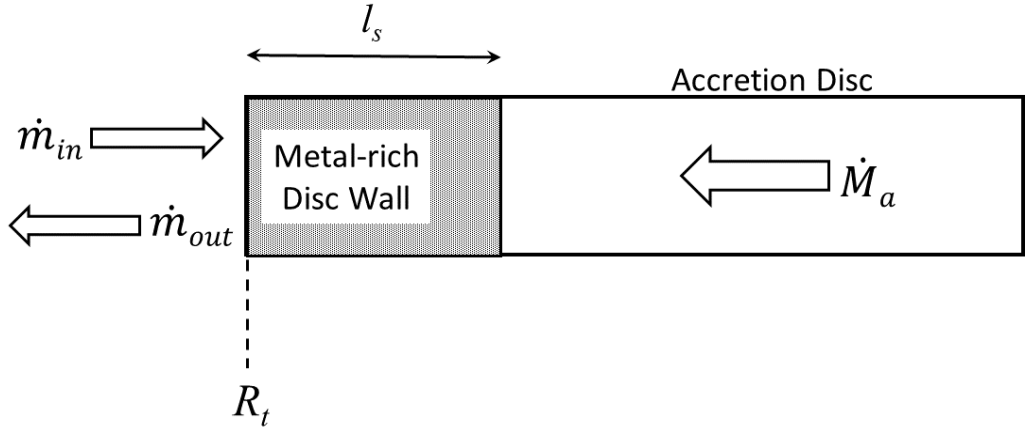

Fig. 24: Mass flows for the inner disc, which acts as a reprocessing environment for CAIs. The disc as a mass accretion rate $\dot{M}_a$, material leaves the inner disc at a mass rate of $\dot{m}_{out}$, while centrifugal ejection processes will inject material back into the inner disc at a mass rate of $\dot{m}_{in}$. The ejected particles will penetrate a stopping distance $l_s$ (Eq. (55)) into the disc. The stopping distance is small relative to the truncation radius, $R_t$, so the inner disc wall becomes enriched with metals relative to the remainder of the accretion disc.

The accretion flow timescale for material to flow through this metal-rich, inner disc region can be deduced from Eq. (23):

$$\tau_a = \frac{2\pi R_t \Sigma l_s}{\dot{M}_a} = 50.4\,\mathrm{yrs}\, \frac{\left(R_t / 0.05\,\mathrm{au}\right)\left(\Sigma / 1.5\times 10^6\,\mathrm{kg\,m^{-2}}\right)\left(l_s / 0.00134\,\mathrm{au}\right)}{\left(\dot{M}_a / 10^{-8}\,\mathrm{M_\odot\,yr^{-1}}\right)}. \qquad (\mathrm{B}.1)$$

We make the assumption that most of the disc material is hydrogen and helium. So the bulk of the material passing through the inner disc wall region satisfies the mass balance equation

$$\frac{\mathrm{d}m_{total}}{\mathrm{d}t} \approx -\frac{m_{total}}{\tau_a} + \dot{M}_a, \qquad (\mathrm{B}.2)$$

where $m_{total}$ is the total mass of material in the inner disc wall region. This equation has the solution:

$$m_{total}(t) \approx m_{total}(t_0)e^{-(t-t_0)/\tau_a} + \dot{M}_a \tau_a \left(1 - e^{-(t-t_0)/\tau_a}\right), \tag{B.3}$$

where $t_0$ is the initial time when the inner wall region starts to fill up with gas and dust. For time intervals greater than $\tau_a$ then

$$m_{total}(\infty) \approx \dot{M}_a \tau_a. \tag{B.4}$$

For a particular "$i$ th" element, the mass balance equation is

$$\frac{dm_i}{dt} \approx -\frac{(1-\chi_i)m_i}{\tau_a} + f_{\odot i} \dot{M}_a, \tag{B.5}$$

with $m_i$ the mass of the element in the inner disc wall, $\chi_i$ is the proportion of $m_i$ that is returned from the accretional flow via the centrifugal ejection mechanism, $f_{\odot i}$ is the solar abundance of the element. Eq. (B.5) has the solution

$$m_i(t) \approx m_i(t_0)e^{-(1-\chi_i)(t-t_0)/\tau_a} + f_{\odot i} \frac{\dot{M}_a \tau_a}{(1-\chi_i)} \left(1 - e^{-(1-\chi_i)(t-t_0)/\tau_a}\right). \tag{B.6}$$

$\Rightarrow$

$$m_i(\infty) \approx f_{\odot i} \frac{\dot{M}_a \tau_a}{(1-\chi_i)}. \tag{B.7}$$

The mass abundance the element in the disc wall, $f_{Wi}$, is

$$f_{Wi} = \frac{m_i(\infty)}{m_{total}(\infty)} \approx \frac{f_{\odot i}}{(1-\chi_i)}. \tag{B.8}$$

We can see that to obtain a factor of ten increase in the mass abundance of a particular element, we require that $\chi_i = 0.9$. In other words, we require that 90% of the element is returned to the inner disc wall.

# Appendix C  Numerical Results for Different Accretion Disc Cases

## Appendix C.1  $3\times10^{-8}$ M$_\odot$yr$^{-1}$, 1 cm particles, $R_{co} > R_t$

For this case, an accretion rate of $3\times10^{-8}$ M$_\odot$yr$^{-1}$ implies that the inner rim of the disc is located at a distance of $R_t = 0.0675$ au from the centre of the star, so the inner rim is larger than the protosolar radius of 2.4 R$_\odot$ = 0.011 au, but smaller than the co-rotation radius at 0.7 au. As a consequence, the keplerian angular speed at $R_t$ is $v_{Kep}(R_t) = 111$ km s$^{-1}$, while the co-rotation speed at the truncation radius is smaller: $v_{co}(R_t) = 104$ km s$^{-1}$. The accretion luminosity, as given by Eq. (6) is approximately 0.33 L$_\odot$.

In this simulation, ten, 1 cm diameter, particles were located at the coordinates $r = 0.99R_t$ and $z = 0$ and were given initial vertical speeds of $v_z$ = 5%, 15%, 25%, 35%, 45%, 55%, 65%, 75%, 85% and 95% of $v_{Kep}(R_t)$. The particles are given an initial angular speed equal to the co-rotation speed (104 km s$^{-1}$) and the resulting projectile paths of the particles are shown in Fig. 25, where it can be seen that the path travelled by each particle is dependent on the initial vertical speed of the particle. The first four particles, which have vertical launch speeds less than or equal to 39 km s$^{-1}$, are not ejected and follow an orbital path between the protosun and the disc. The remaining particles are ejected from the inner disc.

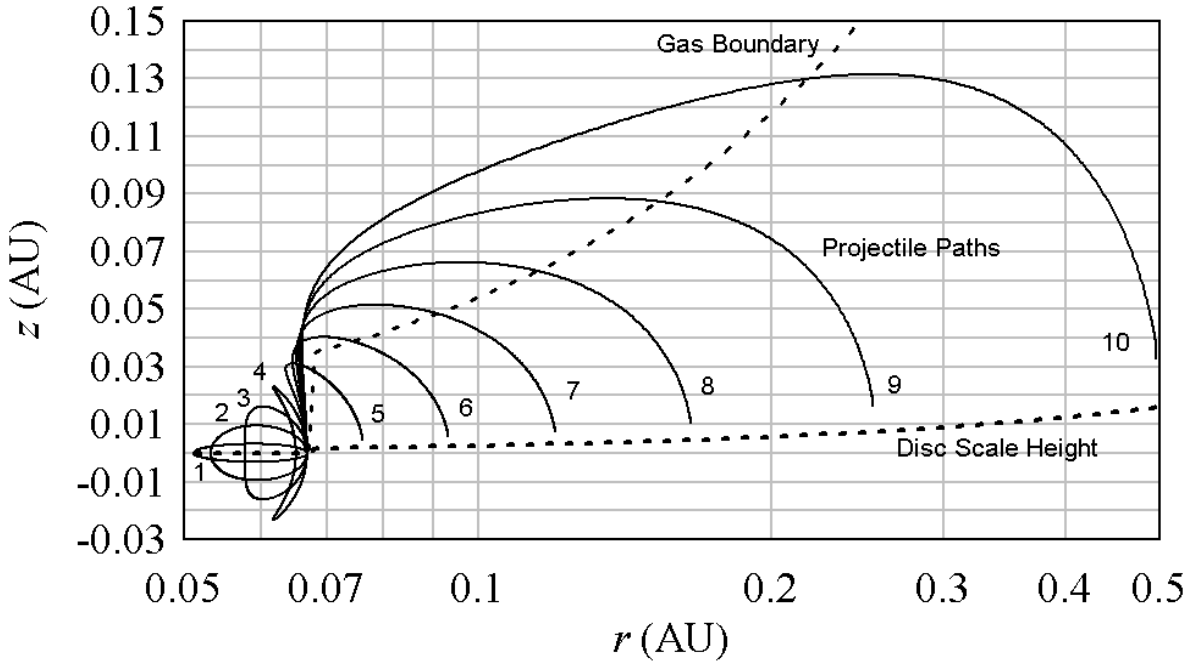

Fig. 25: Projectile paths for the case of $\dot{M}_a = 3\times10^{-8}$ M$_\odot$yr$^{-1}$ for the protosun at an age of $1.2\times10^6$ years. The disc scale height (Eq. (7)) is shown as is the gas boundary, where the density of the gas is effectively zero. The particles are labelled as per the initial $z$ speed, where $v_{1z}$ = 5.6 (km s$^{-1}$), $v_{2z}$ = 16.8, $v_{3z}$ = 28, $v_{4z}$ = 39.1, $v_{5z}$ = 50.3, $v_{6z}$ = 61.4, $v_{7z}$ = 72.6, $v_{8z}$ = 83.8, $v_{9z}$ = 95 and $v_{10z}$ = 106 km s$^{-1}$. The escape speed is 111 km s$^{-1}$. All the particles are launched from the inner truncation radius of $R_t = 0.0675$ au and they have an initial angular speed of 104 km s$^{-1}$ obtained from the co-rotating magnetic field. The particles are subject to gas drag (Eq. (31)) and frictional heating (Eq. (48)) when they enter the disc.

The reason why some of the particles are ejected while others are not, is primarily due to the initial angular speed of the particles which, in this case, is smaller than the keplerian speed. This behaviour can be partly understood from the equation for radial motion:

$$\ddot{r}_p = r_p \dot{\phi}_p^2 - \frac{GM_* r_p}{[r_p^2 + z_p^2]^{3/2}} , \qquad (C.1)$$

where the subscript "p" represents the coordinates of the particle. For outward radial acceleration, we require $\ddot{r}_p > 0$ or

$$r_p \dot{\phi}_p^2 > \frac{GM_* r_p}{[r_p^2 + z_p^2]^{3/2}} , \qquad (C.2)$$

where, for an intuitive physical understanding, we will assume that the initial motion is vertical, i.e., perpendicular to the midplane of the disc and the angular or centrifugal acceleration term ($r_p \dot{\phi}_p^2$) remains constant and with its initial value. Eq. (C.2) can be rewritten to show that

$$z_p \geq z_{eject} \equiv r_p \sqrt{\left(\frac{R_{co}}{r_p}\right)^2 - 1} . \qquad (C.3)$$

Thus, given the initial assumption of vertical motion, a particle will be ejected due to centrifugal acceleration once the altitude of the particle reaches or exceeds the ejection altitude, $z_{eject}$, shown in Eq.(C.3). This behaviour can be seen in Fig. 26, where particles 1 to 4 do not exceed the ejection height and are subject to bound orbits within the inner radius of the disc, $R_t$. The particles that exceed the ejection height move away from the protosun across the face of the disc. The ejection line goes to zero for distances greater than the co-rotation radius. That is, all particles that have angular speeds greater than the keplerian speed will be ejected.

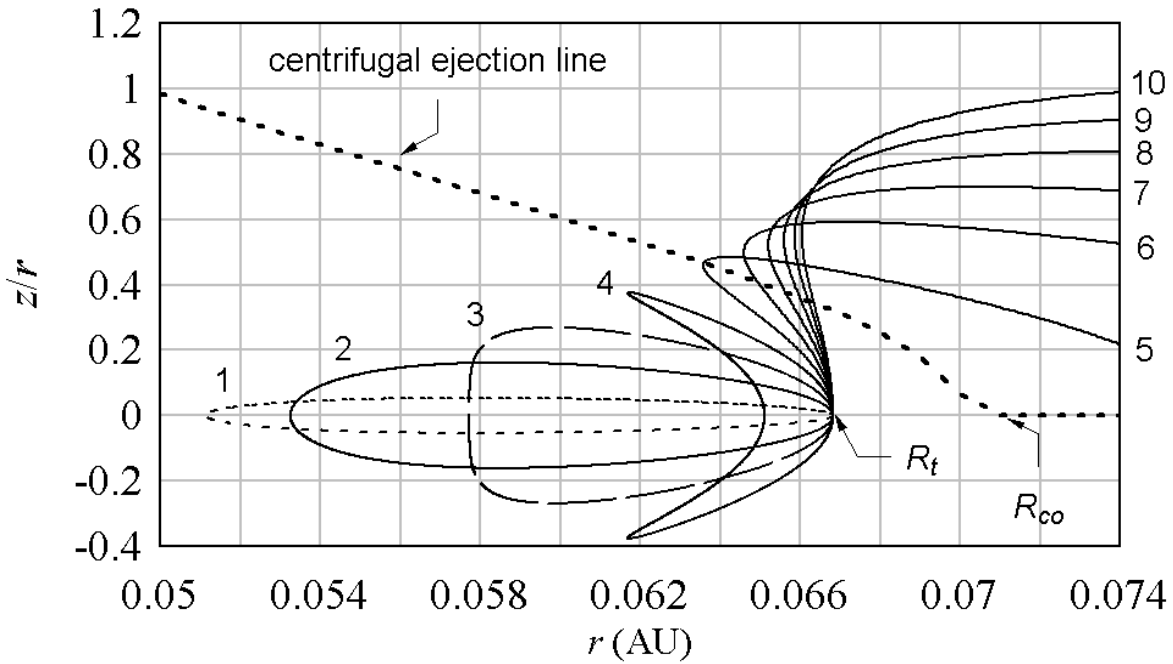

Fig. 26 : A replot of the projectile paths shown in Fig. 25. If the particles go higher than the (approximate) centrifugal ejection line (Eq.(C.3)) then the particles are ejected from the inner regions of the disc. $R_t$ is the inner truncation radius of the disc, while $R_{co}$ is the co-rotation radius.

The temperatures experienced by each particle are shown in Fig. 27. The temperatures of the particles are computed by assuming that they are heated by direct radiation from the protosun (Eq. (46)) and from frictional gas drag (Eq. (48)). The region of the disc through which the particles are traveling is assumed to be optically thin.

The particles start their journey with a temperature of around 1330 K. Particles 1 to 4 are not ejected from the inner disc and suffer periodic temperature changes as their orbits periodically move them closer and further away from the protosun. Particles 5 to 10 are ejected from the inner disc, their effective temperature decreases as they move away from the protosun. However, when they reach the denser regions of the disc then frictional heating can produce a significant increase in particle temperature. For example, the temperature of particle 5 changes from 1,300 K to 1,950 K and back to 1300 K in the space of around two hours. As discussed in section 2.14, it is at this point of the projectile path that may provide a sufficient thermal pulse to produce bowl-shaped CAIs for 1 cm-sized particles.

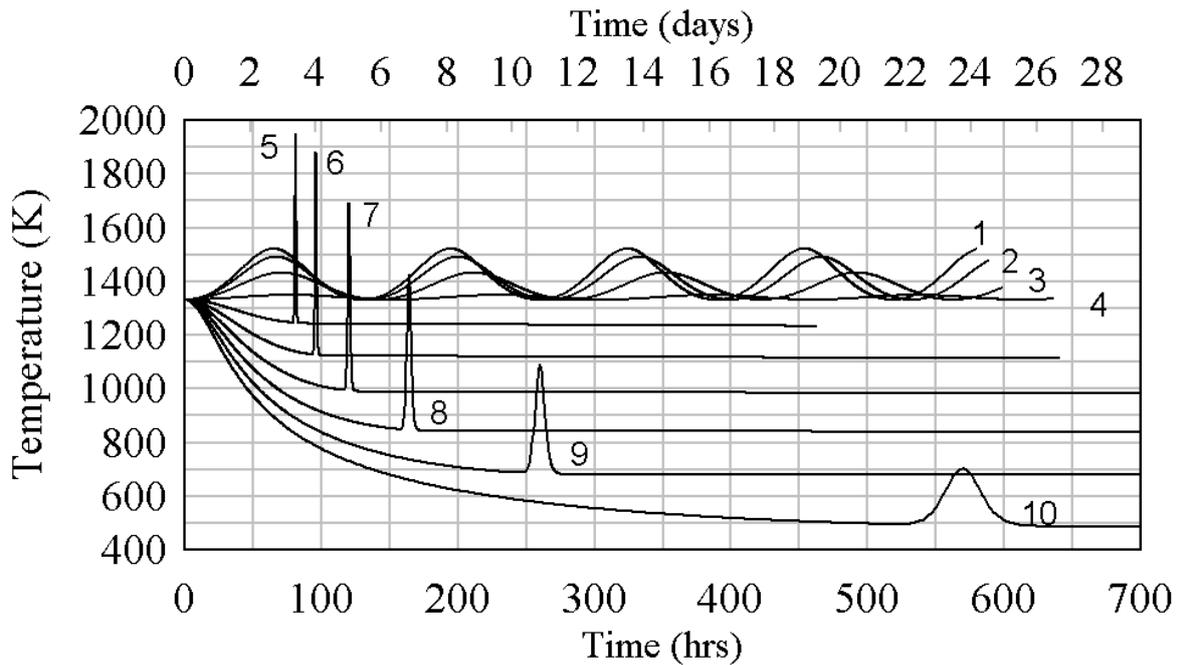

Fig. 27: Particle temperatures as a function of time. The 1 cm CAIs start with a temperature of around 1330K. Particles not ejected from the disc are subject to periodic temperature variations as they orbit the protosun. As the ejected particles move away from the protosun, their temperature decreases until they reenter the denser regions of the disc, where frictional heating produces a spike in particle temperature which lasts for some hours. Significant frictional heating occurs when the CAIs encounter the denser gas regions of the disc at hypersonic speeds. It is assumed that the disc is optically thin.

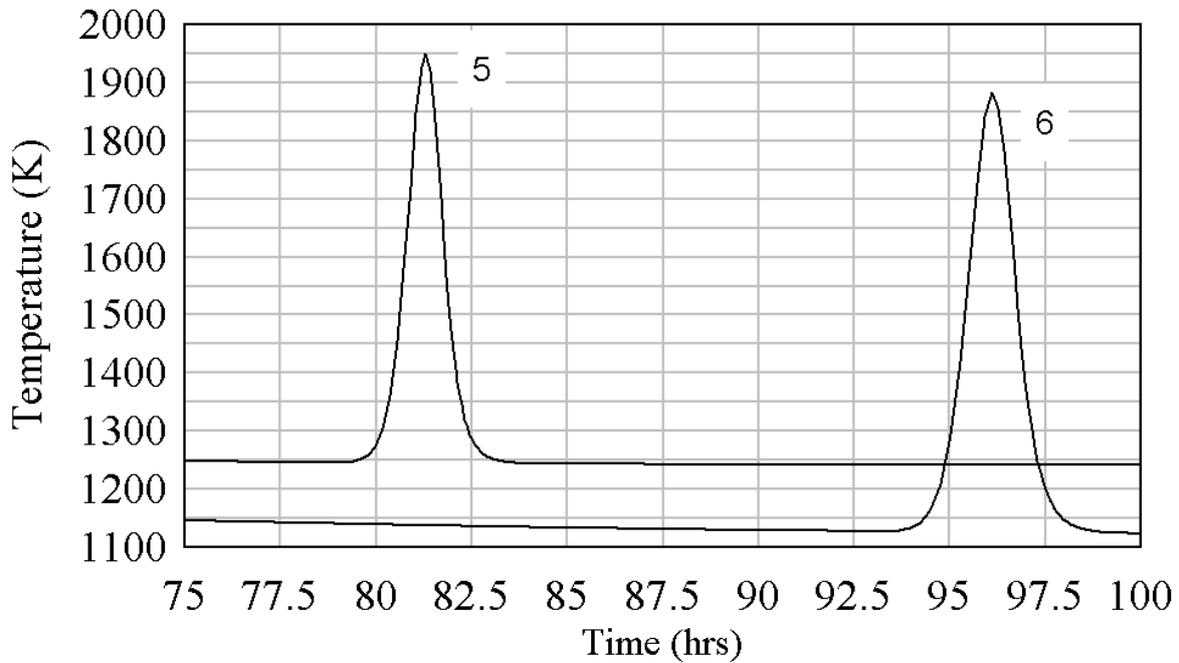

Fig. 28: A magnification of the temperature spikes endured by the particles as they re-enter the accretion disc. For particles 5 and 6, the heating event occurs over the space of a couple of hours.

To compute these frictional heating temperatures, we assume that the gas angular speed is given by the modified Kelperian angular speed for the disc:

$$v_{p\phi} = \frac{r_p \sqrt{GM_*}}{[r_p^2 + z_p^2]^{3/4}} \,. \tag{C.4}$$

As the particles re-enter the disc, they are subjected to increasing gas densities (Fig. 29). The speed of the particle relative to the gas is calculated via the equation

$$v_{pg} = \sqrt{(\mathbf{v}_p - \mathbf{v}_g)^2} = \sqrt{\left(v_{pr} - v_{gr}\right)^2 + \left(v_{p\phi} - v_{g\phi}\right)^2 + \left(v_{pz} - v_{gz}\right)^2} \,. \tag{C.5}$$

These relative speeds are shown in Fig. 30, where we can see that the particles initially travel at hypersonic speeds, relative to the gas, ranging from 33 to 47 km s$^{-1}$.

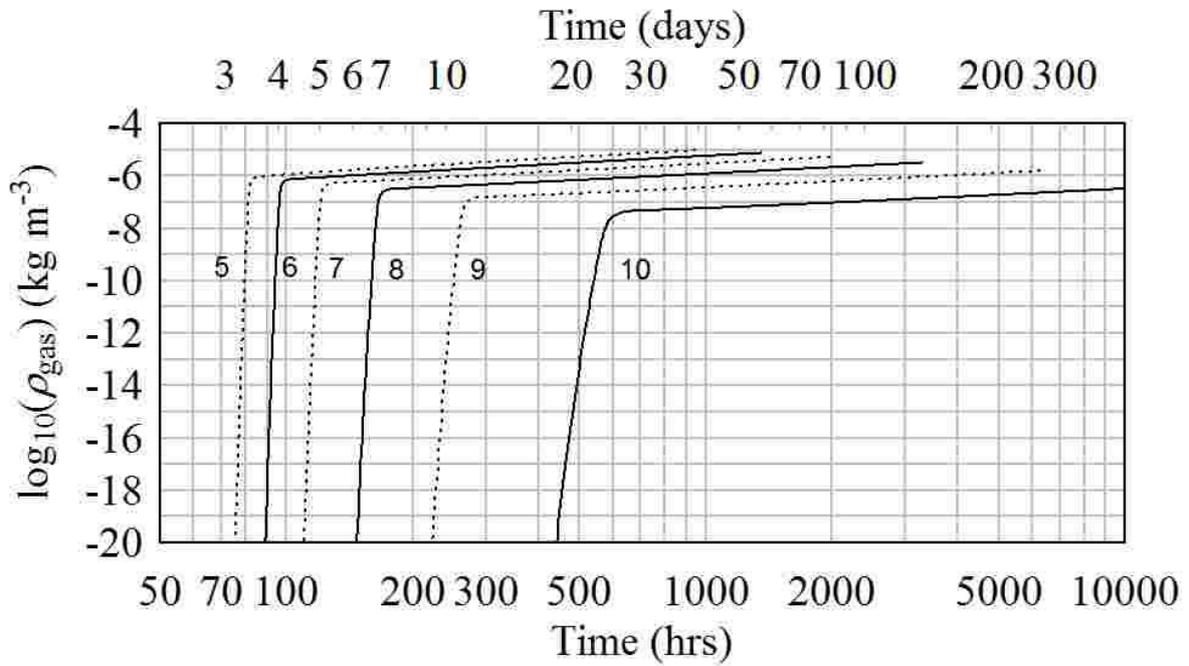

Fig. 29: Gas densities encountered by the particles as they travel from near the protosun and re-enter the disc.

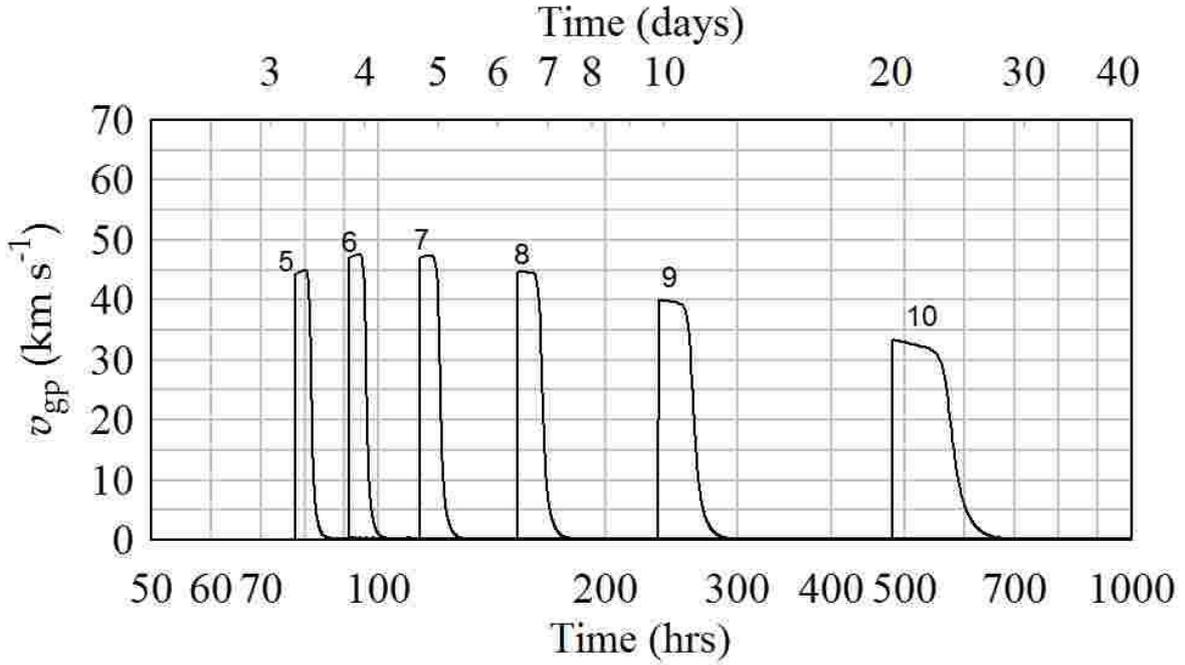

Fig. 30: Speed of the particle relative to the gas. Speeds are only displayed for gas densities in excess of $10^{-15}$ kg m$^{-3}$. The numbers on each curve correspond to the numbers shown on the projectile paths in Fig. 25.

The rates of temperature decline for the particles as they move away from the protosun can be significant (Fig. 31), with some particles experiencing rates approaching -10 K hr$^{-1}$. Such rates are consistent with Eq. (47) and the values deduced by Stolper & Paque (1986).

To determine the distance-averaged rate of temperature decline we use the definition

$$\left\langle \frac{dT}{dt} \right\rangle = \frac{\int_{R_t}^{R_f} \frac{dT}{dt} dr}{R_f} , \qquad (C.6)$$

with $R_f$ the CAI's final distance from the centre of the Sun. In this case, the average rates of temperature decline (Fig. 32) range up to around -4 K hr$^{-1}$.

As can be seen from Fig. 27, the re-entry of the CAIs causes a significant, short-term temperature pulse that can temporarily increase the temperature of the CAI to a temperature of up to 2000 K and then back to equilibrium, effective particle temperature over a timescale of about two hours. The size decrease suffered by the particles is shown in Fig. 33. Fig. 33(a) shows the decrease in size for the particles caught in orbits between the Sun and the inner protosolar disc. Such particles evaporate completely on a timescale of months to years. Fig. 33(b) displays the evaporative loss for the particles ejected from the inner regions of the protosolar disc as they re-enter the protosolar disc.

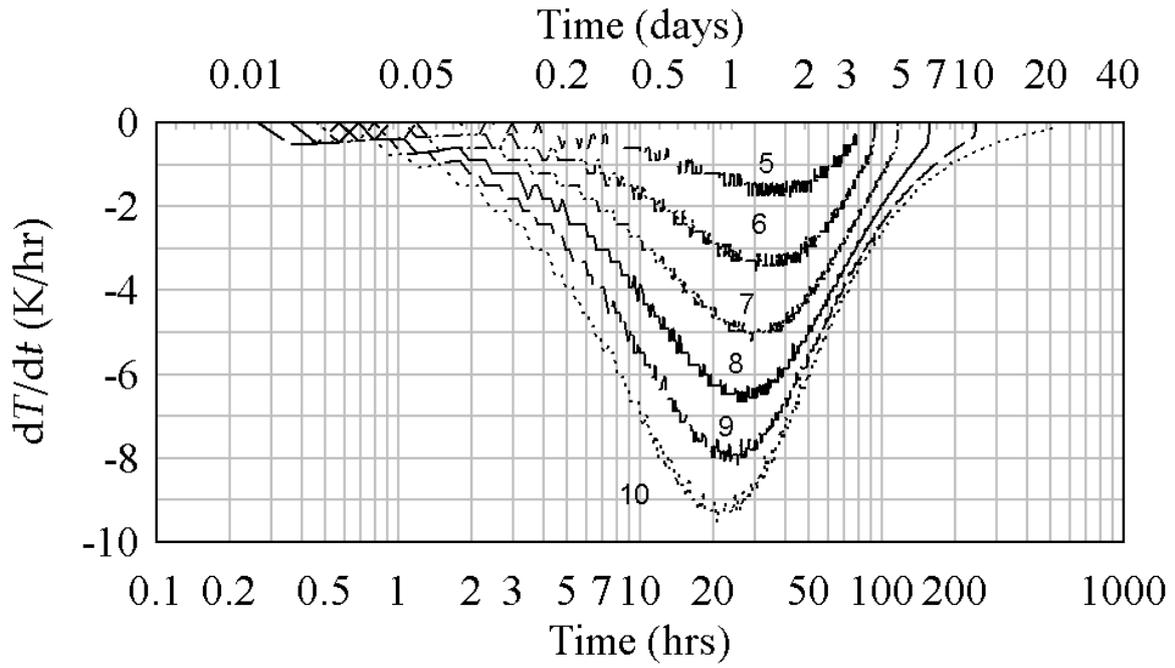

Fig. 31: The rate of cooling as a function of time for the CAI particles ejected from the inner rim of the disc. The particles cool as they move further away from the protosun. These cooling rates are consistent with the experimentally observed cooling rates for type B CAIs (Stolper & Paque 1986) and the deduced cooling rates from Eq. (47).

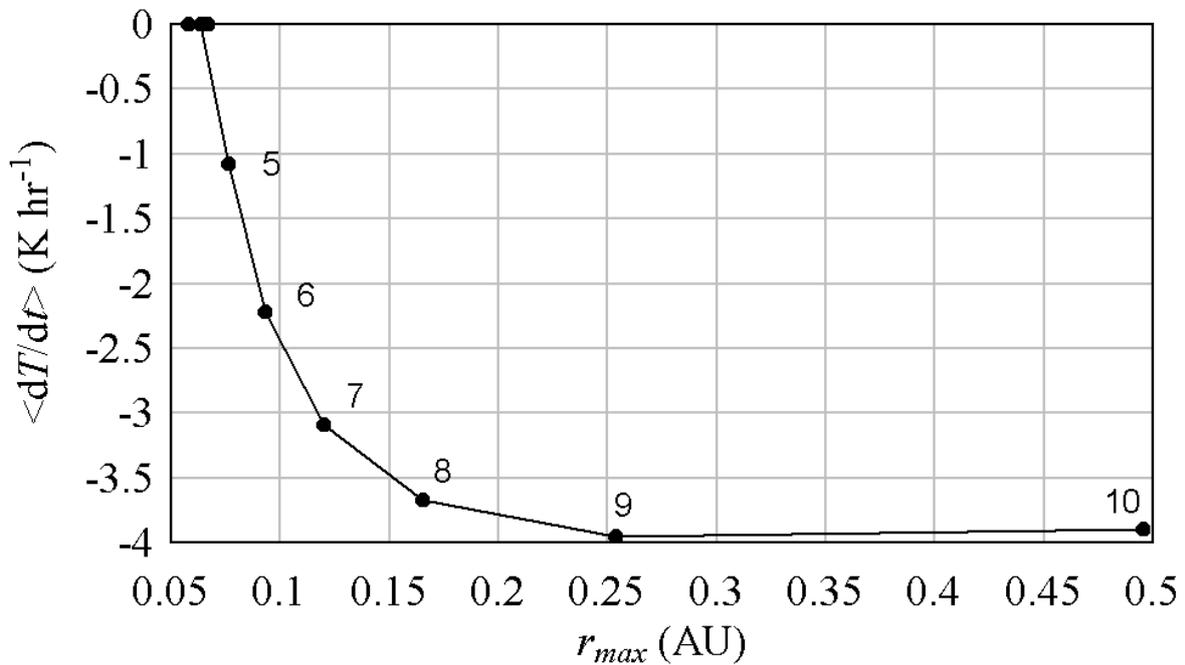

Fig. 32: Average rate of cooling for the CAI particles as a function of the maximum distance travelled.

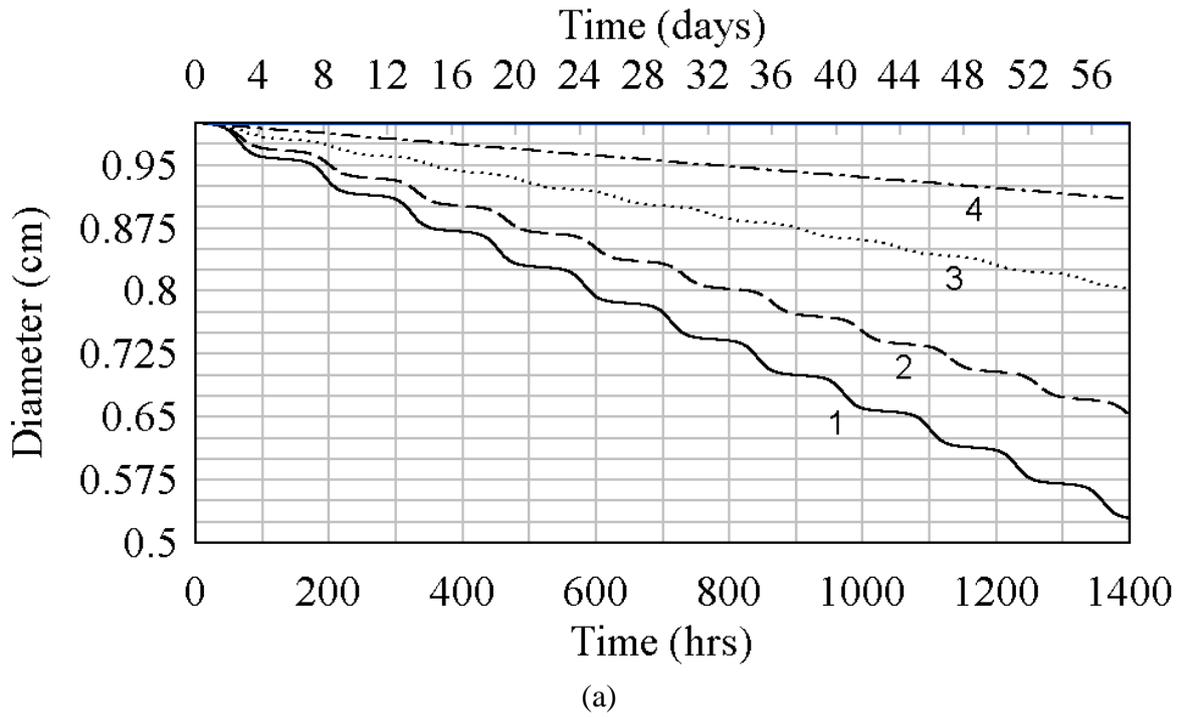

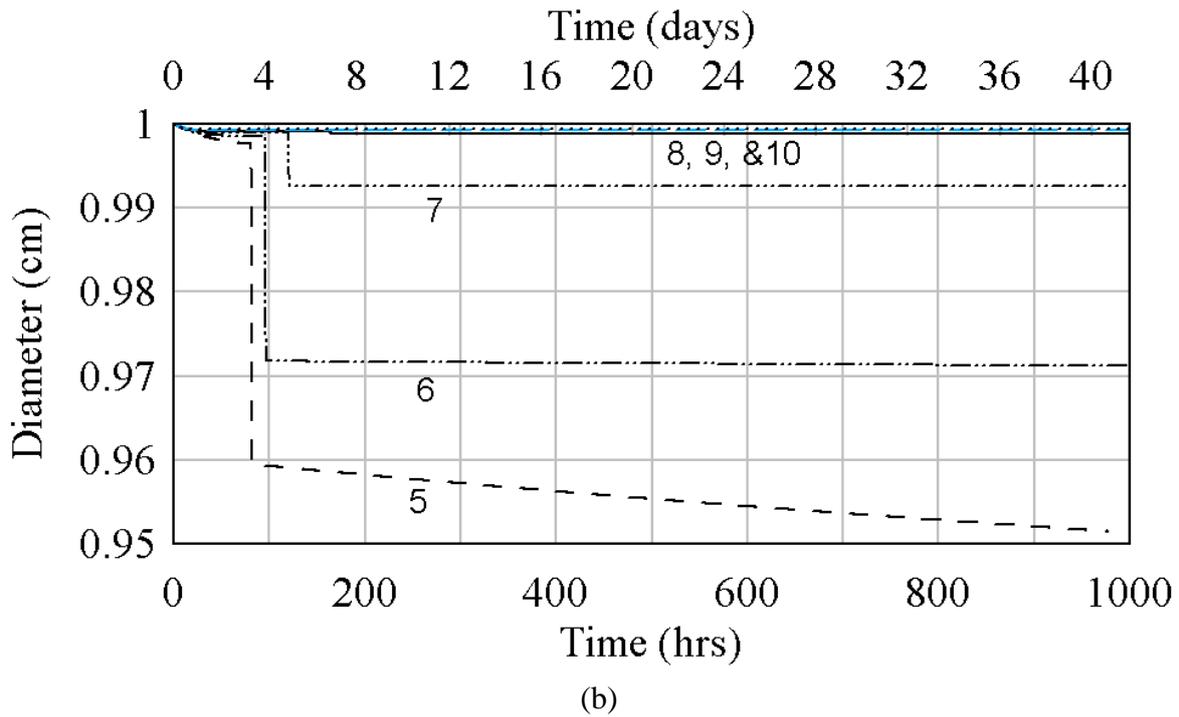

Fig. 33: (a) Evaporation of the CAI on the orbits within the inner truncation radius of the solar accretion disc. These particles will eventually evaporate completely on a timescale of months to years. (b) Evaporative loss due to re-entry for particle 5, 6, and 7. The remaining particles are unaffected by the re-entry heating.

These simulations indicate that if the cm-sized, bowl-shaped CAIs and WLRs were formed via this re-entry mechanism then they formed in a hydrogen rich environment with densities in the range of $10^{-7}$ to $10^{-4}$ kg m$^{-3}$, over a timescale of order one hour with a rate of temperature change of around 1000 K hr$^{-1}$.

## Appendix C.2    $3\times10^{-8}$ M$_\odot$yr$^{-1}$, 1 mm particles, $R_{co} > R_t$.

In this simulation we keep everything the same as in Appendix C.1 except we are now using 1 mm diameter particles. This is due to the observed size distribution of CAIs which range from 100 microns to 1 cm in diameter (MacPherson et al. 2005). As expected, the flight paths and the subsequent temperature cooling rates remain the same (Fig. 25, Fig. 31 and Fig. 32), but the stopping times for the CAIs as they enter the disc are much shorter. As a consequence, the re-entry temperature spikes are significantly smaller (up to 500K smaller) relative to the 1 cm diameter case (Fig. 34). None-the-less, the length of the heating events is of order one hour or more in duration.

Due to the smaller size of the particles, the CAIs on the inner orbits evaporate on a timescale of about a week to a month. However, the ejected particles do not suffer significant evaporation during their re-entry into the protosolar disc (Fig. 35).

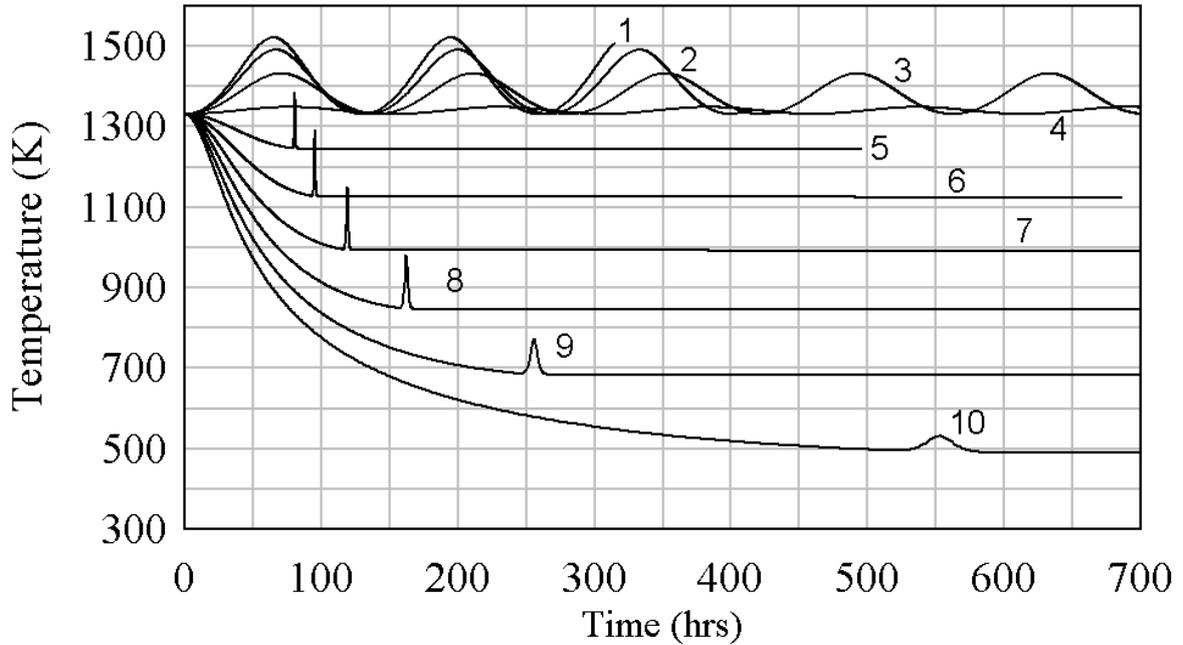

Fig. 34: Particle temperatures as a function of time for the $\dot{M}_a = 3\times10^{-8}$ M$_\odot$yr$^{-1}$ case. The 1 mm CAIs start with a temperature of around 1330K. The particles on inner bound orbits tend to evaporate relatively quickly. For the particles that move away from the protosun, their temperature decreases until they reach the denser regions of the disc, where frictional heating produces a spike in particle temperature of around 100K or less, which lasts for an hour or so. These temperature spikes are significantly smaller compared to those obtained for the 1 cm CAIs (Fig. 27).

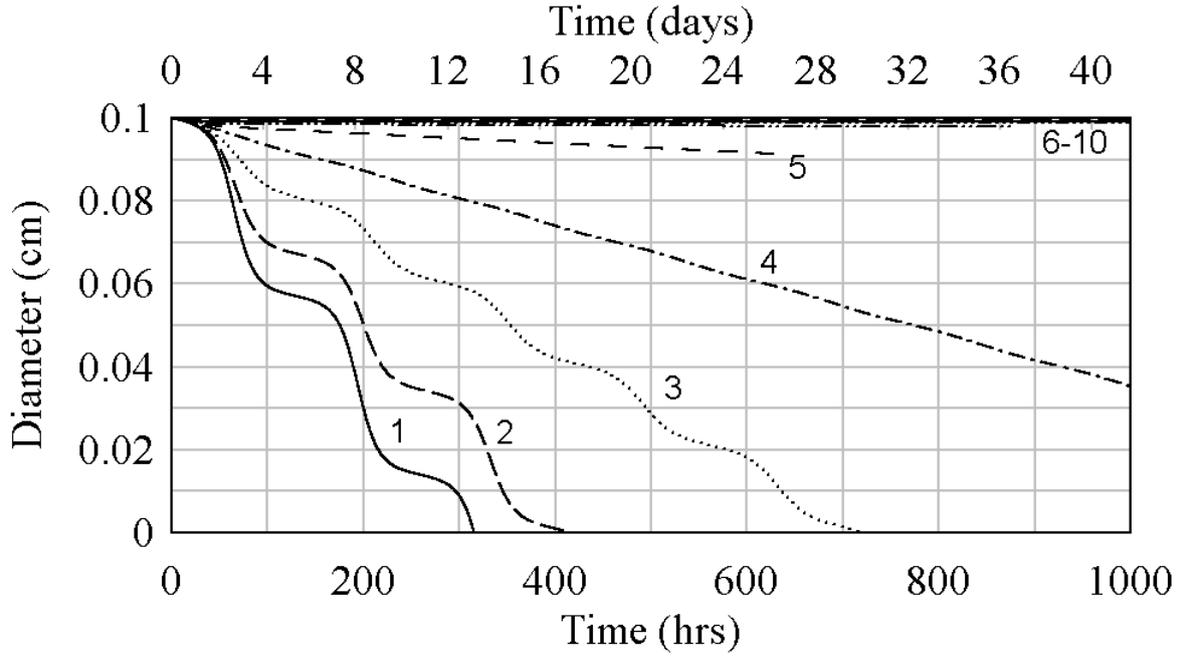

Fig. 35: Evaporation of the CAI on the orbits within the inner truncation radius of the solar accretion disc. These particles will eventually evaporate completely on a timescale of weeks to months. The remaining particles are unaffected by re-entry heating.

## Appendix C.3   $5\times10^{-8}$ M$_\odot$yr$^{-1}$, 1 cm particles, $R_{co} > R_t$.

We now increase the accretion rate from $3\times10^{-8}$ M$_\odot$yr$^{-1}$ to $5\times10^{-8}$ M$_\odot$yr$^{-1}$. This decreases the inner truncation radius from $R_t = 0.0675$ au to $R_t = 0.0583$ au, while the co-rotation radius remains unchanged at 0.07 au. The resulting keplerian speed at the inner truncation radius is 120 km s$^{-1}$, while the co-rotation speed is, approximately, 90 km s$^{-1}$.

It is assumed that the gaseous material at the inner truncation radius will tend to corotate with the stellar magnetic field. As such, we again assume that the ejected particles have an angular speed equal to the co-rotation speed, while their vertical speed, perpendicular to the midplane of the accretion disc, is 5%, 15%, 25%, 35%, 45%, 55%, 65%, 75%, 85% and 95% of the keplerian speed. The particles obtain their hypothetical vertical speed due to the material riding up the stellar magnetic field lines as disc material accretes onto the protosun. The resulting projectile paths are shown in Fig. 36, where we see that most of the particles are not ejected from the inner disc and are instead trapped in orbits that have perihelia which are within the inner truncation radius of the disc. The remaining particles are ejected from the inner accretion disc and eventually re-enter the disc at distances further away from the protosun. The resulting temperatures of the particles are displayed in Fig. 37.

Most of the particles that are not ejected are simply vaporized by the radiation of the protosun as the particles move closer to the protosun. One ejected particle is vaporised on re-entry, while the remaining two ejected particles suffer from partial vaporisation (Fig. 38).

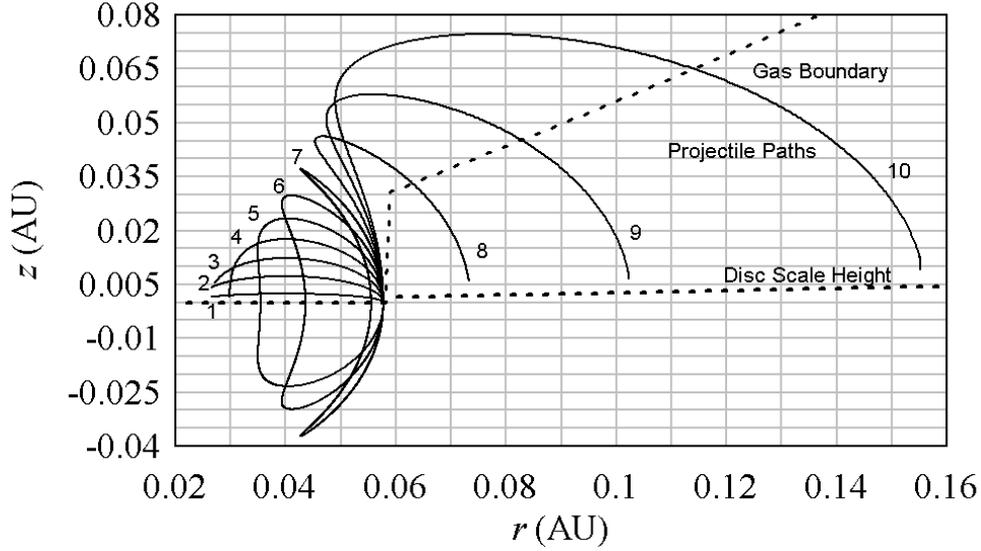

Fig. 36: Projectile paths for the case of $\dot{M}_a = 5\times10^{-8}$ $M_\odot\,yr^{-1}$ for the protosun at an age of $1.2\times10^6$ years. The disc scale height (Eq. (7)) is shown as is the gas boundary, where the density of the gas is effectively zero. The particles are labelled as per the initial $z$ speed, where $v_{1z} = 6$ (km s$^{-1}$), $v_{2z} = 18$, $v_{3z} = 30$, $v_{4z} = 42$, $v_{5z} = 54$, $v_{6z} = 66$, $v_{7z} = 78$, $v_{8z} = 90$, $v_{9z} = 102$ and $v_{10z} = 114$ km s$^{-1}$. The escape speed is 120 km s$^{-1}$. All the particles are launched from just within the inner truncation radius of $R_t = 0.058$ au and they have an initial angular speed of 90 km s$^{-1}$ obtained from the co-rotating magnetic field. Most of the particles are not ejected from the inner disc and they are trapped in orbits that bring them closer to the protosun. The ejected particles are subject to gas drag (Eq. (31)) and frictional heating (Eq. (48)) when they re-enter the disc.

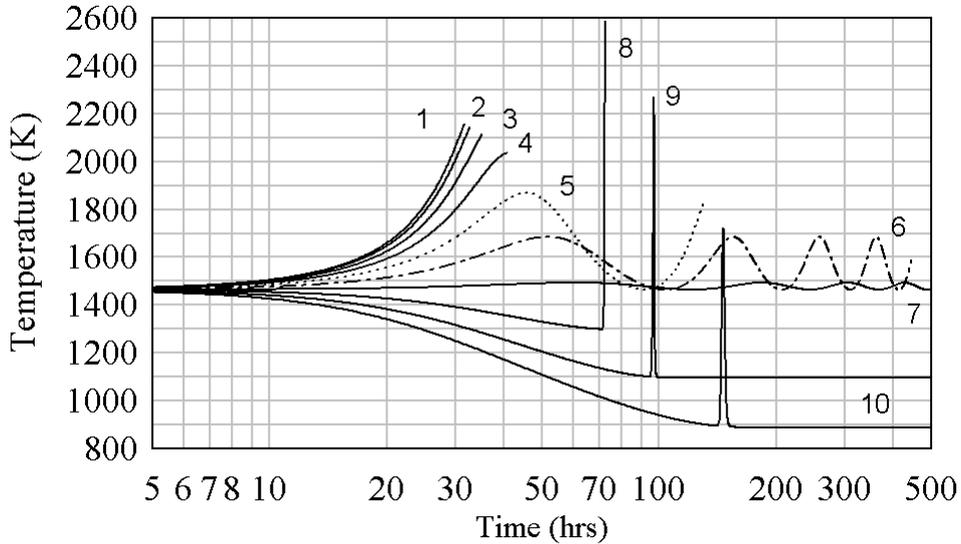

Fig. 37: Particle temperatures as a function of time for the $5\times10^{-8}$ $M_\odot\,yr^{-1}$ case. The 1 cm CAIs start with a temperature of around 1464K. Particles not ejected from the disc are subject to increasing temperatures as they approach the protosun and many are quickly evaporated. As the ejected particles move away from the protosun, their temperature decreases until they reach the denser regions of the disc, where frictional heating produces a spike in particle temperature which may last for an hour or more. Significant frictional heating occurs when the CAIs encounter the denser gas regions of the disc at hypersonic speeds. In this case, one particle is vaporised upon re-entry by the sharp increase in temperature.

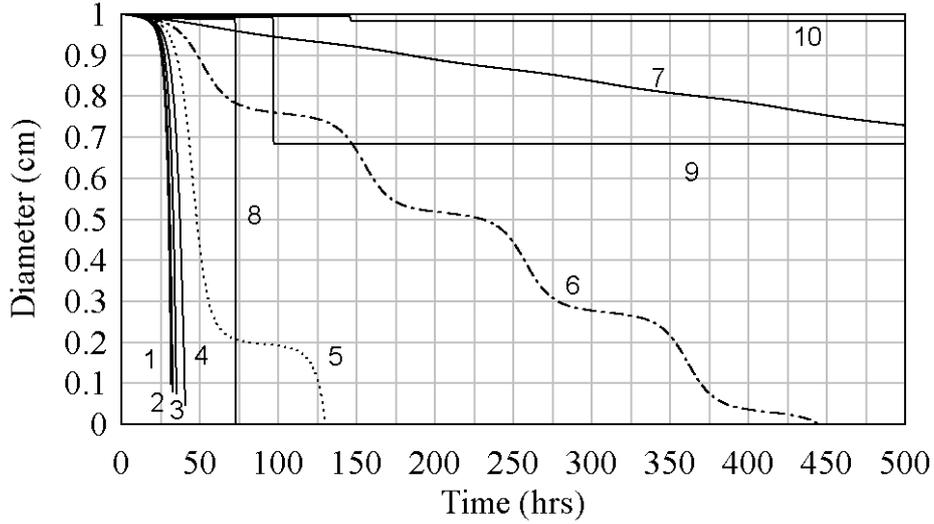

Fig. 38: The diameter of the CAIs as a function of time. Particles 1 to 7 are on orbits within the inner truncation radius of the solar accretion disc. Particles 1 to 4 are destroyed within hours. Particles 5 to 7 are evaporated completely on a timescale of days to months. Particle 8 is ejected from the inner disc and is destroyed upon re-entry. Particle 9 loses over 30% of its diameter upon re-entry, while particle 10 survives re-entry relatively unscathed.

## Appendix C.4    $\dot{M}_a = 5 \times 10^{-8}\,\mathrm{M}_\odot\,\mathrm{yr}^{-1}$, 1 mm particles. $R_{co} > R_t$

For this simulation, we keep all the values the same as shown in Appendix C.3, except we now use 1 mm diameter particles instead of 1 cm particles. All the projectile paths remain the same as in Fig. 36, but the particles on inbound orbits are evaporated faster, while all the outbound particles survive the re-entry into the disc with significantly lower re-entry temperatures (Fig. 39).

All the outbound particles suffer a reduction in size due to the re-entry process. For example, particle 8 reduces in diameter from 1 mm to 0.66 mm, while particles 9 and 10 change from 1 mm diameter to 0.92 mm and 0.94 mm, respectively. Thus the reduction in diameter from 1 cm to 1 mm reduces the re-entry temperature and decreases the relative mass loss due to evaporation upon re-entry.

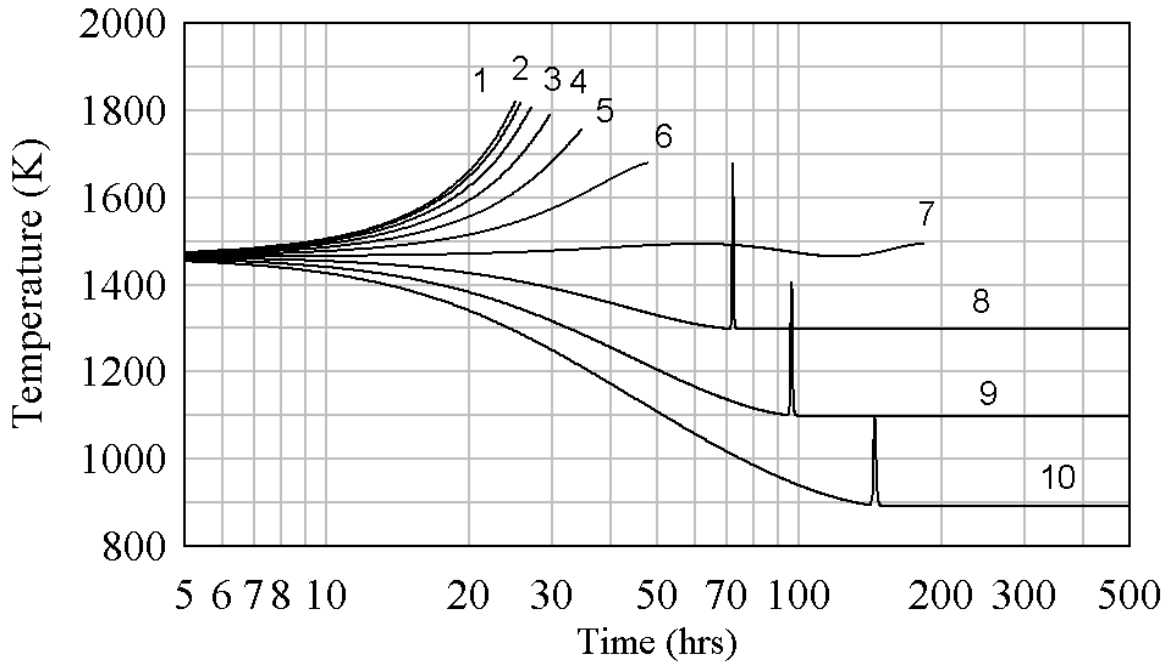

Fig. 39: 1 mm diameter particle temperatures as a function of time for the $5\times10^{-8}$ $M_\odot yr^{-1}$ mass accretion case. The 1 mm CAIs start with a temperature of around 1464K. Particles not ejected from the disc are subject to increasing temperatures as they approach the protosun and are quickly evaporated. All the ejected particles survive their re-entry into the Solar Protoplanetary Disc with re-entry temperatures that are significantly smaller relative to the 1 cm diameter particles.

## Appendix C.5    $2\times10^{-8}$ $M_\odot yr^{-1}$, 1 cm particles, $R_{co} < R_t$

We decreased the accretion rate from $3\times10^{-8}$ $M_\odot yr^{-1}$ to $2\times10^{-8}$ $M_\odot yr^{-1}$. Due to the influence of the solar magnetosphere, this increases the inner truncation radius from $R_t = 0.0675$ au to $R_t = 0.076$ au, while the co-rotation radius remains unchanged at 0.07 au. The resulting keplerian speed at the inner truncation radius is now 106 km s$^{-1}$, while the co-rotation speed is, approximately, 116 km s$^{-1}$. In this case, the co-rotation speed is faster than the local keplerian speed and all the co-moving particles that are located in the co-rotating gas should move on a path away from the protosun.

The particles are initially located on the disc midplane, just within the inner truncation radius of the disc, and are given initial vertical speeds equal to 5%, 15%, 25%, 35%, 45%, 55%, 65%, 75%, 85% and 95% of the keplerian speed. The resulting projectile paths are shown in Fig. 40, where we find that particles 1 and 2 (which have 5 and 15%, respectively, of the keplerian speed) plough directly into the surrounding disc and are immediately vaporised. Particle 10 (given an initial vertical speed equal to 95% of the keplerian speed) is completely ejected from the system, while the remaining particles re-enter into the disc.

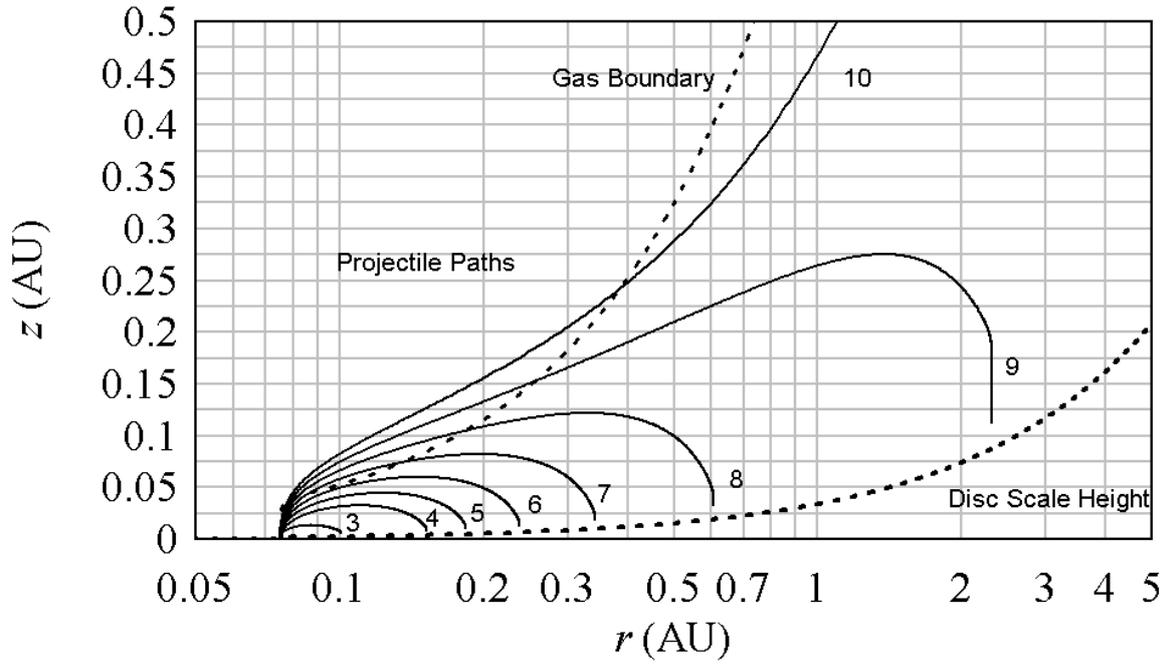

Fig. 40: Projectile paths for the case of $\dot{M}_a = 2 \times 10^{-8}$ $M_\odot \text{yr}^{-1}$ for the protosun at an age of $1.2 \times 10^6$ years. The disc scale height (Eq. (7)) is shown as is the gas boundary, where the density of the gas is effectively zero. The particles are labelled as per the initial $z$ speed, where $v_{1z} = 5.3$ (km s$^{-1}$), $v_{2z} = 15$, $v_{3z} = 26$, $v_{4z} = 37$, $v_{5z} = 47$, $v_{6z} = 58$, $v_{7z} = 58$, $v_{8z} = 79$, $v_{9z} = 89$ and $v_{10z} = 100$ km s$^{-1}$. All the particles were launched from the inner truncation radius of $R_t = 0.076$ au and they have an initial angular speed of 116 km s$^{-1}$ obtained from the co-rotating magnetic field. All the particles are ejected from the inner disc. Particles 1 and 2 are vaporised as they collide with the inner disc soon after initial ejection. Particle 3 is partially vaporized by the same process, but survives to reenter the disc. All the remaining particles reenter the disc without significant damage, apart from particle 10, which is completely ejected from the disc.

The temperature of the particles as a function of time is shown in Fig. 41. All the particles have an initial temperature of 1241K, but particles 1 and 2 suffer gas friction temperatures of over 7000K and are vaporised as they move directly into the disc. Particle 3 survives this initial heat pulse and suffers a much smaller heat pulse upon subsequent re-entry into the denser regions of the gas disc. All the other particle have small re-entry heat pulses, except for particle 10, which is simply ejected from the Solar System. Apart from particles 1 and 2, all the particles undergo minimal or no mass loss during their journey.

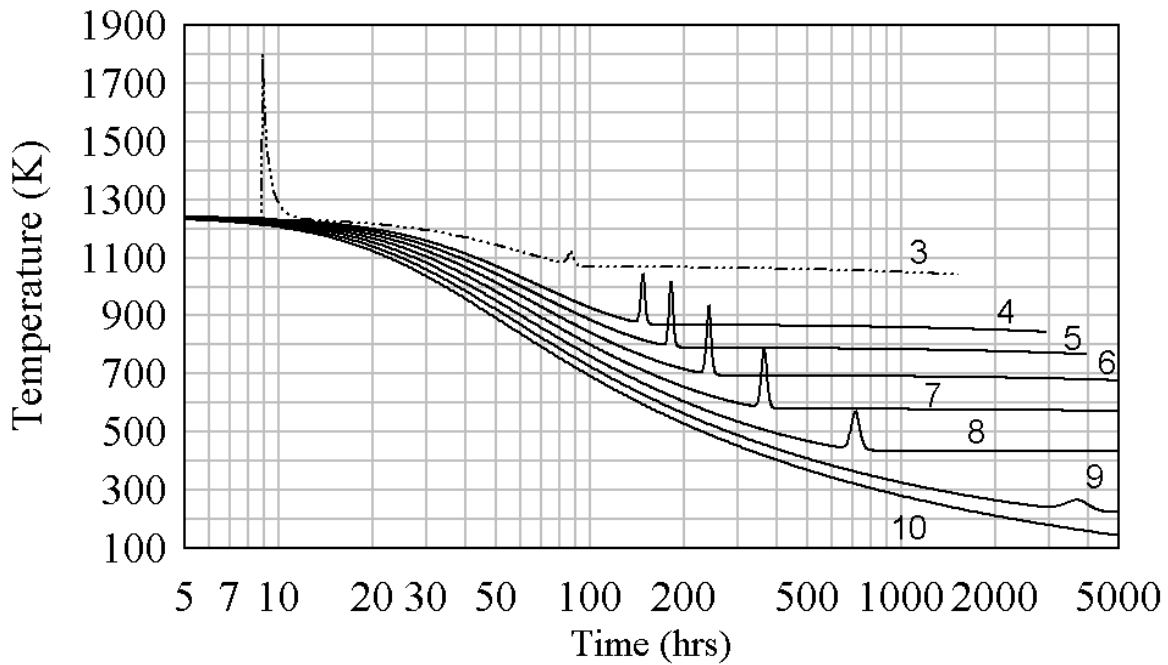

Fig. 41: Particle temperatures as a function of time. The 1 cm CAIs start with a temperature of around 1241K. Particles 1 & 2 are vaporised as they slam directly into the inner disc wall. Particle 3 survives the initial heat pulse as it punches through the inner disc wall and, as with the other surviving particles, cools as it moves away from the protosun. All the particles, apart from particle 10, undergo a heat pulse upon re-entry into the disc. Particle 10 is ejected from the Solar System and cools to interstellar temperatures as it moves away from the protosun.